\documentclass[a4paper,12pt,oneside]{article}
\usepackage[latin1]{inputenc}
\usepackage{amsmath}
\usepackage{amsfonts}
\usepackage{amssymb}
\usepackage{graphicx}
\usepackage{footnote}
\usepackage{float}
\usepackage{subcaption}
\DeclareGraphicsExtensions{.bmp,.png,.pdf,.jpg,.eps}
\usepackage{physics}
\usepackage{mathrsfs}

\usepackage[colorlinks]{hyperref}
\hypersetup{
    colorlinks=true,
    linkcolor=blue,
    filecolor=magenta,  
    urlcolor=[rgb]{0.00,0.00,0.50},    
    citecolor=red,
    }

\usepackage{url}

\advance \textheight by \topskip
\usepackage{amsmath}
\usepackage[english]{babel}
\makeatletter
 \@addtoreset{equation}{section}
\makeatother
\usepackage[latin1]{inputenc}
\usepackage{amsmath}
\numberwithin{equation}{section}

\advance \textheight by \topskip
\usepackage{amsmath}
\usepackage[english]{babel}
\DeclareMathAlphabet\mathbfcal{OMS}{cmsy}{b}{n}
\DeclareMathAlphabet{\boldmathe}{T1}{cmr}{bx}{it}
\newcommand{\mbf}[1]{\boldmathe{#1}}

\def\vr{\mbf{r}}

\def\be{\begin{equation}}
\def\ee{\end{equation}}

\def\R{\mathbb R}

\def\be{\begin{equation}}
\def\ee{\end{equation}}

\def\R{\mathbb R}

\def\vr{\mbf{r}}

\def\be{\begin{equation}}
\def\ee{\end{equation}}

\def\R{\mathbb R}

\usepackage{todonotes}

\begin{document}

\title{\bf Conformal bridge\\ in a cosmic string background }

\author{{\bf  Luis Inzunza and Mikhail S. Plyushchay} 
 \\
[8pt]
{\small \textit{Departamento de F\'{\i}sica,
Universidad de Santiago de Chile, Casilla 307, Santiago,
Chile  }}\\
[4pt]
 \sl{\small{E-mails:   
\textcolor{blue}{luis.inzunza@usach.cl},
\textcolor{blue}{mikhail.plyushchay@usach.cl}
}}
}
\date{}
\maketitle

\begin{abstract}
Hidden symmetries of non-relativistic  $ \mathfrak{so} (2,1) \cong \mathfrak{sl} (2, \R) $ invariant 
 systems  in a cosmic string background are studied using the conformal bridge transformation. 
Geometric properties of this  background are analogous 
to those of a conical surface  with 
a deficiency/excess 
angle 
encoded in the ``geometrical parameter" $\alpha$, 
determined by the linear positive/negative 
mass  density of the string. 
The free particle and the harmonic oscillator
on this background
are shown to be related
by the conformal bridge transformation.
To identify the  
integrals of the free system, we employ a local canonical transformation 
that relates the model with its planar version.
The conformal bridge transformation is then 
used to map the obtained integrals 
to those of the harmonic oscillator on the cone.
Well-defined classical  integrals in both models
 exist only at  $\alpha=q/k$ with $q,k=1,2,\ldots,$ 
 which  for $q>1$ are higher-order
 generators  of  finite nonlinear algebras. 
 The systems are quantized for arbitrary values of $\alpha$; however, 
the well-defined hidden symmetry operators associated  with
spectral degeneracies   only exist
when  $\alpha$ is an integer, that reveals a quantum anomaly.

\vskip.5cm\noindent
\end{abstract}

\section{Introduction}

Special characteristics of classical systems are reflected
 by the
conserved quantities that canonically  generate 
the
symmetry transformations.
At the quantum level, these quantities 
are promoted to the operators 
that 
carry the spectrum information.
In this context, hidden symmetries are associated 
with the
non-obvious integrals of motion that 
mix canonical
coordinates and 
momenta
 in a non-trivial way at the 
classical level, while at the quantum level they are often responsible for the so-called ``accidental degenerations" \cite{Cariglia}. 
Integrals of this type are higher-order functions of the canonical momenta, 
and they may 
generate
nonlinear symmetry algebras \cite{nonlinear2,nonlinear3}. Some examples of integrals related to hidden symmetries
are the Laplace-Runge-Lenz vector for the Kepler-Coulomb problem \cite{Pauli}, 
the Fradkin tensor for the isotropic harmonic oscillator
\cite{Frad}, 
the 
analogs of these integrals
 in a monopole background \cite{PlyWipf,IPW2}, the higher-order symmetry generators of the 
 anisotropic harmonic oscillator with commensurable frequencies \cite{nonlinear2},
and the $ N $ integrals in involution in the Calogero models of $N$ particles 
\cite{Woj,Kuz}. 

On the other hand, it is a quite  general assertion  that 
the systems  
in a curved space-time show special properties 
related to the geometric background itself. 
Some  examples in this direction are the Hawking radiation
\cite{Hawking}, the Unruh effect \cite{Unruh}, 
the conformal invariance of a 
charged particle propagating near the horizon of
the extreme Reissner-Nordstr\"om black hole,
which  
attracted attention in the context 
 of AdS/CFT 
 correspondence \cite{ConformalBH0,ConformalBH1,ConformalBH2,ConformalBH3} 
 (that, in turn, gave rise to a resurgence of interest to the 
 conformal model of de  Alfaro, Fubini and Furlan \cite{AFF}).  
 In a different  but related line of research 
 we also mention here the 
 supersymmetric mechanics models
 on curved spaces  whose construction 
 is based on 
 the WDVV equation 
 formalism 
 \cite{GPL,Kozyrev}. 
 
In the context of the present research, 
the study of classical and quantum dynamics in curved spaces is interesting from the point of view of hidden symmetries.
 An important result in this direction 
 is the proof 
 of the existence of a non-trivial conserved quantity that characterizes 
 the dynamics of a particle moving on a Kerr black hole background. 
 This quantity, known as the Carter integral \cite{Carter}, 
 is responsible for the complete integrability of the system.
Another  interesting result 
 was reported in \cite{Gibbons}, 
where it was shown that spinning particles
in the Kerr-Newman black hole background
are characterized by enhanced supersymmetry with 
additional supercharges
of a  nature different from a square root of the Hamiltonian of the system.
Geometrically, such supercharges are associated 
with  the so-called Yano and conformal Yano tensors \cite{Cariglia2,Frolov,Frolov2,Frolov3}. 
Some other examples related to the particle dynamics in curved spaces 
can be found in references
\cite{tHooft,DesJack,Kay,Furtado,Coelho,Barros,Example1,Kowalski,Example2,Example3,Example4}.

In this article, we address the problem of studying the 
classical and quantum
symmetry
aspects 
of some conformal-invariant non-relativistic particle systems  on a cosmic string background.
These strings are topological 
defects whose creation in the early universe is predicted by quantum field theory arguments \cite{Kibble,Vilen,Vilenkin,Dow},
and which, on the other hand, also appear  in  condensed matter physics
 and wormholes \cite{Visser,Cramer,KatVol,Volovik,Manton}. 
Their  effect is to introduce a conical singularity in the spatial part of the space-time metric \cite{SokSta,Volovik}.

The effects of the presence of the cosmic string was examined in different physical contexts, see, e.g., refs. 
\cite{Aryal,LandCon,Solod,Germano,deMello,Kay,Example1,Example2,Example3,Example4}. 
Our goal
here is to study
the influence of the geometrical properties of this background 
(encoded in a ``geometrical parameter" $\alpha$
 given in terms of the linear mass density of the string)
 on the dynamics of the systems from
  the perspective of   well-defined integrals of motion in
 the phase space when considering classical cases,
 and   well-defined  symmetry operators for 
 the corresponding quantum versions of the systems.
 We are interested in the case of the free particle in the cone, 
 which has the $ \mathfrak {so} (2,1) $  conformal symmetry, 
and in the harmonic oscillators system in the same  geometry, 
which is characterized by 
the $ \mathfrak{sl} (2, \R)\cong \mathfrak {so} (2,1)$ conformal Newton-Hooke
symmetry \cite{Nied2,NewHook,NH1,NH2,IPW}. 
The key construction in
our investigation 
 is the so-called conformal bridge transformation \cite{IPW}, which in general is a mapping that 
allows us to transform the complete set of symmetry generators
from an $ \mathfrak{so} (2,1) $ invariant asymptotically free system, to those of a harmonically confined model 
with the $\mathfrak{sl}(2,\R) $ conformal symmetry. 
The plan is to first characterize
 the free case by identifying 
the corresponding classical and quantum integrals of motion for different values of $ \alpha $,
and then to obtain the complete information on the harmonic oscillator system 
through the  conformal bridge transformation. 

The free dynamics and the 
harmonic oscillator system in conical geometry were already studied in the literature, but under a different 
perspective. In  \cite{tHooft,DesJack}
the classical and quantum scattering on the cone were considered.
There, in particular,  it was noticed  that for the geodesic motion,  there are special
 values of the deficiency angle for  which the particle experiences 
a backward scattering, while for other values
the particle continues to move in the original direction, making several 
revolutions  around  
 the cone vertex. For the harmonic oscillator system, the structure of the wave-functions and the spectrum 
 for arbitrary values of $ \alpha $ were  investigated in
\cite{Furtado,Coelho,Kowalski}. None of these works,
 however, studied the systems in the light of hidden 
 symmetries  which appear for special values of $\alpha$.

A notable result we present here 
is that for the free particle on the cone characterized by the parameter 
$\alpha$,  
there are well-defined integrals of motion in the phase space 
only when  $\alpha$  is a rational number.
In the general case of rational $\alpha$  they are of higher-order in 
momenta and produce a 
finite nonlinear algebra. 
At the quantum level we 
reveal  a quantum anomaly, because although the system can be quantized 
for any real value of $ \alpha$,
the
well-defined hidden symmetry operators in Hilbert physical space can only be constructed 
when this 
geometric parameter is an integer.  
 The same peculiarities
 appear
 for the harmonic oscillator system, whose  
 classical trajectories
 are closed for rational values of $\alpha$, and at the quantum level, 
 only the cases with its integer values are anomaly-free, 
 and   the  hidden symmetry operators 
 reflect then  the corresponding  degeneracies of the spectrum.

\vskip0.1cm
The article is organized as follows.
 We first review the geometry related to a cosmic string background in 
Sec. \ref{SecGeo}  by explicitly showing that
its spatial part  takes the form
of a two-dimensional conical metric, 
whose  
parameter 
is defined 
by the mass density of the string. 
We also show that there is a set of local coordinates which allows us to formally 
relate the conical metric to that of the Euclidean plane.
In Sec. \ref{SecEmotion}, we consider some important aspects of the
non-relativistic  free particle and the harmonic oscillator dynamics in $\R^2$. 
This section serves as the basis from which we can build the symmetry algebra of the 
corresponding versions of these systems in conical geometry by 
using a locally defined canonical transformation. In Sec. \ref{SecConBrid}, we explain
 how the conformal bridge transformation works. 
As an example, it is shown how to relate the Euclidean 
free particle to the planar isotropic harmonic oscillator
by means of this mapping. We also establish there the
nontrivial relation between  the hidden symmetry generators
of these two systems 
and some  sub-algebras generated by them.
In Sec. \ref{SecFreeCone}
we study the free particle on a cosmic string background. 
Using the above-mentioned canonical transformation, we obtain the 
explicit form of the solutions of the equations of motion, 
as well as some conserved quantities which are generally of a formal nature, 
but which serve to construct well-defined integrals in the phase space when the  
geometric parameter is a rational number. At the quantum level, 
we construct the symmetry operators and observe the 
 quantum anomaly by analyzing the action of these 
operators  on the physical eigenstates. 
In Sec. \ref{SecHar}, we employ
 the conformal bridge transformation to get the symmetry generators of the harmonic oscillator 
from the free particle system in the classical and quantum 
 cases. The properties of symmetry algebra for the 
harmonically confined system, as well as the manifestation of the same quantum anomaly
 are immediately obtained
 by using  the same transformation.
Finally, in Sec. \ref{SecDisOut} the 
discussion of the obtained results and outlook are presented.

\section{Cosmic string and conical geometry}
\label{SecGeo}

Cosmic strings are hypothetical
  one-dimensional topological defects which may have
formed in 
the spontaneous symmetry-breaking 
phase transition in the expanding Universe 
\cite{Kibble,Vilen,Vilenkin,Dow}. 
Such defects also are familiar 
in condensed matter  physics  
 \cite{KatVol,Volovik}. 
In this article we are interested in analyzing the 
symmetries of a non-relativistic conformal dynamics, at the classical and 
quantum levels, of a  particle
 in a two-dimensional space with properties of 
 a cosmic string background.
In this section, we
show that this problem is analogous to 
studying the dynamics of a particle in a conical geometry 
 \cite{tHooft,DesJack},
and discuss some of its general properties 
which will be important for  the subsequent  analysis.

Following \cite{SokSta,Vilenkin}, 
the  
solution of the Einstein field  equations 
in (2+1)  dimensions associated with cosmic string 
is described by
the metric 
\begin{eqnarray}
&
dS^2=-c^2 dt^2+ds^2\,,&\label{dS2}\\
&
\label{EinsteinG}
ds^2=\left(1-\frac{8\mu G}{c^2}\ln(\frac{r}{r_0})\right)(dr^2+r^2d\varphi^2)\,,
&
\end{eqnarray}
where  $G$ is the Newton constant,
$c$ is the speed of light, $\mu$ is the linear mass density of the cosmic string,
 and $r_0$ corresponds to its radius. 
By introducing the new coordinate
\begin{eqnarray}
&
r'^2=\left(1-\frac{8\mu G}{c^2}\ln(\frac{r}{r_0})\right)r^2\,,\qquad 
\alpha ^2 dr'^2= \left(1-\frac{8 \mu G}{c^2}\ln(\frac{r}{r_0})\right)dr^2\,,
&\\&
\label{Galpha}
 \alpha=\frac{1}{1-\frac{4\mu G}{c^2}}>0\,,
&
\end{eqnarray}
where  higher-order terms in $\mu G/c^2$ are neglected in the computation of  $dr'^2$, 
the spatial part 
(\ref{EinsteinG})  
of the cosmic string metric (\ref{dS2}) 
takes the form
\begin{eqnarray}
\label{metric3}
ds^2= \alpha^2 dr^2 + r^2 d\varphi^2\,,
\end{eqnarray}
after changing 
the notation $r'\rightarrow r$. 
Eq. (\ref{metric3}) 
corresponds to the two-dimensional metric of a conical geometry.
For $\alpha>1$, 
it can be obtained by reducing 
the three-dimensional Euclidean metric 
\begin{equation}
ds_{E}^2= dr^2+r^2d\varphi^2+dz^2\,,
\end{equation}
to the surface of a cone
$z=\lambda r$, 
and identification 
$\alpha^2=1+\lambda^2$.
Here $\lambda=\cot\beta$, $\beta$ is the aperture angle of the cone,
and  
in 
accordance with (\ref{Galpha}), 
this metric is associated to a cosmic string whose mass density is positive. 
On the other hand, the
 metric (\ref{metric3}) with 
 $0<\alpha<1$
 can be 
 obtained  from 
the $(2+1)$-dimensional Minkowski space 
\begin{equation}
\label{metric2}
ds_{M}^2=-c^2d\tau^2+ dr^2+r^2d\varphi^2\,
\end{equation}
by reduction to the cone surface
 $c\tau=\lambda r$, $0< \lambda< 1$,
  $\alpha^2=1-\lambda^2$. 
  Formally, this  is the reduction 
  to
  the non-causal part of Minkowski space, where 
  particle's 
  velocity 
  is
  greater than $ c $. 
  In
  correspondence with (\ref{Galpha}),
   these spaces are associated with 
  cosmic strings that have  negative mass density
  \cite{Visser}. 
   The case of the conic metric with
   $0<\alpha<1$  
  also  
  describes topological defects in the physics of condensed matter and  wormholes
    \cite{Visser,Cramer,KatVol,Volovik,Manton}. 

From here one sees 
that studying a
free non-relativistic particle
 system in  a 
cosmic string background geometry with a given  value of the parameter $\alpha$
is analogous to solving the problem of the dynamics of a particle on the surface of a cone
described by the action
$
S=-mc^2\int\sqrt{1-\frac{1}{c^2}\left(\frac{ds}{dt}\right)^2}\,dt
$
in the non-relativistic limit $c\rightarrow \infty$.
 In
  the remainder of 
  the
   section, we explore some 
  geometric properties of the conical space
  (\ref{metric3}).  
For this, 
we express the metric $ ds^2 $ in different coordinate systems,
that
 will allow us to 
 clearly see the singularity at the origin, 
 and clarify the corresponding 
  conformal properties of the metric (\ref{dS2}).

First,  we note that
 in Cartesian coordinates, 
 \be
ds^2=\frac{1}{(x^2+y^2)}\left( \alpha^2( xdx +ydy)^2+(xdy-ydx)^2     \right)\,,
\ee
the singularity of the metric 
at the origin $ x = y = 0 $ for $ \alpha \neq 1 $
 becomes apparent.

Introducing a
new radial coordinate $r=r_0e^{\frac{\rho}{\alpha}}$,
 the metric becomes  
\be
\label{Isometric}
ds^2= r_0^2e^{\frac{2\rho}{\alpha}}(d\rho^2+d\varphi^2)\,.
\ee
The variables $\rho$ and $\varphi$ correspond to the isothermal coordinates, 
and we note that when  $\alpha\rightarrow\infty$, the metric  (\ref{Isometric})
transforms into a cylinder's metric. 
From (\ref{Isometric}), as well as from (\ref{metric3}), 
 the invariance of the metric 
 under rotations, 
 $\varphi\rightarrow\varphi+\varphi_0$, is obvious. 
 The corresponding Killing vector 
  is $\frac{\partial}{\partial \varphi}$. 
Metric (\ref{Isometric}) also is conformally invariant 
   under transformations
  $\rho\rightarrow\rho+\rho_0$. In polar coordinates, this  corresponds to dilatation
    in the radial coordinate 
 generated by
  the  
  conformal  Killing vector 
$\frac{\partial}{\partial \rho }=\frac{r}{\alpha}\frac{\partial}{\partial r}\,.$

By introducing  a pair of ``regularized" Cartesian coordinates  
\begin{equation}
\label{Xi}
X_1=\alpha r\cos \Phi \,,\qquad
X_2=\alpha r\sin\Phi\,,
\qquad \Phi=\varphi/\alpha\,,
\end{equation} 
the metric (\ref{metric3}) is transformed into
 \begin{equation}
 \label{EqmetricX}
 ds^2=dX_1^2+dX_2^2\,.
 \end{equation}
This
 formally 
looks like
the metric of the Euclidean plane in Cartesian coordinates, but 
 $0\leq \Phi <2\pi/\alpha$ in  (\ref{Xi}), and the edges $\Phi=0$ and $\Phi=2\pi/\alpha$
 have to be identified, that results in a conical singularity.
This  singularity reveals itself, particularly,  in 
Riemann curvature tensor concentrated at $r=0$:
$\mathcal{R}^{r\varphi}_{r\varphi}=2\pi(1-\alpha^{-1})\delta(X_1)\delta(X_2)$ 
\cite{SokSta,Volovik}.

To clarify further 
the nature of coordinates (\ref{Xi}), consider the complex combination
\be
w=X_1+iX_2=\alpha re^{i\varphi/\alpha}\,,\qquad
ds^2=dwdw^*\,.
\ee
From here it is seen that there may be problems for arbitrary values of $ \alpha $ 
due to the exponential  factor and the 
 associated
branch point.
When  the rational case 
 $\alpha={q}/{k}$ with $q,k=1,2,\ldots$, is considered, 
one can use instead the new coordinates 
$\zeta=\zeta_1+i\zeta_2$,
\begin{equation}\label{zeta}
\zeta= w^{q}=\left(\frac{qr}{k}\right)^q e^{ik\varphi}\quad \Rightarrow \quad 
\zeta_1=\left(\frac{qr}{k}\right)^q \cos(k \varphi)\,,\qquad 
\zeta_2=\left(\frac{qr}{k}\right)^q \sin(k \varphi)\,.
\end{equation}
In their terms  the metric 
reads 
\begin{eqnarray}
ds^2=\frac{d\zeta_1^2+d\zeta_2^2}{q^2(\zeta_1^2+\zeta_2^2)^{1-\frac{1}{q}}}\,.
\end{eqnarray}
For
 $q=1$ $\Rightarrow \alpha=1/k$, it seems that
there is no singularity in the metric,  and 
the
coordinates $\zeta_1$, $\zeta_2$
themselves reveal  no singularity. However,  as it is seen from (\ref{zeta}),
in this case 
 $\zeta_1$ and $\zeta_2$ cover conical space $k$ times
 when $\alpha=1/k$ . The picture is similar to a
 Riemann surface for the function $w=z^{1/k}$
 where we pass from its one sheet to another when angle
 increases in $2\pi$, while here the transition from 
 one sheet to another happens each time when $\varphi$ increases  in $2\pi/k$. 

Note also here that formally 
metric (\ref{EqmetricX}) is invariant under 
translations
$X_{i}\rightarrow X_{i}+a_i$,
$i=1,2$,
produced 
by the vector fields 
\begin{eqnarray}&
\label{Traslations}
\frac{\partial}{\partial X_1}=\frac{1}{\alpha}\cos(\frac{\varphi}{\alpha})\frac{\partial}{\partial r}-
\frac{1}{r}\sin(\frac{\varphi}{\alpha})\frac{\partial}{\partial \varphi}
\,,\qquad 
\frac{\partial}{\partial X_2}=
\frac{1}{\alpha}\sin(\frac{\varphi}{\alpha})\frac{\partial}{\partial r}+\frac{1}{r}\cos(\frac{\varphi}{\alpha})\frac{\partial}{\partial \varphi}\,.&
\end{eqnarray}
The infinitesimal ($\delta_1,\delta_2\sim 0$) form of transformations 
generated by (\ref{Traslations}) is  
\begin{eqnarray}
&
\frac{\partial}{\partial X_1}:\quad \Rightarrow\quad  
r\rightarrow r_1'=r+\frac{\delta_1}{\alpha}\cos(\frac{\varphi}{\alpha})\,,\qquad
\varphi\rightarrow\varphi'_1=\varphi-\frac{\delta_1}{r}\sin (\frac{\varphi}{\alpha})\,,
&\label{Tr1}\\
&
\frac{\partial}{\partial X_{2}}:\quad \Rightarrow\quad  
r\rightarrow r_2'=r+\frac{\delta_2}{\alpha}\sin(\frac{\varphi}{\alpha})\,,\qquad
\varphi\rightarrow\varphi_2'=\varphi+\frac{\delta_2}{r}\cos(\frac{\varphi}{\alpha})\,.
&\label{Tr2}
\end{eqnarray}
In spite of that 
these formal transformations are local isometries, 
one sees their singular nature 
at $r=0$ 
when $\alpha\not=1/k$.
Furthermore,
the corresponding global transformations,
\begin{eqnarray}
\label{Nwdisso1}
&
r^2\rightarrow (r_1')^{2}=r^2+\frac{2\delta_1 r}{\alpha}\cos(\frac{\varphi}{\alpha})+\frac{\delta_1^2}{\alpha^2}\,,\qquad
\varphi\rightarrow\varphi_1'=\alpha\arctan(\frac{\alpha r\sin(\frac{\varphi}{\alpha})}{\alpha r\cos(\frac{\varphi}{\alpha})+\delta_1})\,,&
\\&
\label{Nwdisso2}
r^2\rightarrow (r_2')^{2}=r^2+\frac{2\delta_2 r}{\alpha}\sin(\frac{\varphi}{\alpha})+\frac{\delta_2^2}{\alpha^2}\,,\qquad 
\varphi\rightarrow\varphi_2'=\alpha\arctan(\frac{\alpha r\sin(\frac{\varphi}{\alpha})+\delta_2}{\alpha r\cos(\frac{\varphi}{\alpha})})\,,&
\end{eqnarray}
reveal their singularity  for $\alpha\not=1$.

The space-time metric (\ref{dS2}) with spatial part presented in the form 
(\ref{EqmetricX}) looks like the metric of (2+1)-dimensional 
Minkowski space $dS^2=\eta_{\mu\nu}dX^\mu dX^\nu$, 
$X^0=ct$, $\eta_{\mu\nu}=\text{diag}\,(-1,1,1)$. Locally, it is conformally invariant 
under transformations of the conformal   
$SO(3,2)$ group,
whose classical generators are $P^\mu$, 
$J^{\mu\nu}=X^\mu P^\nu-X^\nu P^\mu$, $K^\mu=2X^\mu (XP)-X^2 P^\mu$
and $D=XP$,
where $P_\mu=\eta_{\mu\nu}P^\nu$ are the momenta canonically conjugate to $X^\mu$.
Taking into account that $P_1=\frac{1}{\alpha}p_r\cos(\varphi/\alpha)-\alpha p_\varphi\sin(\varphi/\alpha)$,
$P_2=\frac{1}{\alpha}p_r\sin(\varphi/\alpha)+\alpha p_\varphi\cos(\varphi/\alpha)$,
where $p_r$ and $p_\varphi$ are the momenta canonically conjugate to $r$ and $\varphi$,
one finds that only the generators of the time translation, $P^0$, 
the spatial rotation, $J^{12}=p_\varphi$,  the dilatations, $D=-X^0 P^0 +r p_r$, 
and special conformal transformations,  $K^0=2
X^0 D -(\alpha^2 r^2-(X^0)^2)P^0$, are globally 
well-defined for 
arbitrary values of the parameter $\alpha$, 
while generators of the spatial translations, $P^i$, Lorentz boosts, $J^{0i}$,
and generators of special conformal transformations, $K^i$, 
are globally well-defined only for $\alpha=1/k$.
After  
  the 
  appropriately taken non-relativistic limit  
\cite{Nied1,Hagen, Barut,LeiPly,HenUnter,Son,BagGop}, as we shall see, the
rotation generator $p_\varphi$ and 
the corresponding analogs of the generators $P^0$, $D$ and $K^0$ 
will play the key role in our  subsequent  analysis.
At the same time, in spite of the globally not well-defined  nature 
(in the general case of the parameter 
$\alpha$ values) of  
generators of the spatial translations and Lorentz boosts,
their corresponding non-relativistic  analogs will be
employed by us for the construction of the globally 
well-defined 
generators of the hidden symmetries.

 \vskip0.1cm
 The
  change of coordinates (\ref{Xi}) 
and the geodesic analysis presented  in Sec. \ref{SecFreeCone} will 
show
 that the non-relativistic dynamics in the cosmic string background
 can be related to the free motion in the Euclidean plane.
Bearing this in mind, 
 instead of jumping directly to the 
analysis of the dynamics in the
  conical geometry, it is  
  appropriate
  to review some important 
 characteristics related to the motion in $\R^2$.

\section{Dynamics in the Euclidean plane}
\label{SecEmotion}
 To understand
the complete symmetry algebra of a given mechanical system 
in a cosmic string (conical)  background,
 both at the classical and quantum levels, it is instructive to remind  the corresponding properties of such 
 a system in the flat Euclidean plane. 
 Later on, we will show that there is a formal 
 canonical transformation related to the change of coordinates (\ref{Xi}) 
 which allows us to connect the dynamics in  conical  geometry 
 with the corresponding dynamics in $\R^2$.
  We are interested in the free particle dynamics as well 
  as the dynamics of the particle in the  harmonic trap, so this section contains all we 
  need to know 
 of these two systems in the Euclidean plane.

\subsection{ The free particle }
\label{AppA}
Here  we present  the complete set of integrals of motion of order not higher than two 
in momenta and display their explicit  Lie algebra for a particle in Euclidean plane.
Next we use the conserved quantities to reconstruct the trajectory of the particle. 
Finally we briefly describe the quantum  theory of the system using the
 polar coordinates.  

The quadratic in momenta and coordinates integrals of motion are
\begin{eqnarray}&
\label{Qmgen1}
H=\frac{1}{2m}p_+p_-\,,\qquad
D=\frac{1}{4}(\chi_+p_-+ p_+\chi_+)\,,\qquad
K=\frac{m}{2}\chi_+\chi_-\,,&\\&
 J_0=\frac{i}{4}(\chi_+p_- - p_+\chi_-)\,,\qquad
 J_\pm=\frac{1}{2}\chi_\pm p_\pm\,,
&\label{Qmgen1+}\\&
T_\pm=\frac{1}{2m}(p_\pm)^2\,,\qquad
S_\pm=\frac{m}{2}(\chi_\pm)^2\,,\label{Qmgen2}
\end{eqnarray}
where 
\be
\label{PyX}
p_\pm=p_1\pm ip_2\,,\qquad\chi_\pm= \chi_1\pm i \chi_2\,
\ee
are the complex combinations of the canonical momenta $p_i$ and the Galilean boost generators 
$\chi_i=x_i-\frac{1}{m}p_i t$.

The Hamiltonian $H$,  the angular momentum 
$p_\varphi=2J_0$, and the
integrals $p_\pm$, $T_\pm$
 are the conserved quantities  not depending explicitly on time $t$. 
The integrals $J_0$, $p_\pm$  and   $\chi_\pm$,
unlike the rest of the listed  integrals,
do not mix coordinates $x_i$ and momenta $p_i$ 
when they act in the phase space
via Poisson brackets.
The integrals $H$, $D$ and $K$ 
correspond to the planar case $\alpha=1$ 
of the non-relativistic  limit of the generators  
$P^0$, $D$ and $K^0$ mentioned  in the previous section, 
while  the Galilean boost generators $\chi_i$ 
appear as the non-relativistic limit of the Lorentz boosts 
$J^{0i}$,
see refs.  \cite{Hagen,Barut,Son}.
 Note  that the 
 integrals $J_\pm$, $T_\pm$ and $S_\pm$,
 being quadratic in $p_i$, 
 correspond here 
 to generators of the hidden symmetries
 \cite{Cariglia}.

The ten generators (\ref{Qmgen1})-(\ref{Qmgen2}) satisfy the following 
Poisson bracket  relations 
of the $\mathfrak{sp}(4,\R)$ algebra,
 \begin{eqnarray}
&\label{Simetryal1}
\{D,H\}=H\,,\qquad\{D,K\}=-K\,,
\qquad \{K,H\}=2D\,,&\\&
\{J_0, J_\pm\}=\mp i J_\pm\,,\qquad
\{J_-, J_+\}= -2 i J_0\,,\label{Simetryal2}
&\\&
\label{Euclidal}
\{J_0, T_\pm\}=\mp iT_\pm\,,\qquad
\{J_0,S_\pm\}=\mp i S_\pm\,,
&\\&
\{H,S_\pm\}=-2J_\pm\,,\qquad
\{H,J_\pm\}=-T_\pm\,,&\\&
\{K,T_\pm\}=2J_\pm\,,\qquad
\{K,J_\pm\}=S_\pm\,,&\\&
\{D,T_\pm\}=T_\pm \,,\qquad \{D,S_\pm\}=-S_\pm\,,&\\&
\{S_\pm, T_\mp\}=\mp 4i(J_0\pm iD)\,,
&\\&
\{J_\pm, T_\mp\}=2H\,,\qquad
\{J_\pm, S_\mp\}=2K\,.\label{Final?}&
\end{eqnarray}
By including the first order generators (\ref{PyX}) with redefinition 
$\xi_\pm=m\chi_\pm$,
we supplement the algebra (\ref{Simetryal1})-(\ref{Final?}) with the
Poisson bracket relations 
\begin{eqnarray}
&\{\xi_\pm,p_\mp\}=2m\,, &\label{Hei2}\\&
\{H,\xi_\pm\}=-p_{\pm}\,,\qquad
\{D,\xi_\pm\}=-\frac{1}{2}\xi_\pm\,,\qquad 
\{J_0,\xi_\pm\}=\mp \frac{i}{2}\xi_\pm \,, &\\&
\{K,p_\pm\}=\xi_{\pm}\,,\qquad\{D,p_\pm\}=\frac{1}{2}p_\pm\,,\qquad 
\{J_0,p_\pm\}=\mp\frac{i}{2}p_\pm\,,&\\&
\{T_\pm,\xi_\mp\}=-2p_\pm\,,\qquad
\{S_\pm,p_\mp\}=2\xi_\pm\,,&\\&
 \{J_\pm,\xi_{\mp}\}=-\xi_{\pm}\,,\qquad
 \{J_\pm,p_{\mp}\}=p_{\pm}\,.\label{Sch4}
&
\end{eqnarray}
The 
not displayed in (\ref{Simetryal1})--(\ref{Sch4}) 
Poisson brackets are equal to zero.
Relations 
(\ref{Hei2})--(\ref{Sch4})
 correspond to the
 ideal sub-algebra generated by $\xi_\pm$ and $p_\pm$,
 with mass $m$ playing a role of the central charge.
 This also is an ideal sub-algebra
 of the Schr\"odinger algebra $\mathfrak{sch}(2)$  \cite{Nied1}, generated 
by   $H$, $D$, $K$, $p_i$, $\xi_i$ and $m$,
 that, in turn, is a sub-algebra of the complete Lie algebra
 (\ref{Simetryal1})--(\ref{Sch4}).

Some remarkable properties of the presented 
symmetry algebra are the following.

\begin{itemize}
\item The algebraic relations (\ref{Simetryal1}) 
correspond to the dynamical $\mathfrak{so}(2,1)
\cong
 \mathfrak{sl}(2,\R)$ conformal algebra.
Its  Casimir element 
is $D^2-HK=-J_0^2=-\frac{1}{4}p_\varphi^2$.
\item Relations 
(\ref{Simetryal2}) correspond to  another  $\mathfrak{sl}(2,\R)$ sub-algebra with 
Casimir element  $-J_0^2+J_+J_-=D^2$.
\item Each  triplet of integrals $(J_0,T_\pm)$, $(J_0,S_\pm)$, 
$(p_\varphi,p_\pm)$ and $(p_\varphi,\xi_\pm)$
generate Euclidean sub-algebra $\mathfrak{e}(2)$.  The corresponding Casimirs 
$T_+T_-$, $S_+S_-$,  $p_+p_-$ and 
$\xi_+\xi_-$ are 
$H^2$, $K^2$, $2mH$ and $2mK$.
\item The two sets
 $(\ell_{+}^{(+)}=\frac{1}{2\gamma}S_+$, $\ell_{-}^{(+)}=\frac{\gamma}{2}T_-$, $\ell_{0}^{(+)}=\frac{1}{2}(J_0+iD))$ 
and $(\ell_{+}^{(-)}=\frac{\gamma}{2}T_+$, $\ell_{-}^{(-)}=\frac{1}{2\gamma}S_-$, $\ell_{0}^{(-)}=\frac{1}{2}(J_0-iD))$ 
 generate 
the $\mathfrak{su}(2)\oplus\mathfrak{su}(2)$ sub-algebra,
where 
$\gamma$ is a constant of dimension of squared length 
that is introduced 
to compensate the corresponding dimensions.
The Casimir elements of these two $\mathfrak{su}(2)$  sub-algebras are 
$C^{(\pm)}=(\ell_0^{(\pm)})^2+\ell_+^{(\pm)}\ell_-^{(\pm)}=0$. Note that 
 $\big(\ell_0^{(+)}\big)^*=\ell_0^{(-)}$,
 $\big(\ell_+^{(\pm)}\big)^*=\ell_-^{(\mp)}$.
\item Each integral is an eigenstate of $D$ 
in the sense of the Poisson bracket relation $\{D,A\}=\lambda A$\,:
$(\lambda=1: H, T_\pm)$, 
$(\lambda=0:  D, J_0, J_\pm)$,
$(\lambda=-1: K,S_\pm)$,
$(\lambda=1/2:  p_\pm)$,
$(\lambda=-1/2:  \xi_\pm)$.
\item Analogously, each integral is an eigenstate of $J_0$:
$(\lambda=1:  J_+, T_+, S_+)$, 
$(\lambda=0:  J_0, H, D, K)$,
$(\lambda=-1:  J_-,  T_-, S_-)$,
$(\lambda=1/2:  p_+, \xi_+)$,
$(\lambda=-1/2:  p_-, \xi_-)$.

\end{itemize}

Introducing linear combinations 
$\mathcal{J}_0=\frac{1}{2}(H+K)$, 
$\mathcal{J}_1=\frac{1}{2}(H-K)$,
and denoting $\mathcal{J}_2=D$,
$\mathcal{J}_\pm=\mathcal{J}_1\pm i\mathcal{J}_2$,
conformal algebra (\ref{Simetryal1}) 
can be presented in the form similar to 
(\ref{Simetryal2}),
and its Casimir takes the form 
$-\mathcal{J}_0^2+\mathcal{J}_+\mathcal{J}_-=-J_0^2$.
The generators 
$\mathcal{J}_\mu$, $\mu=0,1,2$, 
of the conformal algebra
describe the upper sheet of the two-sheeted hyperboloid,
that at $p_\varphi=0$ degenerates into the 
cone with $\mathcal{J}_0\geq 0$.
The generators $J_\mu$ of the 
$\mathfrak{sl}(2,\R)$ algebra (\ref{Simetryal2})
describe a one sheet hyperboloid
that at $\mathcal{J}_2=D=0$ degenerates into a double cone 
with $J_0=2p_\varphi\in \R$.

Using the Casimir of conformal algebra 
(\ref{Simetryal1}) and the explicit form of generators (\ref{Qmgen1}),  one gets 
\begin{eqnarray}
\label{r(t)}
&r^2(t)=\frac{2}{m}(Ht^2+2Dt+K)=\frac{2H}{m}\left(\left(t+\frac{D}{H}\right)^2+\frac{p_\varphi^2}{4H^2}\right)\,,&
\end{eqnarray}
from where we see that at the moment of time  $t_*=-D/H$, the particle is in the ``perihelion" of the 
trajectory,  $r(t_*)\equiv r_*=p_\varphi/\sqrt{2mH}$.  
On the other hand, from  the angular and linear momenta integrals  the 
straight line trajectory is reconstructed in polar coordinates,
\begin{eqnarray}
\label{r(varphi)}&
r(\varphi)=\frac{r_*}{\cos(\varphi-\varphi_*)}\,,
\qquad -\pi/2\leq \varphi-\varphi_* \leq \pi/2\,.&
\end{eqnarray}
By means of (\ref{r(t)}) and (\ref{r(varphi)}) we also find 
\begin{eqnarray}\label{varphifree}
&\varphi(t)=\arctan\left(\frac{1}{2\rho}\left(t+\frac{D}{H}\right)\right)+\varphi_*\,,\qquad
\rho=\frac{ p_\varphi}{2H}\,.&
\end{eqnarray}
{}From $\chi_i$ and $p_i$ 
 one can 
 construct a sort of Laplace-Runge-Lentz
 vector, 
 \begin{eqnarray}
&\chi_i^{\bot}=x_i-p_i \frac{x_jp_j}{p_kp_k}&
 \end{eqnarray}
such that 
$\chi_i^{\bot}p_i=0.$
This vector  (with respect to $p_\varphi$) 
integral specifies  the coordinates of the perihelion, 
$\chi_{1}^{\bot}=r_*\cos(\varphi_*)=x_1(\varphi_*)$, 
$\chi_{2}^{\bot}=r_*\sin(\varphi_*)=x_2(\varphi_*).$
The components $\chi_i^{\bot}$, however,  are not independent integrals since they satisfy
$\chi_{1}^{\bot}=\frac{p_\varphi}{2mH}\,p_2$ and 
$\chi_{2}^{\bot}=-\frac{p_\varphi}{2mH}\,p_1$.

In the quantum case,  it is convenient here 
to use the 
polar coordinates, in which 
the Hamiltonian operator is given by 
\begin{eqnarray}
&\hat{H}=-\frac{\hbar^2}{2m}
\left(\frac{1}{r}\frac{\partial}{\partial r}\left(r\frac{\partial}{\partial r}\right) +
\frac{1}{r^2}\frac{\partial^2}{\partial\varphi}\right)\,.&
\end{eqnarray}
Its eigenstates and eigenvalues are  
\begin{eqnarray}&
\label{freewigen}
\psi_{\kappa, l}^\pm (r,\varphi)=\sqrt{\frac{\kappa}{2\pi}}J_{l}(\kappa r)e^{\pm il  \varphi}\,,\qquad E=\frac{\hbar^2 \kappa^2}{2m}\,, \qquad l=0,1,\ldots\,,&
\end{eqnarray} 
where $J_{\beta}(\zeta)$ are the Bessel functions of the first kind. 
With respect to the inner product 
\begin{eqnarray}
\label{EuIner}&
\bra{\Psi_1}\ket{\Psi_2}=
\int_0^{\infty}r dr\int_{0}^{2\pi}d\varphi \Psi^*_1\Psi_2\,,&
\end{eqnarray}
eigenstates (\ref{freewigen}) 
satisfy the orthogonality relation 
$\braket{\psi_{\kappa,l}^\pm}{\psi_{\kappa',l'}^\mp}=\delta_{ll'}\delta(\kappa-\kappa')
$,  
and due to the property of the Bessel 
functions $J_{-l}(\eta)=(-1)^{l}J_{l}(\eta)$, one has 
$\psi_{\kappa, -l}^\pm=(-1)^l\psi_{\kappa, l}^\mp$.

The basic first order differential operators of the system in polar coordinates are 
\begin{eqnarray}
&
\label{prpphi}
\hat{p}_{r}=-i\hbar\left(\frac{\partial}{\partial r}+\frac{1}{2r}\right)\,,\qquad
\hat{p}_\varphi=-i\hbar\frac{\partial}{\partial\varphi}\,,&
\end{eqnarray}
where $\hat{p}_r$ is symmetric with respect to (\ref{EuIner}), but not self-adjoint. We use them to construct 
the well-defined operators 
\begin{eqnarray}
\label{Pii}&
\hat{p}_{\pm}
=
-i\hbar  e^{\pm i\varphi}\left[
\frac{\partial}{\partial r}
\pm i\frac{1}{ r}\frac{\partial}{\partial\varphi} 
\right]\,,\qquad
p_{+}^\dagger=\hat{p}_{-}\,,&
\end{eqnarray}
which are the quantum version of the classical integrals $p_\pm$.  
Their action  on eigenstates can be found  by using the recurrence relations
$\frac{2\beta}{\zeta}J_{\beta}(\zeta)=J_{\beta-1}(\zeta)+J_{\beta+1}(\zeta)\,,$ 
$2\frac{d}{d\zeta}J_{\zeta}(\zeta)=J_{\beta-1}(\zeta)-J_{\beta+1}(\zeta)\,,$
and  we get
\begin{eqnarray}
&
\hat{p}_\pm\psi_{\kappa,l}^\pm (r,\varphi)=
 i \hbar\kappa \psi_{\kappa,l+1}^\pm (r,\varphi)\,,\qquad 
\hat{p}_\pm\psi_{\kappa,l}^\mp (r,\varphi)=- i \hbar\kappa
\psi_{\kappa,l-1}^\mp (r,\varphi)\,.
&
\end{eqnarray} 
These relations 
 show that the operators $\hat{p}_\pm$ 
change the angular momentum quantum number of the wave-function
 without changing the energy, that reflects  
 the infinite degeneracy of 
the energy levels. 

Quantum version of other integrals is obtained 
by the substitution  $p_\pm \rightarrow\hat{p}_\pm$ 
and using  the Weyl (symmetric) ordering
in  (\ref{Qmgen1})-(\ref{Qmgen2}). 
 In the general case we 
 have dynamical integrals $\hat{A}(t)$
which include the  explicit dependence on time,
 $\frac{d}{dt} \hat{A}(t)=\frac{\partial}{\partial t} \hat{A}(t)
 +\frac{1}{i\hbar}[\hat{A}(t),\hat{H}]=0$.
For them we have 
\be
\label{DynamocalAction}
\hat{A}(t)\Psi(r,\varphi,t)=e^{-\frac{it\hat{H}}{\hbar}}\hat{A}|_{t=0}\Psi(r,\varphi,t=0)\,,
\ee   
where $\Psi(r,\varphi,t)$ is a solution of the time dependent Schr\"odinger equation.
In particular, the quantum generators of dilatations and special conformal transformations, 
\begin{eqnarray}
&\label{QD}
\hat{D}=\frac{1}{4}(\hat{\chi}_+\hat{p}_-+\hat{p}_+\hat{\chi}_-)= 
-i\frac{\hbar}{2}\left(r\frac{\partial}{\partial r}+1\right)- \hat{H}t\,,
&\\&
\label{QK}
\hat{K}=\frac{1}{2}m \hat{\chi}_-\hat{\chi}_+= \frac{1}{2}m
r^2-2\hat{D}t-\hat{H}t^2\,,
\end{eqnarray}
are examples of dynamical symmetry operators. Together with the Hamiltonian, they generate the 
quantum 
$\mathfrak{sl}(2,\R)$ algebra 
\begin{eqnarray}
&\label{freeparticleso2}
[\hat{D},\hat{H}]=i\hbar \hat{H}\,,\qquad[\hat{D},\hat{K}]=-i\hbar\hat{K}\,,\qquad 
[\hat{K},\hat{H}]=2i\hbar \hat{D}\,. &
\end{eqnarray}
On a Hilbert subspace with fixed value of the quantum number $l=0,1,\ldots$, the eigenstates (\ref{freewigen}) 
correspond to an irreducible 
infinite dimensional representation of 
conformal $\mathfrak{sl}(2,\R)$ algebra (\ref{freeparticleso2})
of the discrete type series $D^+_j$ 
characterized by 
the Casimir operator value 
$ -\hat{\mathcal{J}}_0^2 +\hat{\mathcal{J}}_1^2
+\hat{\mathcal{J}}_2^2=\hat{D}^2-\frac{1}{2}(\hat{K}\hat{H}+\hat{H}\hat{K})=
-\hbar^2 j(j-1)$ with $j=\frac{1}{2}(l+1)$,
and   eigenvalues of 
the compact generator $\hat{\mathcal{J}}_0$ to be
$j+n$, $n=0,1,\ldots$.
The present free particle's  Hilbert space, in which 
the non-compact $\mathfrak{sl}(2,\R)$ generator
$\hat{\mathcal{J}}_0+\hat{\mathcal{J}}_1=\hat{H}$
is diagonal, corresponds to the so-called parabolic
realization of  $D^+_j$  representation \cite{sl2R}.

For quantum  analog of the $\mathfrak{sl}(2,\R)$ algebra 
(\ref{Simetryal2}), the Casimir operator is 
$\hat{J}_0(-\hat{J}_0+\hbar)+\hat{J}_+\hat{J}_-=\hat{D}^2+\frac{1}{4}\hbar^2$. 
Operator $\hat{D}$ is self-adjoint with respect to the scalar product 
(\ref{EuIner}), and at $t=0$ its eigenfunctions are 
$\Psi_\lambda(r)=r^{2i\lambda -1}/\sqrt{2}\pi$, 
$\hat{D}\Psi_\lambda=\lambda \Psi_\lambda$, $\lambda\in \R$,
$\bra{\Psi_\lambda}\ket{\Psi_\lambda'}=\delta(\lambda-\lambda')$.
One sees then that  the quantum  analog of the $\mathfrak{sl}(2,\R)$
algebra  (\ref{Simetryal2}) at fixed value of $\lambda$ corresponds here
to the principal continuous series representation 
characterized by the Casimir invariant value $\hat{C}=\hbar^2(\lambda^2+1/4)\geq \hbar^2/4$,
in which the eigenvalues of the compact generator $\hat{J}_0$  are
$j_0=\hbar j$ with  $j=0, \pm1,\pm2,\ldots$ on the subspace 
with even values of $l$, and 
$j=\pm1/2,\pm 3/2,\ldots$ on the subspace with odd values of $l$
\cite{Barg}.

The difference between  representations generated by 
$\hat{\mathcal{J}}_\mu$ and $\hat{J}_\mu$ operators 
is coherent with the difference of the above-mentioned corresponding 
classical hyperboloids \cite{sl2R}.

\subsection{The isotropic harmonic oscillator}
\label{AppB}
Here we consider some general properties of the harmonic oscillator in the Euclidean plane. 
First,  we construct the symmetry generators and the 
symmetry algebra. 
In the next step, we use these integrals of motion to 
algebraically reproduce the orbit 
of the particle from the integrals of motion, 
and finally, we review the quantum picture of the model.

The classical Hamiltonian of the planar isotropic harmonic oscillator system
\begin{eqnarray}&
H_{\text{os}}=H_1+H_2=
\frac{p_r^2}{2 m}+\frac{p_{\varphi}^2}{2mr^2}+\frac{m\omega^2}{2}r^2\,&
\end{eqnarray}
can be understood in Cartesian coordinates $x_j$, $j=1,2$,   as the sum  
of the two independent one-dimensional 
harmonic oscillator Hamiltonians 
$H_j=\frac{1}{2m}(p_j^2+ m^2\omega^2 x_j^2)$ 
with the same 
frequencies  and masses,   while
in the polar coordinates
it can be considered as a two-dimensional generalization 
of the de Alfaro Fubini and Furlan conformal mechanics model
\cite{AFF}. 

The quantities  
\begin{eqnarray}
\label{acartecian}&
a_j^\pm=\frac{1}{\sqrt{2}}e^{\mp i\omega t}\left( \sqrt{m\omega}x_j\mp i\frac{p_j}{\sqrt{m\omega}} \right)\,,\qquad j=1,2\,,&
\end{eqnarray}
being classical analogs of the quantum ladder operators 
multiplied by $e^{\mp i \omega t}$, are the basic dynamical integrals 
expressed in Cartesian coordinates.
Their linear combinations  
\begin{eqnarray}
\label{gen1}
&
b_1^-= \frac{1}{\sqrt{2}}(a_1^--ia_2^-)= \frac{1}{2}e^{i(\omega t-\varphi)}
\left(
\sqrt{m\omega}r
+\frac{p_\varphi}{\sqrt{m\omega}r} +i\frac{p_r}{\sqrt{m\omega}}\right)\,,\qquad b_1^+=(b_1^-)^*\,,
 &\\&
b_2^-= \frac{1}{\sqrt{2}}(a_1^-+ia_2^-)= \frac{1}{2}e^{i(\omega t+\varphi)}
\left(
\sqrt{m\omega}r
-\frac{p_\varphi}{\sqrt{m\omega}r} +i\frac{p_r}{\sqrt{m\omega}}\right)\,,\qquad b_2^+=(b_2^-)^*\,,
\label{gen1+}
\end{eqnarray}
are more convenient, however,  when we work in the polar coordinates.
They can be produced by a particular 
 classical canonical transformation, or 
 the corresponding  unitary transformation at the quantum level, 
that we consider  below. 
The ten second-order in these basic integrals 
symmetry generators  are
\begin{eqnarray}
&
\mathcal{J}_0=\frac{1}{2} b_j^+b_j^-=
\frac{1}{2\omega}H_\text{os}\,,\qquad 
\mathcal{L}_2=\frac{1}{2}(b_1^+b_1^--b_2^+b_2^-)=\frac{1}{2}p_\varphi\,,\qquad 
\mathcal{L}_\pm=b_1^\pm b_2^\mp\,,
\label{intnot}&\\&
\mathcal{J}_\pm =b_1^\pm b_2^\pm=\frac{1}{2}\left((a_1^\pm)^2+ (a_2^\pm)^2\right)\,,\qquad
\mathcal{B}_j^\pm =(b_j^\pm)^2\,. \label{gen2}&
\end{eqnarray}
Unlike the free particle case, only  four integrals (\ref{intnot}) here  do not depend 
explicitly on time.
The integrals (\ref{intnot}) and  (\ref{gen2}) satisfy the following non-zero Poisson bracket relations 
\begin{eqnarray}
\label{Sl2Ros}
&
\{\mathcal{J}_0,\mathcal{J}_\pm\}=\mp i \mathcal{J}_\pm\,,\qquad
\{\mathcal{J}_-,\mathcal{J}_+\}=-2i\mathcal{J}_{0}\,, &\\&
\label{Su(2)OS}
\{\mathcal{L}_2,\mathcal{L}_\pm\}=\mp i\mathcal{L}_\pm\,,\qquad
\{\mathcal{L}_+,\mathcal{L}_-\}=-i 2 \mathcal{L}_2\,, &\\&
\{\mathcal{J}_\pm,\mathcal{L}_\mp\}=\pm i\mathcal{B}_2^\pm\,,\qquad 
\{\mathcal{J}_\pm,\mathcal{L}_\pm\}=\pm i\mathcal{B}_1^\pm &\\&
\{\mathcal{J}_0, \mathcal{B}_a^\pm\}=\mp i\mathcal{B}_a^\pm\,,\qquad
\{\mathcal{J}_\mp,\mathcal{B}_2^\pm \}=\mp2i\mathcal{L}_{\mp}\,,\qquad 
\{\mathcal{J}_\mp,\mathcal{B}_1^\pm \}=\mp2i\mathcal{L}_{\pm}\,, &\\&
\{\mathcal{L}_{2}, \mathcal{B}_1^\pm \}=\mp i \mathcal{B}_1^\pm \,,\qquad
\{\mathcal{L}_{2}, \mathcal{B}_2^\pm \}=\pm i \mathcal{B}_2^\pm \,,\qquad 
\{\mathcal{L}_{\pm}, \mathcal{B}_1^\mp \}=\pm 2i \mathcal{J}_\mp\,,&\\
&\{\mathcal{L}_{\pm}, \mathcal{B}_2^\pm \}=\mp 2i \mathcal{J}_\pm\,,\quad
\{\mathcal{B}_1^-,\mathcal{B}_1^+\}=-4i\left(\mathcal{J}_0+\mathcal{L}_2\right)\,,\quad
\{\mathcal{B}_2^-,\mathcal{B}_2^+\}=-4i\left(\mathcal{J}_0-\mathcal{L}_2\right)\,.\qquad
\label{Sl2Ros+}
 &
\end{eqnarray}

The brackets involving the  basic integrals are
\begin{eqnarray}&\{b_{i}^-,b_{j}^+\}=-i\delta_{ij}\,,\label{Ideal1}&\\&
\{\mathcal{J}_{0},b_{j}^\pm\}=\mp \frac{i}{2}b_{j}^\pm\,,\qquad 
\{\mathcal{J}_{\mp},b_{1}^\pm\}=\mp ib_{2}^\mp\,,\qquad
\{\mathcal{J}_{\mp},b_{2}^\pm\}=\mp ib_{1}^\mp\,,
&\\&
 \{\mathcal{B}_{i}^\pm,b_{j}^\mp\}= \pm 2i\delta_{ij}b_{j}^\mp\,,\qquad
  \{\mathcal{L}_{2},b_{1}^\pm\}=\mp\frac{i}{2}b_{1}^\pm\,,\qquad
  \{\mathcal{L}_{2},b_{2}^\pm\}=\pm\frac{i}{2}b_{2}^\pm\,,&\\&
  \{\mathcal{L}_{\pm},b_1^\mp\}=\pm ib_2^\mp \,,\qquad
 \{\mathcal{L}_{\pm},b_2^\pm\}=\mp i b_1^\pm \,,&\\&
 \{\mathcal{J}_\pm,b_j^\pm\}=
 \{\mathcal{L}_\pm,b_1^\pm\}= \{\mathcal{L}_\pm,b_2^\mp\}=0\,.
 \label{FinalOssym}
  &
\end{eqnarray}
Some properties
of the  Lie  algebra  (\ref{Sl2Ros})--(\ref{FinalOssym}) are the following.
\begin{itemize}
\item Dynamical integrals $b_j^\pm$ generate an ideal sub-algebra. 
\item Relations  (\ref{Sl2Ros}) correspond to the conformal  $\mathfrak{sl}(2,\R)$ symmetry. 
The Casimir is $-\mathcal{J}_0^2+\mathcal{J}_+\mathcal{J}_-=-\mathcal{L}_2^2$.
\item The not-depending explicitly on time  
 integrals of motion  $\mathcal{L}_2$ and $\mathcal{L}_\pm=\mathcal{L}_3\pm i \mathcal{L}_1$ 
 generate the  $\mathfrak{su}(2)$ algebra (\ref{Su(2)OS}) with the Casimir  invariant
 $\mathcal{L}_1^2+\mathcal{L}_2^2+\mathcal{L}_3^2=\mathcal{J}_0^2=\frac{1}{4\omega^2}H_\text{os}^2$. 
 \item Each  triplet of integrals $(\mathcal{L}_2,\mathcal{B}_j^-)$, $(\mathcal{L}_2,\mathcal{B}_j^+)$, 
$(p_\varphi,b_{j}^-)$ and $(p_\varphi,b_j^+)$
generate Euclidean sub-algebra $\mathfrak{e}(2)$.  The corresponding Casimirs 
$\mathcal{B}_1^-\mathcal{B}_2^-$, $\mathcal{B}_1^+\mathcal{B}_2^+$,  
$b_{1}^-b_{2}^-$, and $b_{1}^+b_{2}^+$
are $\mathcal{J}_-^2$, $\mathcal{J}_+^2$, $\mathcal{J}_-$ and 
$\mathcal{J}_+$, respectively.
\item The sets of integrals  
($\mathscr{J}_\pm^{(+)}=\mathcal{B}_1^\pm/\sqrt{2}$, $\mathscr{J}_0^{(+)}=\mathcal{J}_0+\mathcal{L}_2$), 
and  
 ($\mathscr{J}_\pm^{(-)}=\mathcal{B}_2^\pm/\sqrt{2}$, $\mathscr{J}_0^{(-)}=\mathcal{J}_0-\mathcal{L}_2$), 
 generate the 
 $\mathfrak{sl}(2,\R)\oplus\mathfrak{sl}(2,\R) $ algebra.
 Note that generators $\mathscr{J}_0^\pm$ can be 
  reinterpreted  as 
  the Landau problem's
   Hamiltonians in the symmetric gauge, with 
 positive/negative magnetic field  of the magnitude 
  $B=\frac{2cm\omega}{q}$, where $q$ and $c$ correspond to
 the electric charge of the particle and the speed of light
 \cite{IPW}. 
\item Each integral is an eigenstate of $i\mathcal{J}_{0}$ 
in the sense of $\{i\mathcal{J}_0,A\}=\lambda A$\,:
$(\lambda=\pm 1: \mathcal{J}_\pm, \mathcal{B}_j^\pm)$,
$(\lambda=0:  \mathcal{J}_0, \mathcal{L}_2, \mathcal{L}_\pm)$,
$(\lambda=\pm 1/2:  b_j^\pm)$. 
\item Analogously, each integral is an eigenstate of $i\mathcal{L}_2$:
$(\lambda=0:  \mathcal{L}_{2}, \mathcal{J}_0, \mathcal{J}_\pm)$, 
$(\lambda=\pm 1:  \mathcal{L}_\pm, \mathcal{B}_j^\pm)$,
$(\lambda=\pm 1/2:  b_1^\mp, b_2^\pm)$.
\end{itemize}
Using the dynamical integrals  $\mathcal{J}_\pm$ and integral 
$\mathcal{J}_0$,  
we define the generators of the Newton-Hooke conformal symmetry 
\cite{Nied2,NewHook,NH1,NH2,IPW}
 \begin{eqnarray}&
\mathcal{K}=\frac{1}{2\omega}(\mathcal{J}_+
+\mathcal{J}_-+2\mathcal{J}_0)=
\frac{1}{2}mr^2 \cos(2\omega t)
-\frac{1}{2\omega }rp_r\sin(2\omega t)
+
\frac{1}{\omega^2 }H_\text{os}\sin^2(\omega t)\,,\qquad
&\\&
\mathcal{D}=\frac{1}{2i}(\mathcal{J}_-
-\mathcal{J}_+)
=
\frac{1}{2}rp_r \cos(2\omega t)-
\frac{1}{2\omega}( H_{\text{os}}-m\omega^2  r^2)\sin(2\omega t)\,.
\end{eqnarray}
Together with $H_\text{os}=2\omega \mathcal{J}_0$ 
they satisfy the Poisson bracket relations
\begin{eqnarray}
\label{NHalgebra}
\{\mathcal{D},H_{os}\}=H_\text{os}-2\omega^2\mathcal{K}\,,\qquad
\{\mathcal{D},\mathcal{K}\}=-\mathcal{K}\,,\qquad
\{H_\text{os},\mathcal{K}\}=2\mathcal{D}\,,
\end{eqnarray}
and in the limit $\omega\rightarrow 0$ take the form 
of the free particle integrals (\ref {Qmgen1}).
From here, we obtain
\begin{eqnarray}\label{rDKH}
&r^2(t)=\frac{1}{m\omega^2}
\Big(H_\text{os}+2\omega \mathcal{D}\sin(2\omega t)+
(2\omega^2\mathcal{K}-H_\text{os})\cos(2\omega t)\Big)\,.&
\end{eqnarray}
 Taking into account the equivalent form 
$\mathcal{D}^2+\omega^2\mathcal{K}^2-\mathcal{K}H_\text{os}=-\frac{1}{4}p_\varphi^2$
for  the Casimir of the conformal 
$\mathfrak{sl}(2,\R)$ symmetry, equation  (\ref{rDKH})
allows us to find the  radial turning points 
\begin{eqnarray}&
r_\pm^{2}=\frac{H_{\text{os}}}{m\omega^2}\left(1\pm \delta \right)\,, 
\qquad 
\delta=\sqrt{1-\frac{\omega^2p_{\varphi}^2}{H_{\text{os}}^2}}\,,\label{rminmax}&
\end{eqnarray}
which also can be found directly from the Hamiltonian 
by applying the condition $ p_r = 0 $.

  On the other hand, by using the explicit form of the integrals $\mathcal{L}_1$ and $\mathcal{L}_3$ in polar coordinates,  
 \begin{eqnarray}
&
 \mathcal{L}_1=\sin(2\varphi)\left(\frac{1}{2\omega}H_\text{os}-
 \frac{1}{2m\omega r^2}p_\varphi^2 \right)-\cos(2\varphi)\frac{1}{2m\omega r}p_rp_\varphi\,,&\\&
 \mathcal{L}_3=\cos(2\varphi)\left(\frac{1}{2\omega}H_\text{os}-
 \frac{1}{2m\omega r^2}p_\varphi^2 \right)+\sin(2\varphi)\frac{1}{2m\omega r}p_rp_\varphi\,,&
 \end{eqnarray}
one deduces the elliptic  
trajectory for the isotropic planar harmonic oscillator,
 \begin{eqnarray}
\label{r2(varphi)}&
r^2(\varphi)=\frac{r_0^2}{1+\delta \cos(2(\varphi-\varphi_*))}\,,\qquad
r_0^2=\frac{p_\varphi^2}{mH_\text{os}}\,,&
 \end{eqnarray}
where $\varphi=\varphi_*$ corresponds to the angular position of one of the two 
``perihelia" of the trajectory. Eq. (\ref{r2(varphi)}) has a form of elliptic trajectory in Kepler 
problem but with $r(\varphi)$ and $\varphi$ there changed for $r^2(\varphi)$  and 
$2\varphi$ here.
By means of (\ref{rDKH}) and (\ref{r2(varphi)}) we also get 
\begin{eqnarray}&
\varphi(t)=
\arctan(\frac{r_+}{r_-}\tan(\omega (t-t_{*})))+\varphi_*\,,&
\end{eqnarray}
where $t_*$ indicates the moment of time when the particle is in the corresponding perihelion.

Now let's take a look at the quantum case. We first consider the Cartesian coordinates representation, 
then we present the picture in  the polar coordinates representation, and finally we
show how these 
two representations are related to each other by a unitary transformation.
\vskip0.2cm 
\noindent 
\underline{\emph{Cartesian coordinates representation }}
\vskip0.2cm 
\noindent The quantum versions of (\ref{acartecian}) (at $t=0$) are given by 
\begin{eqnarray}&
\hat{a}_i^\pm=\sqrt{\frac{m \omega }{2\hbar}}\left(x_i \mp  \frac{\hbar}{m\omega}\frac{\partial}{\partial x_i} \right)\,,\qquad
[\hat{a}_i^-,\hat{a}_j^+]=1\,.& 
\end{eqnarray}
In terms of them, we construct the set of operators 
(no summation in  the repeated index),
\begin{eqnarray}
&
\hat{\mathcal{J}}_{0}^{i}=\frac{1}{2\omega\hbar}\hat{H}_{i}= \frac{1}{4}(\hat{a}_i^+\hat{a}_i^-+\hat{a}_i^-\hat{a}_i^+)\,,\qquad
\hat{\mathcal{J}}^{i}_\pm= \frac{1}{2}(\hat{a}_i^\pm)^2\,, \qquad i=1,2\,,
&\\&
\hat{\mathcal{L}}_1= \frac{1}{2}(\hat{a}_1^+\hat{a}_2^-+\hat{a}_2^+\hat{a}_1^-)\,,
\qquad 
\hat{\mathcal{L}}_2= \frac{i}{2}(\hat{a}_1^+\hat{a}_2^--\hat{a}_2^+\hat{a}_1^-)=\frac{1}{2\hbar}\hat{p}_\varphi\,,
&\\&
\hat{\mathcal{L}}_3= \frac{1}{2}(\hat{a}_1^+\hat{a}_1^-- \hat{a}_2^+\hat{a}_2^-)= 
\hat{\mathcal{J}}_0^{1}-\hat{\mathcal{J}}_0^{2}\,,\qquad\hat{\mathcal{A}}_\pm = \hat{a}_{1}^\pm \hat{a}_{2}^\pm \,. &
\end{eqnarray}
These ten operators satisfy the Lie algebra which (up to a unitary transformation) corresponds
 to the quantum version of the classical 
algebra (\ref{Sl2Ros})-(\ref{FinalOssym}). In this  
 representation, the set of physical eigenstates and the spectrum of the system are 
\begin{eqnarray}
&
\psi_{n_1,n_2}(x_1,x_2)=\psi_{n_1}(x_1)\psi_{n_2}(x_2)\,,\qquad E_{n_1,n_2}=\hbar(n_1+n_2+1)\,,\quad
n_{1,2}=0,1,\ldots,\qquad
&\\&
\psi_{n_i}(x_i)= \frac{1}{\sqrt{2^n n!}}\left(\frac{m\omega}{\pi\hbar}\right)^{\frac{1}{4}}
H_{n_i}\left(\sqrt{\frac{m\omega}{\hbar}}x_i\right)e^{-\frac{m\omega}{2\hbar}x_i^2}\,,
&
\end{eqnarray}
where $H_{n_i}$ is the Hermite polynomial of order $n_i$.
These wave-functions diagonalize simultaneously the  operators  
$\hat{\mathcal J}_0^{i}$,
$i=1,2$,
 and, as a consequence,  $\hat{H}_\text{os}$ and $\hat{\mathcal{L}}_3$. 
\vskip0.25cm
\noindent 
\underline{\emph{Polar coordinates representation}}
\vskip0.25cm
\noindent 
The quantum versions of the classical dynamical 
 integrals (\ref{gen1}),(\ref{gen1+}), (\ref{intnot}) at $t=0$, and of the 
 integrals (\ref{gen2}) are 
\begin{eqnarray}
&
\hat{b}_1^-= \frac{1}{\sqrt{2}}(\hat{a}_1^--i\hat{a}_2^-)= \frac{1}{2}e^{-i\varphi}\sqrt{\frac{m \omega }{\hbar }}
\left(
r+
\frac{\hbar }{m\omega}\left(\frac{\partial}{\partial r}
-\frac{i}{r}\frac{\partial}{\partial \varphi} \right)\right)\,, 
\qquad \hat{b}_1^+=(\hat{b}_1^-)^\dagger\,, &\\&
\hat{b}_2^-= \frac{1}{\sqrt{2}}(a_1^-+ia_2^-)=\frac{1}{2}e^{-i\varphi}\sqrt{\frac{m \omega }{\hbar }}
\left(
r+
\frac{\hbar }{m\omega}\left(\frac{\partial}{\partial r}
+\frac{i }{r}\frac{\partial}{\partial \varphi}\right) 
\right) 
\,, \qquad \hat{b}_2^+=(\hat{b}_2^-)^\dagger\,,
&\\
&
\hat{\mathcal{J}}_0=\frac{1}{2\hbar\omega}\hat{H}_\text{os}=\frac{1}{4}([\hat{b}_1^+,\hat{b}_1^-]_{{}_+}+
[\hat{b}_2^+,\hat{b}_2^-]_{{}_+})\,,\qquad
 \hat{\mathcal{L}}_2= \frac{1}{2}(\hat{b}_1^+\hat{b}_1^--\hat{b}_2^+\hat{b}_2^-)
 =\frac{1}{2\hbar}\hat{p}_\varphi\,,&\\&
\hat{\mathcal{L}}_\pm= \hat{b}_1^\pm \hat{b}_2^\mp=\hat{\mathcal{L}}_3\pm i\hat{\mathcal{L}}_1=
\frac{m\omega}{4\hbar }e^{\pm 2i\varphi}\left(
r^2+\frac{\hbar^2}{m^2\omega^2}\left(\frac{1}{r^2}\frac{\partial^2}{\partial \varphi^2}-
\frac{\partial^2}{\partial r^2}\pm i\frac{2}{r}
\frac{\partial^2}{\partial r\partial \varphi}\right)\right)
\,,&\\&
\hat{\mathcal{J}}_\pm = \hat{b}_1^\pm\hat{b}_2^\pm=
-\frac{m\omega}{4\hbar }\left(\hat{H}_\text{os}-m\omega^2 r^2 \pm \hbar\omega(r\frac{\partial}{\partial r}+1)
\right)\,,\qquad 
\hat{\mathcal{B}}_j^\pm =(\hat{b}_j^\pm)^2\,, &
\end{eqnarray}
where $[,]_{{}_+}$
 is the anti-commutator of the operators. 
In this representation we diagonalize simultaneously 
$\hat{H}_\text{os}=2\hbar\omega\hat{\mathcal{J}}_0$ and $\hat{p}_{\varphi}=2\hbar\hat{\mathcal{L}}_2$.
The eigenstates and spectrum are 
\begin{eqnarray}
&
\psi_{n_r,l}^\pm(r,\varphi) =\left(\frac{m\omega}{\hbar}\right)^{\frac{1}{2}}
\sqrt{\frac{n_r!}{2\pi\Gamma(n_r+ l+1)}}\,
\zeta^{ l}L_{n_r}^{(l)}(\zeta^2)
e^{-\frac{\zeta^{2}}{2} \pm i l\varphi}\,,\qquad 
\zeta=\sqrt{\frac{m\omega}{\hbar}}r\,, &\\&
E_{n,l}=\hbar\omega(2n_r+ l+1)\,,\qquad
n_r\,,l=0,1,\ldots\,,&
\end{eqnarray}   
where $L_{n_r}^{(l)}$ are the Laguerre polynomials.
\vskip0.25cm
\noindent 
\underline{\emph{Unitary  transformation}}
\vskip0.25cm
\noindent 
The two representations corresponding to Cartesian and polar coordinates
can be related  by the unitary operator
\begin{eqnarray}
&
\label{UnitaryRepChange}
\hat{U}=\exp(-i\frac{2\pi}{3}\frac{1}{\sqrt{3}}(\hat{\mathcal{L}}_1+\hat{\mathcal{L}}_2+\hat{\mathcal{L}}_3) )\,,
&
\end{eqnarray}
 which produces  the rotation of the $\mathfrak{su}(2)$ generators in the  \emph{
``ambient
three-dimensional  space}"
 for the angle $-2\pi/3$, 
 \be
\hat{U}\hat{\mathcal{L}}_1 \hat{U}^{\dagger}=\hat{\mathcal{L}}_3\,,\qquad
\hat{U}\hat{\mathcal{L}}_2 \hat{U}^{\dagger}=\hat{\mathcal{L}}_1\,,\qquad
\hat{U}\hat{\mathcal{L}}_3 \hat{U}^{\dagger}=\hat{\mathcal{L}}_2\,.
 \ee
Then, from 
$
\hat{\mathcal{L}}_3\psi_{n_1,n_2}=\frac{n_1-n_2}{2}\psi_{n_1,n_2}$ 
we obtain 
$\hat{\mathcal{L}}_2 \hat{U}\psi_{n_1,n_2}=\frac{n_1-n_2}{2}\hat{U}\psi_{n_1,n_2}\,,
$
and $\hat{U}\psi_{n_1,n_2}$ are the states that diagonalize 
the Hamiltonian $\hat{H}_\text{os}=2\hbar\omega\hat{\mathcal{J}}_0$ and 
the operator 
$\hat{p}_{\varphi}=2\hbar\hat{\mathcal{L}}_2$ in the 
polar coordinates representation. 
Acting on operators $\hat{a}_i^\pm$, the unitary transformation produces  
\begin{eqnarray}
\label{UnitTranChange}
&
\hat{U}\hat{a}_j^\pm\hat{U}^\dagger= e^{\pm i\frac{\pi}{4}}\hat{b}_j^\pm\,,
&
\end{eqnarray}  
that corresponds to the spinor nature of the basic 
integrals with respect to the action of the $\mathfrak{su}(2)$  generators 
in the 
ambient
three-dimensional  space.
\vskip0.1cm

In the next section we show how these two different physical systems, the free particle and harmonic oscillator in $\R^2$, 
are related to each other by the 
conformal bridge transformation \cite{IPW}. This will serve as
a precedent 
for 
the procedure that we will use to extract all the information for a
system with the harmonic potential in  conical geometry from the free 
dynamics  on 
the same geometric background.

\section{The conformal bridge transformation}
\label{SecConBrid}

In the general case, a mechanical system is governed by a symmetry algebra which encodes its peculiarities. At the classical level, these 
symmetries can be related with the geometric 
properties of the trajectory, 
and at the quantum level they encode  the information related with the energy 
spectrum. 

According to Dirac \cite{Dirac}, 
a given symmetry algebra (up to isomorphisms) can represent different mechanical systems. 
In this section we show how the two forms of dynamics 
associated with the conformal algebra $\mathfrak{so}(2,1) \cong \mathfrak{sl}(2,\R)$ 
are related to each other by means of a particular mapping,
 the so-called conformal bridge transformation \cite{IPW}. 
  This also will allow us 
  to establish the interesting relation between the corresponding 
  symmetry sub-algebras of the free particle
  and harmonic oscillator systems and 
  understand   the change of  nature of some of them
  due to a non-unitarity of the conformal bridge 
  transformation.

Consider the classical $\mathfrak{so}(2,1)$ algebra 
\be
\{D,H\}=H\,,\qquad\{D,K\}=-K\,,
\qquad \{K,H\}=2D\,,\ee
without specifying the form of the generators themselves. 
Now, let us introduce
the following complex linear combinations of them,
\be
\label{sl2Rgen}
\mathcal{J}_0=\frac{1}{2}(\omega^{-1}H+\omega K)\,,\qquad
\mathcal{J}_\pm=-
\frac{1}{2\omega}( H- \omega^2 K \pm i2\omega D)\,,
\ee
where $ \omega $ is a constant  that is introduced   to compensate the dimensions of the generators.
 These new complex quantities satisfy the classical $\mathfrak{sl}(2,\R) $ algebra 
\be
\{\mathcal{J}_0,\mathcal{J}_\pm\}=\mp i \mathcal{J}_\pm\,,\qquad
\{\mathcal{J}_-,\mathcal{J}_+\}=-2i\mathcal{J}_{0} \,.
\ee
Independently, both algebraic structures can represent different physical systems. 
In the case where $ H $ is the Hamiltonian of a certain
asymptotically free  model with conformal symmetry (the free particle, for example),
 we have that $ K $ and $ D $ are dynamical integrals that explicitly depend on $ t $. 
 From here one has
\begin{eqnarray}
&K=T_H(t)(K_0)\,,\qquad K_0=T_H(-t)(K)=K|_{t=0}\,, &\\&
D=T_H(t)(D_0)\,,\qquad D_0=T_H(-t)(D)=D|_{t=0}\,,
&
\end{eqnarray}
where $ T_{H}(\pm t) $ denotes a Hamiltonian flow, which is a canonical transformation 
generated  by the Hamiltonian itself. 
The flow generated by a  phase space function $F$  is
\begin{eqnarray}&\label{TransCan}
\exp(\gamma F)\star f(q,p):=f(q,p)+\sum_{n=1}^\infty 
\frac{\gamma^n}{n!}\{F,\{\ldots,\{F,f\underbrace{\}\ldots\}\}}_{n}=:T_F(\gamma)(f)\,.&
\end{eqnarray}
In the same way, the compact generator $ 2 \omega \mathcal{J}_0 $ 
(ignoring the dependence on $ t $ of $F$  in the definition (\ref{sl2Rgen})) can be interpreted as the 
Hamiltonian of a harmonically trapped system (the harmonic oscillator, for example) 
with frequency $ \omega $, and the quantities $ \mathcal{J}_\pm $ are its dynamical integrals that satisfy 
\be
\mathcal{J}_\pm=T_{2\omega \mathcal{J}_0}(\tau)(\mathcal{J}_\pm|_{t=0})\,,\qquad
\mathcal{J}_\pm|_{t=0}=T_{2\omega \mathcal{J}_0}(-\tau)(\mathcal{J}_\pm)
\ee
Both forms of dynamics are related to each other  by the complex canonical transformation 
\be
\label{ClasicalBrige}
\mathscr{T}(\tau,\beta,\delta,\gamma,t)= T_{2\omega \mathcal{J}_0}(\tau)\circ T_{\beta\delta\gamma} \circ T_H(-t)\,,
\ee
where 
\begin{eqnarray}
&\label{Tabg0}
T_{\beta\delta\gamma}:=T_{K_0}(\beta)\circ T_{H}(\delta) \circ T_{D_0}(\gamma)=
 T_{K_0}(\delta)\circ T_{D_0}(\gamma) \circ T_{H}(2\delta)\,,
&\\&
\delta=\frac{i}{2\omega}\,,\qquad
 \beta=-i\omega \,,\qquad \gamma=-\ln 2\,.
&\label{betadeltagamma}
\end{eqnarray}
In this composition, the first transformation $T_{H}(-t)$ removes
 the $t$ dependence in the dynamical integrals $D$ and $K$.
The second transformation
relates these $t=0$ generators  with the generators of the $\mathfrak{sl}(2,\R)$ algebra $\mathcal{J}_0$ 
and $\mathcal{J}_\pm$ taken at $\tau=0$.  
The last transformation  $T_{2\omega \mathcal{J}_0}(\tau)$ restores the $\tau$ dependence. 
Explicitly  one has
\begin{eqnarray}
&\label{Conformal1}
\mathscr{T}(\tau, \beta,\delta,\gamma,t)(H)=-\omega\mathcal{J}_-\,, \qquad
\mathscr{T}(\tau, \beta,\delta,\gamma,t) (D)=-i\mathcal{J}_0\,, &\\&
\label{Conformal2}
 \mathscr{T}(\tau, \beta,\delta,\gamma,t)(K)=\frac{1}{\omega}\mathcal{J}_+\,.
&
\end{eqnarray}
The  corresponding inverse transformation is given by  
\begin{eqnarray}
&
(\mathscr{T}(\tau, \beta,\delta,\gamma,t))^{-1}=T_H(t)\circ (T_{\beta\delta\gamma})^{-1}\circ T_{2\omega \mathcal{J}_0}(-\tau)\,, &\\&
(T_{\beta\delta\gamma})^{-1}=T_{D_0}(-\gamma) \circ T_{H}(-\delta) \circ T_{K_0}(-\beta)\,.
&
\end{eqnarray}
This transformation is 
a generalization of
 the classical version of the quantum conformal bridge transformation introduced in \cite{IPW}, 
and corresponds to an automorphism of algebra since it relates 
the Wick rotated non-compact generator $ iD $ and 
compact  generator $ \mathcal{J}_0 $ to each other, 
the real-valued  non-compact generators $H$ and $K$ with the non-compact 
complex-valued
generators $-\omega\mathcal{J}_-$ and $\frac{1}{\omega}\mathcal{J}_+$, respectively. 

In the particular case of the free particle  
in the Euclidean space the generators of the  $ \mathfrak{so}(2,1) $ 
algebra are specified in  (\ref{Qmgen1}).
Then the generators of the  $ \mathfrak{sl}(2,\R)$ algebra
 resulting from applying the transformation 
 (\ref{ClasicalBrige})
  are given by the planar  harmonic oscillator generators $\mathcal{J}_0$ and $\mathcal{J}_\pm$
   defined  in (\ref{intnot})-(\ref{gen2}). 
 The transformation also allows us to map the remaining symmetry generators 
of the free particle system into the 
corresponding  generators of the  harmonic oscillator. 
In particular, by direct application of the transformation 
to the generators $p_\pm$ and $\xi_\pm$ we get 
\begin{eqnarray}
&\label{Conformal3}
\mathscr{T}(\tau, \beta,\delta,\gamma,t)(p_-)=-i\sqrt{2m\omega}\,b_1^-\,,\qquad
\mathscr{T}(\tau, \beta,\delta,\gamma,t)(p_+)=-i\sqrt{2m\omega}\,b_2^-\,,
&\\&\label{Conformal4}
\mathscr{T}(\tau, \beta,\delta,\gamma,t)(\xi_{+})=\sqrt{\frac{2m}{\omega}}\,b_1^+\,,\qquad
\mathscr{T}(\tau, \beta,\delta,\gamma,t)(\xi_-)=\sqrt{\frac{2m}{\omega}}\,b_2^+\,,
&
\end{eqnarray}
where the relations (\ref{Ideal1})-(\ref{FinalOssym}) were used. 
In the same vein, 
 we list the effects of the transformation on the remaining second order generators,
\begin{eqnarray}
&
\mathscr{T}(\tau, \beta,\delta,\gamma,t)(J_0)=\mathcal{L}_2\,,\qquad
\mathscr{T}(\tau, \beta,\delta,\gamma,t)(J_\pm)=-i\mathcal{L}_\pm\,,&\label{Conformal5}
\\&
\mathscr{T}(\tau, \beta,\delta,\gamma,t)(S_+)=\frac{1}{\omega}\mathcal{B}_1^+\,,\qquad
\mathscr{T}(\tau, \beta,\delta,\gamma,t)(T_-)=-\omega \mathcal{B}_1^-\,,\label{Conformal6}
&\\&
\mathscr{T}(\tau, \beta,\delta,\gamma,t)(S_-)=\frac{1}{\omega}\mathcal{B}_2^+\,,\qquad
\mathscr{T}(\tau, \beta,\delta,\gamma,t)(T_+)=-\omega \mathcal{B}_2^-\,.\label{Conformal7}
&
\end{eqnarray}

In the following table we summarize the correspondence between generators and some subalgebras in both systems,
\begin{table}[H]
\begin{center}
\begin{tabular}{| c | c |}
\hline
Free Particle & Harmonic Oscillator \\ \hline
$\mathfrak{so}(2,1)$\,: $(H,\,D,\,K)$& $\mathfrak{sl}(2,\R)$\,: $(\mathcal{J}_0,\,\mathcal{J}_\pm)$ \\
$\mathfrak{sl}(2,\R)$\,: $(J_0,\,J_\pm)$& $\mathfrak{su}(2)$\,: $(\mathcal{L}_2,\mathcal{L}_\pm)$ \\
$\mathfrak{e}_2$\,: $(J_0,\,T_\pm)$ & $\mathfrak{e}_2$\,: $(\mathcal{L}_2,\,\mathcal{B}_{j}^-)$\\
$\mathfrak{e}_2$\,: $(J_0,\,S_\pm)$ & $\mathfrak{e}_2$\,: $(\mathcal{L}_2,\,\mathcal{B}_{j}^+)$\\
$\mathfrak{e}_2$\,: $(p_\varphi,\,p_\pm)$ & $\mathfrak{e}_2$\,: $(\mathcal{L}_2,\,b_{j}^-)$\\
$\mathfrak{e}_2$\,: $(p_\varphi,\,\xi_\pm)$ & $\mathfrak{e}_2$\,: $(\mathcal{L}_2,\,b_{j}^+)$\\
$\mathfrak{su}(2)\oplus \mathfrak{su}(2)$\,: 
$(\ell_0^{(\pm)},\,\ell_\pm^{(\pm)})$& $\mathfrak{sl}(2,\R)\oplus \mathfrak{sl}(2,\R)$\,:
 $(\mathscr{J}_0^{(\pm)},\mathscr{J}_\pm^{(\pm)})$ \\
\hline
\end{tabular}
\caption{\small{Correspondence between some subalgebras and their generators. }}
\end{center}
\end{table}
\noindent
\vskip-0.75cm
Note that the quantity $J_0=\mathcal{L}_2$ is the only object which is invariant under the transformation. This happens because 
the angular momentum  is the only conserved quantity which Poisson commutes with all the 
$\mathfrak{sl}(2,\R)$ conformal symmetry generators of the free particle.
On the other hand, the second $\mathfrak{sl}(2,\R)$ 
subalgebra
is changed for the $\mathfrak{su}(2)$ symmetry  algebra 
after the transformation, and this is due to 
the imaginary unit appearing in the second equation in 
(\ref{Conformal5}). 
In the same vein, the two copies of the $\mathfrak{su}(2)$
algebra of the free particle with generators related 
by the complex conjugation are mapped into the two copies
of the $\mathfrak{sl}(2,\R)$ algebra associated with the Landau problem
due to the appearance  of the minus sign 
on the right hand side in the transformations of $T_\pm$ 
in (\ref{Conformal6}) and (\ref{Conformal7}).

The quantum analogue of the conformal bridge transformation corresponds to the non-unitary transformation produced by the operators 
\begin{eqnarray}
&
\hat{\mathfrak{S}}(t,\tau)=e^{-\frac{i}{\hbar}2\omega \hat{\mathcal{J}}_0\tau}e^{-\frac{\omega}{\hbar} \hat{K}_0} 
e^{\frac{\hat{H}}{2\hbar \omega}}
e^{\frac{i}{\hbar}\ln(2)\hat{D}_0}e^{\frac{i}{\hbar}\hat{H}t}\,,&\label{QCB1}\\&
\hat{\mathfrak{S}}^{-1}(t,\tau)=e^{-\frac{i}{\hbar}\hat{H} t}e^{-\frac{i}{\hbar}\ln(2)\hat{D}_0}e^{-\frac{\hat{H}}{2\hbar \omega}}
e^{\frac{\omega}{\hbar}\hat{K}_0}e^{\frac{i}{\hbar}2\omega \hat{\mathcal{J}_0}\tau}\,.
&\label{QCB2}
\end{eqnarray}
In this context, the Hamiltonian 
flux is changed by employing the 
 Baker-Campbell-Hausdorff formula, 
and in correspondence with 
 the established relation 
 between  the two-dimensional free particle and the planar isotropic 
harmonic oscillator in the Euclidean space, the classical relations  
(\ref{Conformal1})-(\ref{Conformal2}) and  (\ref{Conformal3})-(\ref{Conformal7}) 
are preserved at the quantum level\footnote{We have considered dimensionless operators 
for the harmonic oscillator at the quantum level in the previous section. To recover 
these generators by the conformal bridge transformation 
we must compensate 
the multiplicative constants that appear in the quantum versions of the above-mentioned 
relationships. The relations involving 
the  momenta operators linearly (quadratically) are multiplied by $\hbar^{1/2}$ ($\hbar$). 
This is 
taken into account in equations (\ref{QCB1}), (\ref{QCB2}).
}.

In what concerns to  eigenstates, in the general case
the transformation implies that 
\begin{eqnarray}
&\label{EigenEstates}
\hat{D}\ket{\lambda}=i\hbar \lambda \ket{\lambda}\quad \Rightarrow\quad \hat{\mathcal{J}}_0(\hat{\mathfrak{S}}\ket{\lambda})=\lambda\hat{\mathfrak{S}}\ket{\lambda}\,,
&\\&
\label{CoherentEstates}
\hat{H}\ket{E}=E\ket{E}\quad \Rightarrow\quad \hat{\mathcal{J}}_-(\hat{\mathfrak{S}}\ket{E})=-\frac{E}{\hbar 
\omega}\hat{\mathfrak{S}}\ket{E}\,.&
\end{eqnarray}
This means that the formal eigenstates 
of the dilatation operator with imaginary eigenvalue are mapped to the 
energy eigenstates of the harmonically confined system, 
while 
 the asymptotic plane wave eigenstates of 
the
asymptotically free Hamiltonian
$\hat{H}$ correspond to coherent states of 
the system with the 
additional harmonic potential term, which, in turn, are eigenstates of 
the  quadratic lowering operator. 
Note that to have the physically acceptable solutions, the states 
$\hat{\mathfrak{S}}\ket{\lambda}$ must be normalizable, and to ensure
this,
the  three requirements must be met. First, 
the series  \begin{eqnarray}
&
  e^{\frac{\hat{H}}{2\hbar \omega}}\ket{\lambda}=\sum_{n=0}^{\infty}\frac{1}{n!(2\hbar \omega)^{n}} (\hat{H})^n\ket{\lambda}\,,&
  \end{eqnarray}
 has  to  reduce to  a finite number of terms, i.e., $\ket{\lambda}$ should 
be the Jordan states\footnote{The Jordan states are given by wave functions
that
satisfy $P(\hat{H})\Omega_{\lambda}=\psi_{\lambda}$, where $\hat{H}\psi_{\lambda}=\lambda\psi_{\lambda}$ and $P(\eta)$ 
represents a polynomial 
\cite{CJP,InzPly3}.
Here
we consider Jordan states satisfying  
the relations $(\hat{H})^\ell\Omega_{\lambda}=\lambda\psi_{\lambda}$ with $\lambda=0$ for a certain 
natural $\ell$.} of $\hat{H}$ corresponding to  zero energy \cite{IPW}. 
Second, the function  $\braket{\vr}{\lambda}$ must not have singularities in the corresponding 
operators domain. 
And finally, these functions must be single-valued with respect to the angular coordinate.

In conclusion of this section, let us apply 
the inverse conformal bridge 
(similarity) transformation
to
 the unitary 
 operator
 (\ref{UnitaryRepChange}). This yields  us  
 the non-unitary operator
\begin{eqnarray}&
\label{Wop}
\hat{W}=\exp(-i\frac{2\pi}{3}\frac{1}{\sqrt{3}\hbar}(\hat{J}_0+i(\hat{J}_1-i\hat{J}_2)) )\,,&
\end{eqnarray}
and from equations (\ref{UnitTranChange}) one obtains the transformation relations
\begin{eqnarray}
&
\hat{W}(\hat{\xi}_1)\hat{W}^{-1}=\frac{1}{\sqrt{2}}e^{i\frac{\pi}{4}}\hat{\xi}_+\,,\qquad
\hat{W}(\hat{\xi}_2)\hat{W}^{-1}=\frac{1}{\sqrt{2}}e^{i\frac{\pi}{4}}\hat{\xi}_-\,,&\label{T1}\\&
\hat{W}(\hat{p}_1)\hat{W}^{-1}=\frac{1}{\sqrt{2}}e^{-i\frac{\pi}{4}}\hat{p}_+\,,\qquad
\hat{W}(\hat{p}_2)\hat{W}^{-1}=\frac{1}{\sqrt{2}}e^{-i\frac{\pi}{4}}\hat{p}_-\,.
&\label{T2}
\end{eqnarray}
Equations (\ref{T1}), (\ref{T2})
provide  us with  the 
similarity  transform of the Cartesian operators 
$\hat{\xi}_j$ and $\hat{p}_j$
into the complex operators 
$\hat{\xi}_\pm$ and  $\hat{p}_\pm$ used
 in the  polar coordinates representation for the quantum free particle. 
The conformal symmetry generators 
of the free particle, $\hat{H}$, $\hat{D}$ and $\hat{K}$, 
commute with the operator  (\ref{Wop}). 

\vskip0.1cm
In the next section we finally pass over to the 
study of the free 
particle
dynamics in conical geometry. 
Since the system possesses the conformal $\mathfrak{sl}(2,\R)$  and rotational symmetries  
for any 
value of $\alpha$,  then it will be possible to apply
the conformal bridge transformation
 to analyze  the dynamics of 
 the harmonic oscillator in the same  geometry.

\section{Free motion in a cosmic string background}
\label{SecFreeCone}
As we have shown in Sec. \ref{SecGeo},
the study of the dynamics in a cosmic string background 
is analogous to 
analyzing
 the motion of a particle in conical geometry. A classical system in this space is governed 
by the  action 
\begin{eqnarray}&\label{Lagrangian}
I=\int L dt\,,\qquad 
L=\frac{m}{2}g_{ij}\frac{dx_i}{dt}\frac{dx_j}{dt}-V(\vr)=\frac{m}{2}\left(\alpha^2 \dot{r}^2+r^2\dot{\varphi}^2\right)-V(\vr)\,.&
\end{eqnarray}

In this section, we study the case of the 
free motion ($ V (\vr) = 0 $) from the perspective of its symmetries. 
We will show that  
when  $\alpha $ is a rational number, it is possible to construct higher-order 
globally well-defined classical integrals of motion
that can be identified as generators of hidden symmetries \cite{Cariglia}.
Then 
we study the system at the quantum level,
and show that the only cases in which these conserved 
quantities can be promoted to the well-defined symmetry
operators
correspond to 
integer values of $ \alpha $.
Thus, we reveal  here a kind of a quantum anomaly
in the case of
rational, non-integer values of $ \alpha $.

\subsection{Classical case}
\label{SubsecClasFree}
The classical dynamics of a free particle in a
conical geometry is governed by the Hamiltonian
\begin{eqnarray}
\label{FreConH}
&H^{(\alpha)}=\frac{1}{2m}\left(\frac{p_r^2}{\alpha^2}+\frac{p_{\varphi}^2}{r^2}\right) \,. &
\end{eqnarray}
Remarkably, the canonical transformation
\begin{eqnarray}
\label{Canonicaltrnas}
&r\rightarrow \alpha r\,,\qquad
p_{r}\rightarrow\frac{p_r}{\alpha}\,, \qquad
\varphi\rightarrow\frac{\varphi}{\alpha}\,,\qquad
p_{\varphi}\rightarrow\alpha p_{\varphi}\,,&
\end{eqnarray}
applied 
to  the Hamiltonian of 
the Euclidean free particle  
gives us the
Hamiltonian (\ref{FreConH}). 
Also  note that when we apply this transformation to the usual Cartesian coordinates 
we get the ``regularized" Cartesian coordinates (\ref{Xi}). From the analysis
of Sec. \ref{SecGeo}
related to  
those coordinates
it is clear that the canonical transformation
(\ref{Canonicaltrnas})
is well-defined only locally.

\begin{figure}[H]
\begin{center}
\hskip1.5cm
\begin{subfigure}[c]{0.28\linewidth}
\includegraphics[scale=0.45]{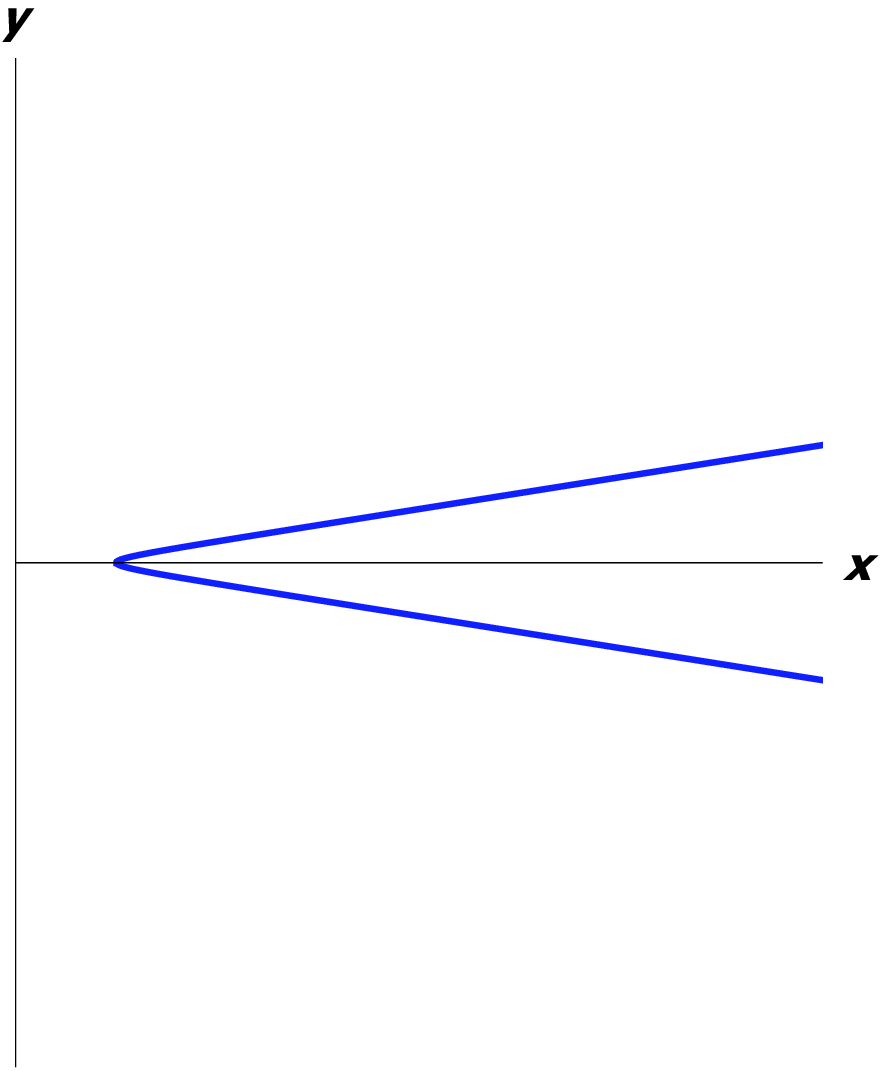}
\caption{\small{$\alpha=1/10$}}
\end{subfigure}
\begin{subfigure}[c]{0.28\linewidth}
\includegraphics[scale=0.45]{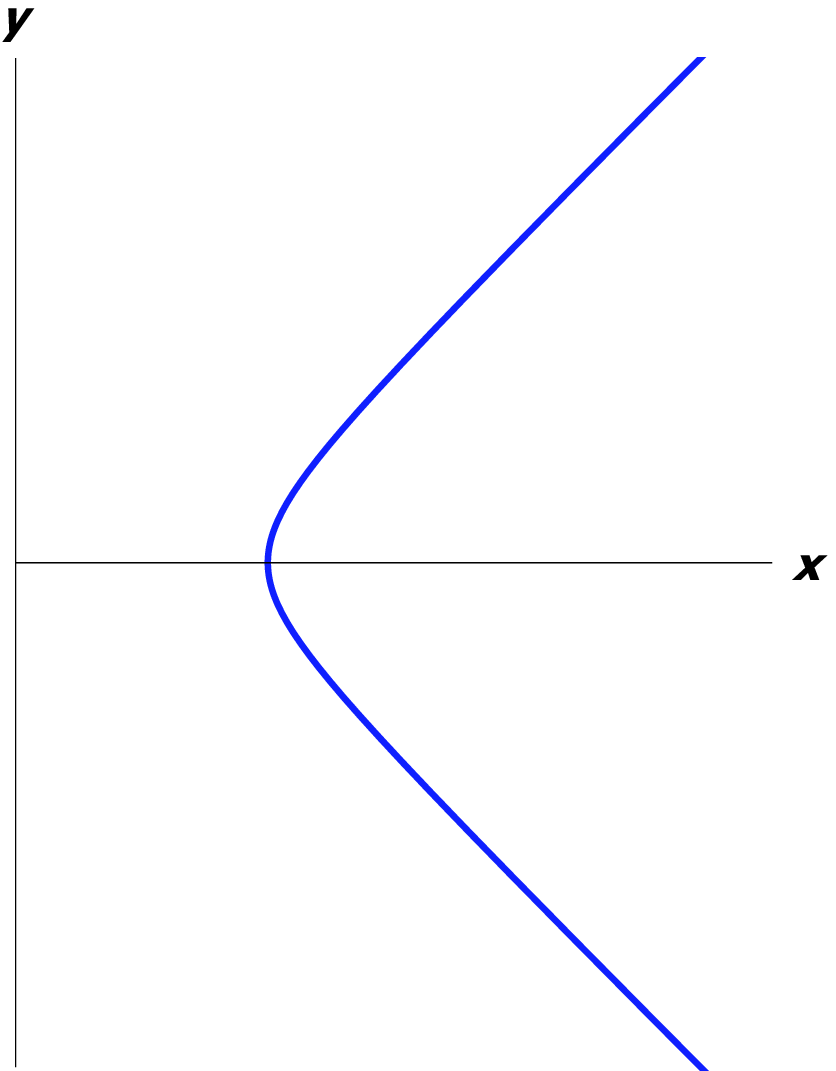}
\caption{\small{$\alpha=1/2$}}
\end{subfigure}
\begin{subfigure}[c]{0.28\linewidth}
\includegraphics[scale=0.45]{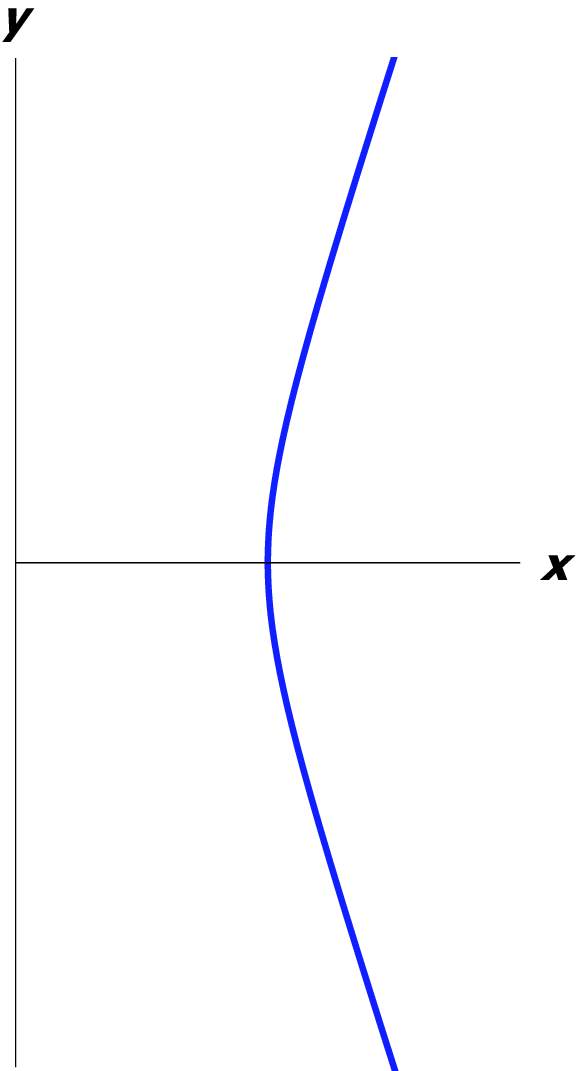}
\caption{\small{$\alpha=4/5$}}
\end{subfigure}
\vskip0.5cm
\hskip0.5cm
\begin{subfigure}[c]{0.25\linewidth}
\includegraphics[scale=0.45]{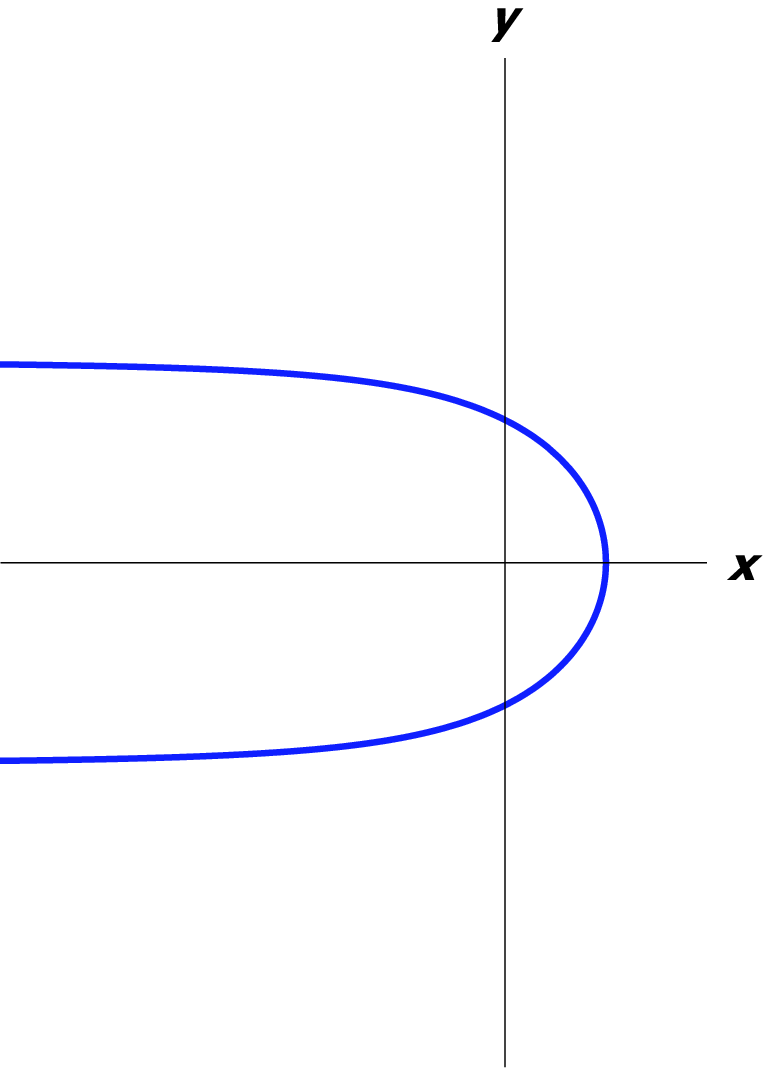}
\caption{\small{$\alpha=2$}}
\end{subfigure}
\begin{subfigure}[c]{0.2\linewidth}
\includegraphics[scale=0.45]{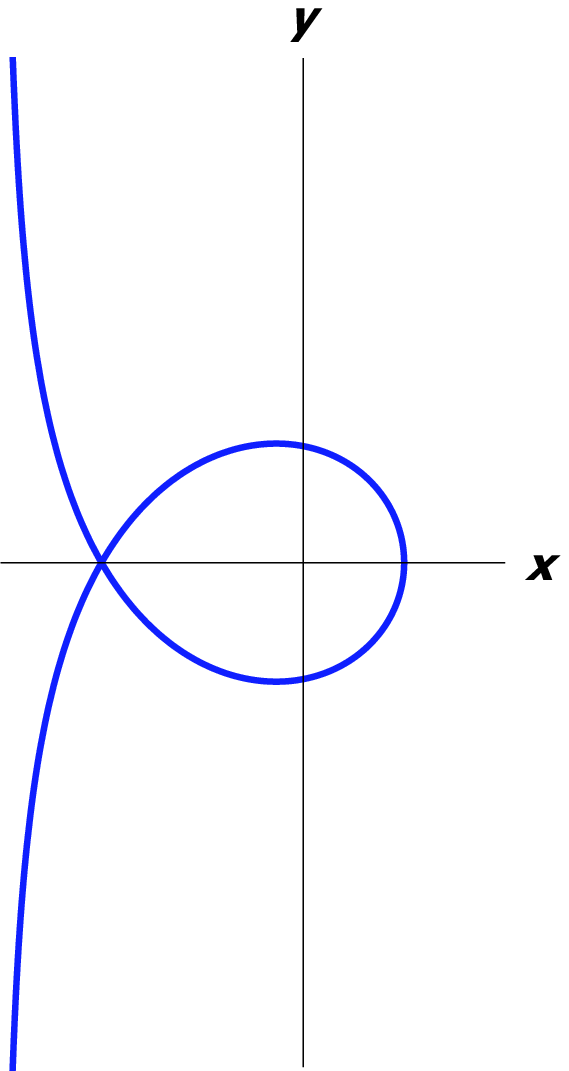}
\caption{\small{$\alpha=3$}}
\end{subfigure}
\begin{subfigure}[c]{0.2\linewidth}
\includegraphics[scale=0.45]{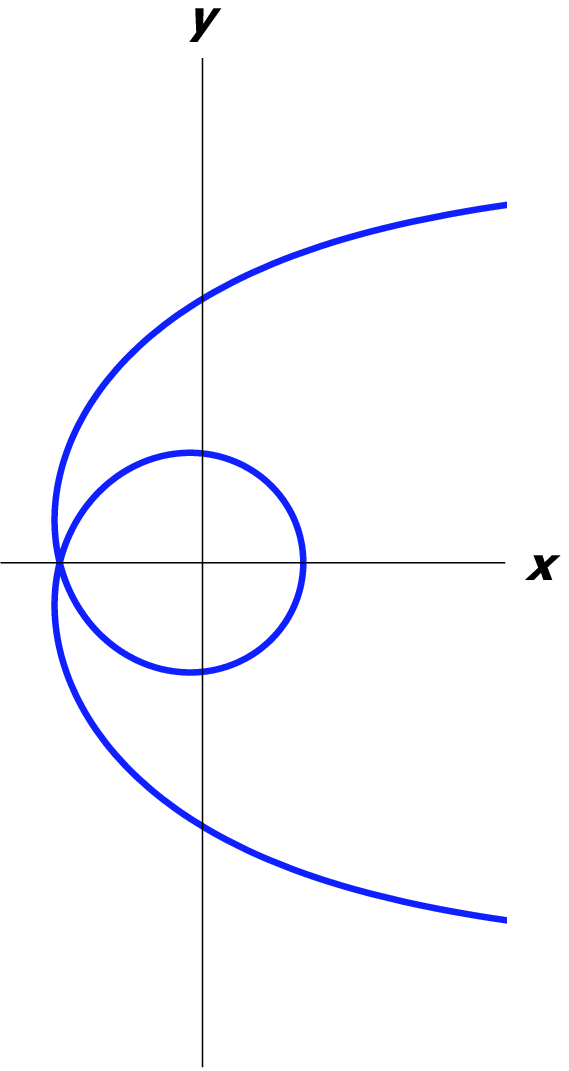}
\caption{\small{$\alpha=4$}}
\end{subfigure}
\begin{subfigure}[c]{0.25\linewidth}
\includegraphics[scale=0.45]{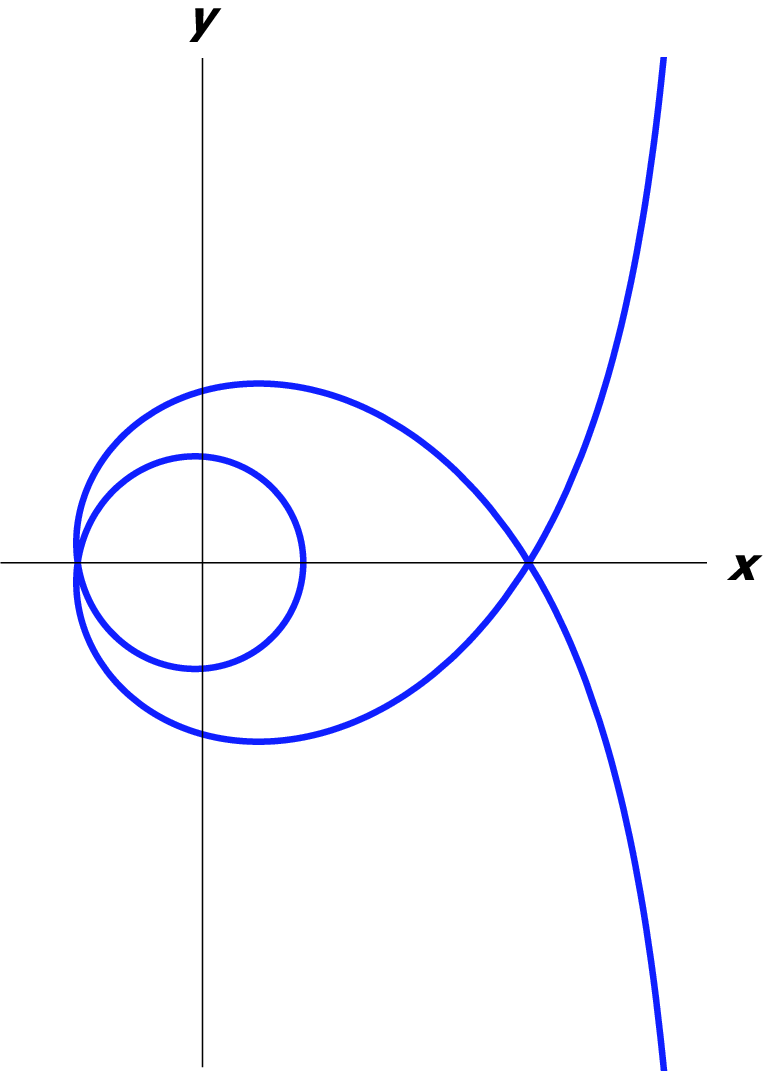}
\caption{\small{$\alpha=5$}}
\end{subfigure}
\end{center}
\caption{\small{Some examples of the geodesics  in the conical geometry.
The axes are oriented so that the perihelion is on the positive side of the $ x $-axis
($\varphi_*=0$).
From the first three figures one sees that for $0<\alpha<1$, the dynamics is somewhat similar to
that in the Kepler-Coulomb  problem in the case of repulsive potential.
 For the shown 
integer  values of $\alpha$ the $t=-\infty$ and $t=+\infty$ straight line asymptotes 
coincide with  a vertical line
for odd values of $\alpha$, and are parallel horizontal lines 
for even values of $\alpha$.}
 }
\label{figure1}
\end{figure}

In spite of the indicated deficiency,
we can use the canonical transformation (\ref{Canonicaltrnas}) to reconstruct 
 the solutions of the equations of motion 
of the system 
 (\ref{FreConH}). The trajectory equation, 
 as well as the time dependence of the radial and angular variables
are immediately  obtained   from the corresponding 
relations (\ref{r(t)}), (\ref{r(varphi)})  and (\ref{varphifree})
 for the free  motion in the plane, and
  are given by
\begin{eqnarray}
\label{Trayalpha}
&r(\varphi)=\frac{r_*}{\cos\left((\varphi-\varphi_*)/\alpha\right)}\,,\qquad
r_*=\frac{p_\varphi}{\sqrt{2mH^{(\alpha)}}}\,,
\qquad -\frac{\pi}{2}\alpha\leq \varphi-\varphi_* \leq \frac{\pi}{2}\alpha\,.&\\
&r^2(t)=\frac{2}{\alpha^2m}\left(H^{(\alpha)}t^2+2Dt+K\right)\,,\quad
\varphi(t)=\alpha\arctan\left(\frac{1}{2\rho}(t+\frac{D}{H^{(\alpha)}})\right)+\varphi_*\,,\label{Trayalpha+}&
\end{eqnarray}
where $\rho=\alpha\frac{p_\varphi}{2H^{(\alpha)}}\,.$
Here $H^{(\alpha)}$, 
$D$, $K$
and $p_\varphi$ 
are the integrals being generators of the $\mathfrak{sl}(2,\R)\oplus\mathfrak{u}(1)$ symmetry of the 
system (\ref{FreConH}), 
see Eqs. (\ref{ClasicalH}), (\ref{ClasicalH2}) below,
and we see that the scattering angle is $\varphi_\text{scat}= \varphi(+\infty)-
\varphi(-\infty)=\alpha\pi$ under assumption $p_\varphi>0$.
Some pictures of the trajectory for different values of $\alpha$ are displayed in Fig \ref{figure1}. 

\vskip-0.5cm
From these figures one sees  that when $ \alpha $ is an even number,
$\alpha=2\ell$, the particle experiences a backward scattering, 
that corresponds to the scattering angle $2\pi\ell$, while when
$ \alpha $ is odd, $\alpha=2\ell+1$, the scattering angle is $2\pi\ell +\pi$, and the 
particle continues asymptotic motion 
in the initial direction 
after realizing $\ell$  revolutions  around the origin being the vertex of the cone. 
This was already observed in ref. \cite{DesJack}\footnote{Note 
that our parameter $\alpha$ corresponds to $\alpha^{-1}$ in 
notations of \cite{DesJack}, where only the case of positive mass
density of the cosmic string (that corresponds to $\alpha>1$ values
of our parameter) was considered.}.

\begin{figure}[H]
\begin{center}
\hskip0.75cm
\begin{subfigure}[c]{0.45\linewidth}
\includegraphics[scale=0.6]{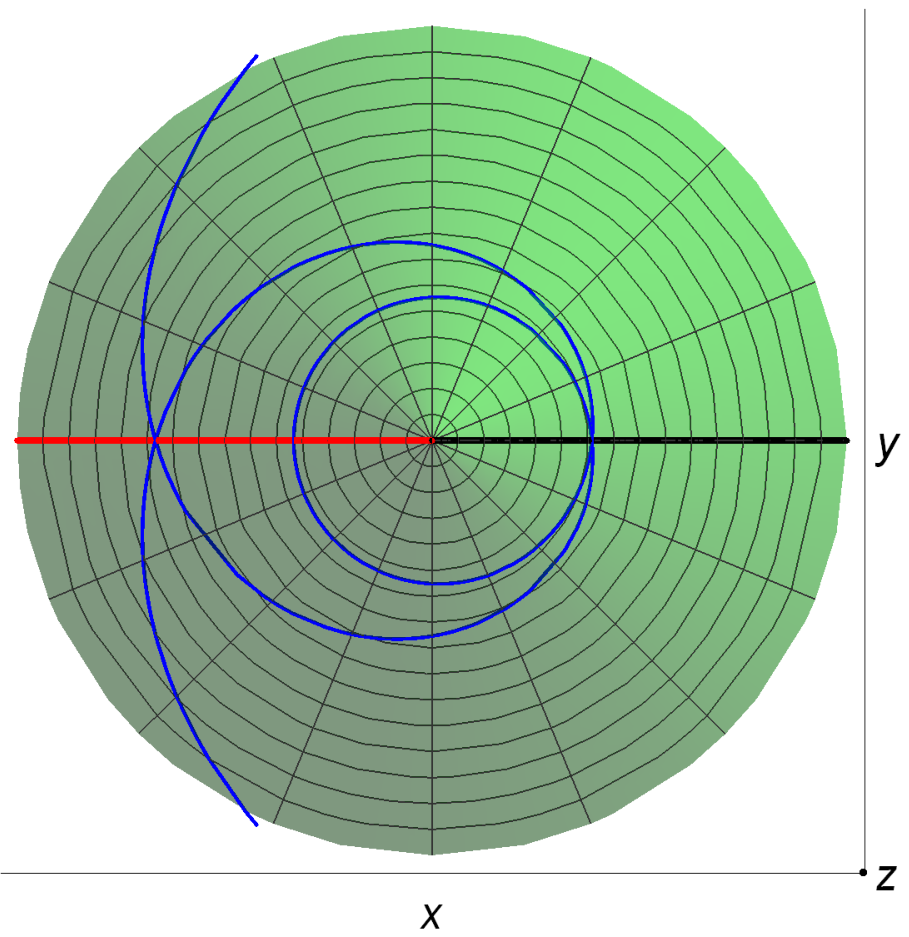}
\caption{\small{Cutting the cone.}}
\label{Fig1a}
\end{subfigure}
\begin{subfigure}[c]{0.4\linewidth}
\includegraphics[scale=0.15]{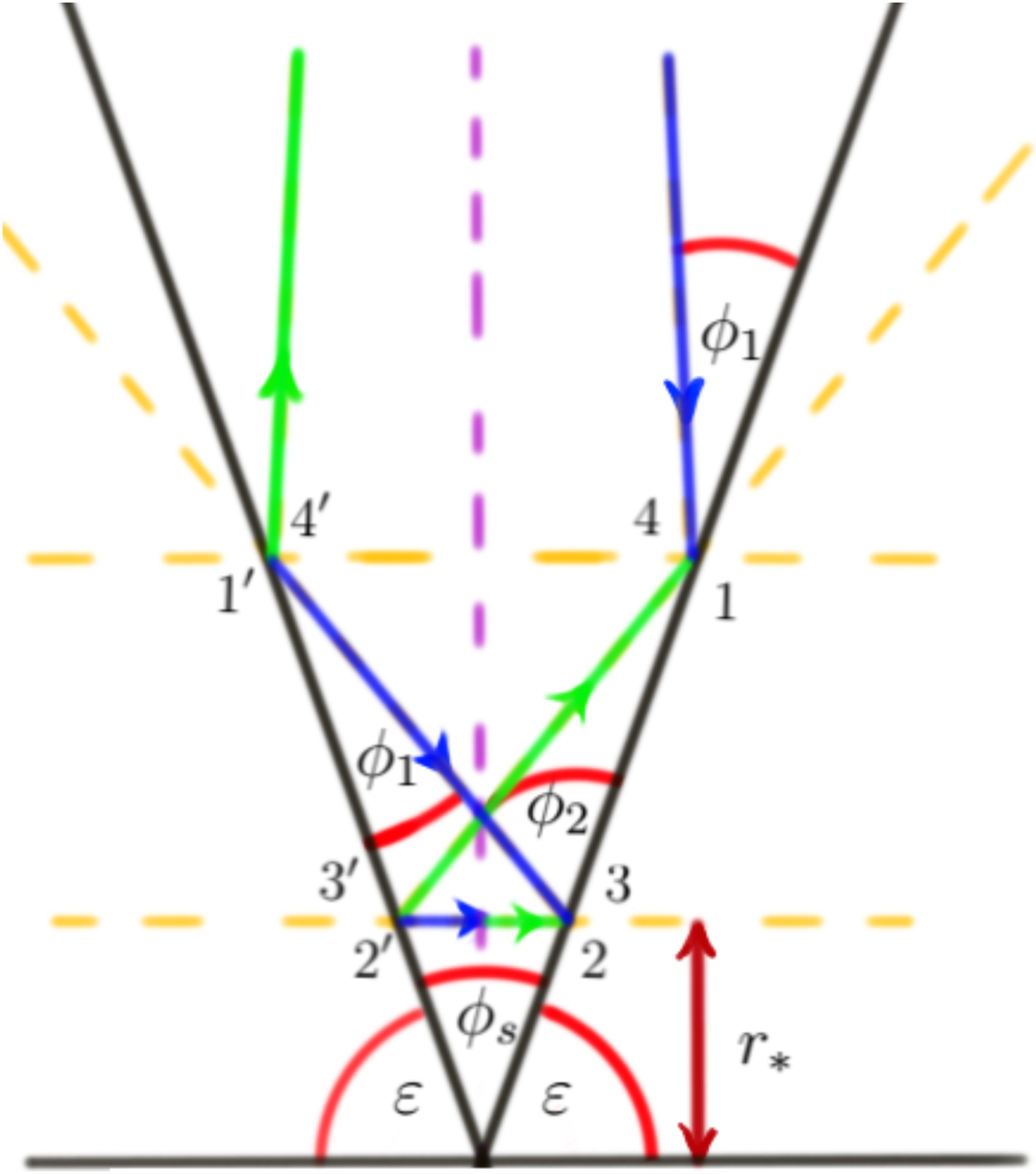}
\caption{\small{Rectilinear geodesic on the cut and flattened cone.}}
\label{Fig1b}
\end{subfigure}
\vskip0.25cm
\hskip0.5cm
\begin{subfigure}[c]{0.7\linewidth}
\includegraphics[scale=0.2]{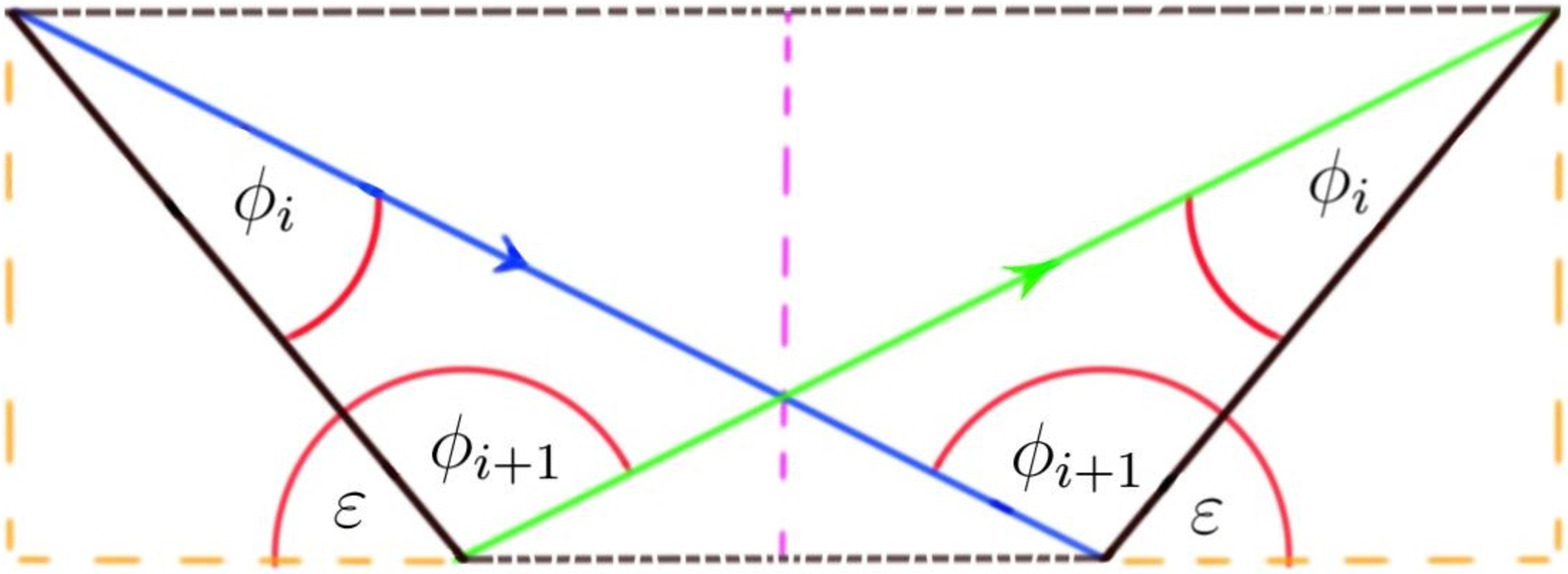}
\caption{\small{Incidence and reflection angles at the edges of the angular sector.}}
\label{Fig1c}
\end{subfigure}
\end{center}
\caption{ \small{
Rectilinear geodesic motion in the cone.\\
 Fig. \ref{Fig1a}: 
To construct the flattened cone, 
we
cut it along the straight line  (shown in black), which is opposite (with respect to the cone symmetry axis) 
to the straight line (shown in red) going from the cone's vertex to the point of perihelion. 
As a result we obtain a flattened sector 
of the angle $\varphi_s=\frac{2\pi}{\alpha}$, with identified edges. \\
Fig. \ref{Fig1b}: The rectilinear character of geodesic in the obtained angular sector 
with identified edges.
The incoming part of the trajectory (from infinity to 
the point of perihelion $r_*$) is represented in blue, while the outgoing part 
(from $r_*$ to infinity) is shown in green.  
 When the particle arrives at the right edge in a point $i$
at  an angle  of incidence $ \phi_i $, 
it emerges  from the left edge in the point $i'$ at the same angle, 
and continues its rectilinear path to the point 
$(i+1)$, and so on. 
In the general case  the particle ``hits" the edges several times. 
 Note that by the construction, the lowest triangle on the diagram, appearing when  $\alpha>2$, 
 with the vertex corresponding to the 
 cone's vertex and opposite side going through the perihelion $r_*$, 
 is always isosceles. This  implies  that the last 
 angle of incidence equals $ \varepsilon=\pi\left(\frac{1}{2}-\frac{1}{\alpha}\right)$.\\
 Fig. \ref{Fig1c}: The shown  geometric structure appears 
 several times in accordance with the number of 
 nonzero incidence angles. 
 Using the geometric properties of the triangles, one has that $ \phi_{i+ 1} = \phi_{i}-2\varepsilon + \pi $, and 
 making use of  the fact 
 that the last angle corresponds to $\varepsilon$, we obtain $ \phi_{n-\ell}=\pi\left(\frac{1}{2} - \frac{2\ell+1}{\alpha}\right)$ 
 where $n$ represents the number of incidence angles (the number of ``reflections" at the edges for $\alpha>2$)
 and  $ \ell = 0,1, \ldots, n-1$. From the condition $\phi_{1}>0$ we get that $2(2n-1)<\alpha$.}}
\label{figure2}
\end{figure}

To reveal the locally rectilinear  
character of geodesics in the case $\alpha>1$,  
it is convenient to consider the  shape of trajectories in a cut and  flattened cone 
represented by the angular sector $-\frac{\pi}{2}\alpha\leq \varphi\leq \frac{\pi}{2}\alpha$
(wedge)
in which the symmetric points on the edges  $\varphi=-\frac{\pi}{2}\alpha$ and 
$\varphi=\frac{\pi}{2}\alpha$ 
correspond to the cut line and must be identified.
This is illustrated by  Fig. \ref{figure2}. 

\vskip-0.25cm

From the diagrams shown  in Fig. \ref{figure2} 
one reveals  the following special features in the case of integer $\alpha$:
 \begin{itemize}
 \item[1.] If we look for the case in which the asymptotes of the trajectory are parallel to symmetry 
 axis (going through the cone's vertex and the perihelion),
  we must impose the condition $ \phi_1 + \varepsilon = \pi/2 $ (where $ \varepsilon=\phi_ {n}$), 
  which can be fulfilled if $ \alpha = 4n $, with $ n = 1,2, \ldots $.
 \item[2.] When $\alpha=2(2n+1)$, one has that the first incidence
  angle is given by $\pi/(2n+1)$, which coincides with $\varphi_s$. 
 This implies that  the asymptotes of the path are parallel to the
 edges.  
This  also includes the case $\alpha=2$ for which $n=0$ and the trajectory 
 suffers no reflections at the edges.

 \item[3.] When  $\alpha=4n-1$, the  first incidence   
 angle corresponds to $\frac{1}{4n-1}\frac{\pi}{2}=\frac{\varphi_s}{4}$. 
 When 
 $\alpha=4n+1$, the first incidence
 angle is given by $\frac{3}{4n+1}\frac{\pi}{2}=\frac{3}{4}\varphi_s$. 
 \end{itemize}

Now, let's pass over  to constructing the symmetry generators of the model
by  
applying the transformation (\ref{Canonicaltrnas}) to those of the Euclidean free particle
analyzed
  in Sec. \ref{SecEmotion}. 
We
 start with the complex combinations of the 
locally defined translations
and the Galilean boost generators, 
\begin{eqnarray}
&\label{Pi+-}
\Pi_\pm=\Pi_1 \pm i \Pi_2= \left(\frac{p_r}{\alpha}\pm i \frac{p_\varphi}{r}\right)e^{\pm i\frac{\varphi}{\alpha}}\,,&\\&
\label{X+-}
\Xi_\pm=\Xi_1 \pm i\Xi_2=
\left[
\alpha m r-t \left(\frac{p_r}{\alpha}\pm i \frac{p_\varphi}{r}\right)\right]e^{\pm i\frac{\varphi}{\alpha}}\,.
&
\end{eqnarray} 
Note that the presence of the factor $ \alpha^{-1} $ in the integrals (\ref{Pi+-}) and (\ref{X+-}) implies that they
are globally
well-defined functions on the phase space only in the case of natural values of $\alpha^{-1}=k=1,2,\ldots$, and otherwise they are formal 
objects.
In spite of this deficiency,
these quantities  will serve as the basis for constructing the 
globally well-defined in 
the phase space  symmetry generators of the system. 
For
arbitrary values of $\alpha$ we have 
\begin{eqnarray}&
\label{ClasicalH}
H^{(\alpha)}=\frac{1}{2m}\Pi_+\Pi_-\,,\qquad
D=\frac{1}{4m}(\Xi_+\Pi_-+ \Pi_+\Xi_+)= \frac{rp_r}{2}-H^{(\alpha)}t\,,&\\&
K=\frac{1}{2m}\Xi_+\Xi_- =\frac{m}{2}\alpha^2r^2-2Dt-H^{(\alpha)}t^2\,, \qquad 
J_0= \frac{i}{4m}(\Xi_+\Pi_- - \Pi_+\Xi_+)=\frac{\alpha}{2} p_\varphi\,,
\label{ClasicalH2}
&
\end{eqnarray} 
which are the well-defined angular-independent generators of the $\mathfrak{sl}(2,\R)\oplus\mathfrak{so}(2)$
symmetry,
\begin{eqnarray}
&
\{D,H^{(\alpha)}\}=H^{(\alpha)}\,,\qquad\{D,K\}=-K\,,
\qquad \{K,H^{(\alpha)}\}=2D\,, &\\&
\{J_0,H^{(\alpha)}\}=\{J_0,D\}=\{J_0,K\}=0\,,&
\end{eqnarray}
with the  $\mathfrak{sl}(2,\R)$ subalgebra Casimir element  $D^2-H^{(\alpha)}K=-J_0^2$.

For rational values $ \alpha =q/k $,
$ q, k = 1, 2,\ldots $, local integrals
(\ref{Pi+-})-(\ref{X+-}) 
can be used to construct the higher-order 
globally well-defined
integrals
\begin{eqnarray}
&
\label{ClasicalO}
\mathcal{O}_{\mu,\nu}^\pm=(\Xi_\pm)^\mu (\Pi_\pm)^\nu\,,
\qquad 
\mu=0,1,\ldots,q,\qquad  \nu=q-\mu\,, &\\&
\label{ClasicalS}
\mathcal{S}_{\mu',\nu'}^\pm=(\Xi_\pm)^{\mu'} (\Pi_\pm)^{\nu'}\,,
\qquad 
\mu'=0,1,\ldots,2q, \qquad \nu'=2q-\mu' \,.&
\end{eqnarray}
There are $ 2 (q + 1) $  conserved quantities of the type $\mathcal{O}^\pm_{\mu,\nu} $
and $ 2 (2q + 1) $   conserved quantities of the type $ \mathcal{S}^\pm_{\mu',\nu'}$, 
which have the  angular dependence  of the form $e^{\pm i k\varphi}$  and 
$e^{\pm i2 k\varphi}$, respectively.
So, these 
are globally well-defined functions in the phase space.

 In the Euclidean case 
$ q = k = 1 $, 
 the integrals (\ref{ClasicalO}) correspond to 
 the generators of the two-dimensional 
 Heisenberg algebra  (\ref{Hei2}), 
 while
 the integrals (\ref{ClasicalS}) 
 constitute   the set of  the second-order 
 integrals
 $ J_0 $, $ J_\pm $, $ T_\pm $, $ S_\pm $
defined by  Eqs. (\ref{Qmgen1+}), (\ref{Qmgen2}).
  In the case of $ q = 1 $ and $ k = 2,3, \ldots $, 
 i.e. when $\alpha$ is a unit fraction $1/k$ with $k>1$,
the   integrals 
$\Pi_\pm$ and $\Xi_\pm$ are 
well-defined functions in the corresponding 
phase space, and the set 
of integrals (\ref{ClasicalO}), (\ref{ClasicalS}) is similar to that
of the free particle in Euclidean plane. 
Together with the conformal symmetry generators $H^{(\alpha)}$, $D$ and $K$,
they generate 
the same 
Lie
 algebra 
  as for the free particle in the plane
since the canonical transformation (\ref{Canonicaltrnas}) 
does not change the Poisson bracket relations.

To characterize the symmetry algebra for $q>1$, which will be nonlinear in the general case, 
it is necessary to calculate some Poisson bracket relations, and for this we use the identity
$
\{A,B^n\}=n\{A,B\}B^{n-1}
$
together with the Poisson bracket
relations 
\begin{eqnarray}
\label{FormalClaAl1}
&\{\Xi_\pm,\Pi_\mp\}=2m\,,&\\&
\{H^{(\alpha)},\Xi_\pm\}=-\Pi_{\pm}\,,\qquad
\{D,\Xi_\pm\}=-\frac{1}{2}\Xi_\pm\,,\qquad 
\{J_0,\Xi_\pm\}=\mp \frac{i}{2}\Xi_\pm \,, &\\&
\{K,\Pi_\pm\}=\Xi_{\pm}\,,\qquad\{D,\Pi_\pm\}=\frac{1}{2}\Pi_\pm\,,\qquad 
\{J_0,\Pi_\pm\}=\mp\frac{i}{2}\Pi_\pm\,.
\label{FormalClaAl2}
&
\end{eqnarray}
For integrals $\mathcal{O}_{\mu,\nu}^\pm$ we have then
\begin{eqnarray}
&
\{J_0,\mathcal{O}_{\mu,\nu}^\pm\}=\mp i\frac{\mu+\nu}{2}\mathcal{O}_{\mu,\nu}^\pm\,,\qquad
\{D,\mathcal{O}_{\mu,\nu}^\pm\}=\frac{\nu-\mu}{2}\mathcal{O}_{\mu,\nu}^\pm\,,\label{DconOClasic}
&\\&
\{H^{(\alpha)},\mathcal{O}_{\mu,\nu}^\pm\}=-\mu\mathcal{O}_{\mu-1,\nu+1}^\pm\,,\quad
\{K,\mathcal{O}_{\mu,\nu}^\pm\}=\nu\mathcal{O}_{\mu+1,\nu-1}^\pm\,,
\quad\{\mathcal{O}_{\mu,\nu}^\pm,\mathcal{O}_{\lambda,\sigma}^\pm\}=0\,,
&
\end{eqnarray}
and by using the Jacobi identity we get 
\begin{eqnarray}
\label{Jacobi1}
&\{J_0,\{\mathcal{O}_{\mu,\nu}^+,\mathcal{O}_{\lambda,\sigma}^-\}\}=0\,,\qquad
\{D,\{\mathcal{O}_{\mu,\nu}^+,\mathcal{O}_{\lambda,\sigma}^-\}\}=\frac{\nu-\mu+\sigma-\lambda}{2}\{\mathcal{O}_{\mu,\nu}^+,\mathcal{O}_{\lambda,\rho}^-\}\,,&\\&
\label{Jacobi3}
\{H^{(\alpha)},\{\mathcal{O}_{\mu,\nu}^+,\mathcal{O}_{\lambda,\sigma}^-\}\}=-\lambda\{\mathcal{O}_{\mu,\nu}^+,\mathcal{O}_{\lambda-1,\sigma+1}^-\}-\mu\{\mathcal{O}_{\mu-1,\nu+1}^+,\mathcal{O}_{\lambda,\sigma}^-\})\,,
&\\&
\label{Jacobi4}
\{K,\{\mathcal{O}_{\mu,\nu}^+,\mathcal{O}_{\lambda,\sigma}^-\}\}\}=\sigma\{\mathcal{O}_{\mu,\nu}^+,\mathcal{O}_{\lambda+1,\sigma-1}^-\}+
\nu\{\mathcal{O}_{\mu+1,\nu-1}^+,\mathcal{O}_{\lambda,\sigma}^-\}\,.
&
\end{eqnarray}
From 
relations
(\ref{Jacobi1})
 it follows that the
 Poisson brackets
$ \{\mathcal{O}_{\mu,\nu}^+,\mathcal{O}_{\lambda,\sigma}^-\} $ must be functions of the 
globally well-defined generators 
$ m $, $ D $, $ J_0 $, $ H^{(\alpha)} $ 
and $ K $ since they are the only generators 
Poisson-commuting 
with $ J_0 $.
 We do not treat these 
functions
 as new, independent 
 generators, 
 but consider them as coefficients of a nonlinear algebra. On the other hand, 
 all the generators $\mathcal{O}_{\mu,\nu}^\pm$ are eigenstates of the generators $D$ and $J_{0}$, 
 in the sense of the Poisson bracket relations
 $\{I,A\}=\lambda A$, $I=D,J_0$.
By taking their 
 Poisson brackets with $H^{(\alpha)}$ and $K$, we generate the complete list 
 of symmetry generators of this kind. 
Thus,
 the set $ \mathcal{U}_1 = \{H^{(\alpha)}, D, K, J_0, \mathcal{O}_{\mu,\nu}^\pm\} $ 
generates  a finite  nonlinear algebra.

In the same way, one can 
see
that the integrals $\mathcal{S}_{\mu',\nu'}^\pm $ satisfy 
the
 relations similar to those shown above
for $\mathcal{O}_{\mu,\nu}^\pm$ but 
with 
the Greek indices
changed
  for their primed versions.
  This implies
    that the set of generators 
$ \mathcal{U}_2 = \{H^{(\alpha)}, D, K, J_0, \mathcal{S}_{\mu ',\nu'}^\pm \} $ also produces a finite nonlinear algebra.
To study what happens when we mix both 
sets of generators, we consider the relations
\begin{eqnarray}
\label{Jacobi5}
&
\{D,\{\mathcal{S}_{\mu', \nu'}^\pm,\mathcal{O}_{\mu,\nu}^\mp\}\}=\frac{\mu-\nu-\mu'-\nu'}{2}\,
\{\mathcal{S}_{\mu',\nu'}^\pm,\mathcal{O}_{\mu,\nu}^\mp\}\,,&\\&
\label{Jacobi6}
\{J_0,\{\mathcal{S}_{\mu',\nu'}^\pm,\mathcal{O}_{\mu,\nu}^\mp\}\}=\mp i \frac{\mu+\nu}{2}
\{\mathcal{S}_{\mu',\nu'}^\pm,\mathcal{O}_{\nu,\nu}^\mp\}\,,
&
\end{eqnarray} 
where we have used the Jacobi identity one more time. 
Comparing the last relation with the first equation in (\ref{DconOClasic}),  we deduce that in the general case,
the integrals $\{\mathcal{S}_{\mu',\nu'}^\pm,\mathcal{O}_{\mu,\nu}^\mp\} $
 must be functions of the integrals 
from the set  $ \mathcal{U}_1 $.
This means   
 that the set of integrals $\mathcal{O}_{\mu,\nu}^\mp$ 
 corresponds to generators of 
 an ideal nonlinear subalgebra.

\subsection{Quantum case}

The quantum Hamiltonian of the system is constructed by using the Laplace-Beltrami operator,
\begin{eqnarray}
\label{free-Hamiltonian}&
\hat{H}^{(\alpha)}=-\frac{\hbar^2}{2m}\frac{1}{\sqrt{g}}\frac{\partial}{\partial x^i}\sqrt{g}g^{ij}\frac{\partial}{\partial x^j}
=-\frac{\hbar^2}{2m}
\left(\frac{1}{\alpha^2r}\frac{\partial}{\partial r}\left(r\frac{\partial}{\partial r}\right) +\frac{1}{r^2}\frac{\partial^2}{\partial\varphi}\right)\,.&
\end{eqnarray} 
The corresponding eigenstates and spectrum are given by 
\begin{eqnarray}
\label{FreeParticleE}&
\psi_{\kappa,l}^\pm (r,\varphi)=\sqrt{\frac{\kappa}{2\pi \alpha}}J_{\alpha l}(\kappa r)e^{\pm il 
 \varphi}\,,\qquad E=\frac{\hbar^2\kappa^2}{2m\alpha^2}\,, \qquad l=0,1,\ldots\,.&
\end{eqnarray} 
These eigenstates satisfy the orthogonality relation $\braket{\psi_{\kappa,l}^\pm}{\psi_{\kappa',l'}^\mp }=\delta_{ll'}\delta(\kappa-\kappa')$, 
with respect to 
the scalar product
\begin{eqnarray}
\label{scalarp}&
\braket{\Psi_1}{\Psi_2}=
\int_V \Psi_1^*\Psi_2  \sqrt{g}  dV=
\int_0^{\infty}\alpha r dr\int_{0}^{2\pi}d\varphi \Psi_1^*\Psi_2\,.&
\end{eqnarray}
To address the problem of analyzing the quantum symmetry of the system,  we consider  the quantum versions 
of the formal integrals $\Pi_\pm$ and $\Xi_\pm$, which have been
 the basis of the algebraic construction in the classical case. 
We start with 
\begin{eqnarray}&
\hat{\Pi}_{\pm}=\hat{\Pi}_1\pm i\hat{\Pi}_2=
e^{\pm \frac{i\varphi}{2\alpha}}\left(
\frac{1}{\alpha} \hat{p}_r \pm \frac{i}{r} \hat{p}_\varphi
\right)e^{\pm \frac{i\varphi}{2\alpha}}
=
-i\hbar \frac{1}{\alpha}e^{\pm i\frac{\varphi}{\alpha}}\left(
\frac{\partial}{\partial r}
\pm i\frac{\alpha}{ r}\frac{\partial}{\partial\varphi} 
\right)\,,&
\end{eqnarray}
where $\hat{p}_r$ and $\hat{p}_\varphi$ 
are the radial and angular momentum operators introduced in (\ref{prpphi}). Operators
$\hat{\Pi}_\pm$
are formal since their action on the eigenstates produces  
\begin{eqnarray}
\label{FormalAction1}
&
\hat{\Pi}_\pm\psi_{\kappa,l}^\pm (r,\varphi)= i 
\frac{\hbar\kappa}{\alpha}\sqrt{\frac{\kappa}{2\pi \alpha}}J_{\alpha l+1}(\kappa r)e^{\pm i(l+\frac{1}{\alpha })\varphi}\,,&\\&
\label{FormalAction2}
\hat{\Pi}_\pm\psi_{\kappa,l}^\mp (r,\varphi)=- i \frac{\hbar\kappa}{\alpha}
\sqrt{\frac{\kappa}{2\pi \alpha}}J_{\alpha l-1}(\kappa r)e^{\pm i(l-\frac{1}{\alpha})\varphi}\,,
&
\end{eqnarray} 
from where we explicitly see that they cannot be physical operators for arbitrary values of $\alpha$ 
since  
in the general 
case they can produce the functions outside the Hilbert space generated by
the states $ \psi_{\kappa,l}^\pm$. 

We also consider the quantum versions of the Galilean boosts (\ref{X+-}), corresponding to the operators 
\begin{eqnarray}
\label{Quantum-Galilean boost}
&
\hat{\Xi}_\pm=\alpha m r  e^{\pm i\frac{\varphi}{\alpha}}  \pm  t\hat{\Pi}_\pm =
e^{\pm i\frac{\varphi}{\alpha}} \left(m \alpha r -i\hbar \frac{\kappa}{\alpha}t\left[
\frac{\partial}{\partial(\kappa r)}
\pm i\frac{\alpha}{\kappa r}\frac{\partial}{\partial\varphi} 
\right]\right)\,.
&
\end{eqnarray}
Due to 
appearance of  $\hat{\Pi}_\pm$ in (\ref{Quantum-Galilean boost}),  
the operators $\hat{\Xi}_\pm$  inherit all the problems of the former operators.

Let us carefully study the case $ \alpha =q/k $. As in the classical analysis of the previous subsection, 
we can consider the powers of the $\hat{\Pi}_\pm$ and $\hat{\Xi}_\pm$
operators. The action of 
$ (\hat{\Pi}_\pm)^q $  
on the 
Hamiltonian eigenstates produces
\begin{eqnarray}
&
\label{sklAction1}
(\hat{\Pi}_\pm)^q \psi_{\kappa,l}^\pm (r,\varphi)= \left(i \frac{k\hbar\kappa}{q}\right)^{q}\sqrt{\frac{k\kappa}{2\pi q}}
J_{\frac{q}{k} (l+k)}(\kappa r)e^{\pm i(l+k)\varphi}= 
\left(i \frac{k \hbar \kappa}{q}\right)^{q} \psi_{\kappa,l+k}^\pm (r,\varphi)\,,&\\&
\label{sklAction2}
(\hat{\Pi}_\pm)^q \psi_{\kappa,l}^\mp (r,\varphi)=\left(- i \frac{k\hbar\kappa}{q}\right)^q \sqrt{\frac{k\kappa}{2\pi q}}
J_{\frac{q}{k}( l-k)}(\kappa r)e^{\pm i(l-k)\varphi}\,.
&
\end{eqnarray} 

Last equation means that the functions 
 $$(\hat{\Pi}_\pm)^{q}\psi_{\kappa,j}^{\mp}(r,\varphi)\propto  J_{-\frac{q|k-j|}{k}}(\kappa r)e^{\mp i|k-j|\varphi} \,,\qquad j=1,\ldots ,k-1\,,$$
are outside the Hilbert space 
since the index of the Bessel function is negative and 
non-integer.
On the other hand, 
in the special case $k=1$ we have 
\begin{eqnarray}
&
\label{slAction1}
(\hat{\Pi}_\pm)^q \psi_{\kappa,l}^\pm (r,\varphi)= \left(i \frac{\hbar\kappa}{q}\right)^{q}\sqrt{\frac{\kappa}{2\pi q}}
J_{q (l+1)}(\kappa r)e^{\pm i(l+1)\varphi}= \left(i \frac{\hbar\kappa}{q}\right)^{q} \psi_{\kappa,l+1}^\pm (r,\varphi)\,,&\\&
\label{slAction2}
(\hat{\Pi}_\pm)^q \psi_{\kappa,l}^\mp (r,\varphi)=\left(- i \frac{\hbar\kappa}{q}\right)^q \sqrt{\frac{\kappa}{2\pi q}}
J_{q( l-1)}(\kappa r)e^{\pm i(l-1)\varphi}=\left(- i \frac{\hbar\kappa}{q}\right)^q \psi_{\kappa,l-1}^\mp (r,\varphi)\,.
&
\end{eqnarray} 
Now, using the fact that $ql=\mathfrak{n}$ is a positive integer  number and  
$
J_{\mathfrak{-n}}(\zeta)=(-1)^{\mathfrak{n}}J_\mathfrak{n}(\zeta)\,,$
one can show that 
$\psi_{\kappa,\mp l}^\pm=(-1)^{ls}\psi_{\kappa,\pm l}^\mp$.
This
implies that the operators $(\hat{\Pi}_\pm)^q$ 
always produce physical eigenstates
in the case of integer values $\alpha=q$.
They are a kind of the spectrum generating operators
which allow us to change the quantum 
number $l$ of the states in $\pm 1$ without
changing their energies.

 As in the classical case, 
formal operators $\hat{\Pi}_\pm$ and $\hat{\Xi}_\pm$ are the basis in the construction 
of the symmetry algebra, and we can built the quantum version
 of the   
generators of the $\mathfrak{so}(2,1)\oplus \mathfrak{u}(1)$
symmetry taking place 
 for arbitrary values of $\alpha$. We have 
\begin{eqnarray}
&\label{QD}
\hat{H}^{(\alpha)}=\frac{1}{2m}\hat{\Pi}_+\hat{\Pi}_-\,,\qquad
\hat{D}=\frac{1}{4m}(\hat{\Xi}_+\hat{\Pi}_-+\hat{\Pi}_+\hat{\Xi}_-)= 
\frac{\hbar}{2i}\left(r\frac{\partial}{\partial r}+1\right)- \hat{H}^{(\alpha)}t\,,
&\\&
\label{QK}
\hat{K}=\frac{1}{2m}\hat{\Xi}_-\hat{\Xi}_+= \frac{m}{2}\alpha^2r^2-2\hat{D}t-\hat{H}^{(\alpha)}t^2\,,\qquad 
\hat{J}_0= \frac{1}{2}(\hat{\Xi}_1\hat{\Pi}_2-\hat{\Xi}_2\hat{\Pi}_1)=\frac{\alpha}{2}\hat{p}_\varphi\,.
\end{eqnarray}
These operators satisfy  the commutation relations
\begin{eqnarray}
&
[\hat{D},\hat{H}^{(\alpha)}]=i\hbar \hat{H}^{(\alpha)}\,,\qquad[\hat{D},\hat{K}]=-i\hbar\hat{K}\,,\qquad 
[\hat{K},\hat{H}^{(\alpha)}]=2i\hbar \hat{D}\,,\label{QuantumSO21}&\\&
[\hat{J}_0,\hat{D}]=[\hat{J}_0,\hat{H}^{(\alpha)}]=[\hat{J}_0,\hat{K}]=0\,.
\end{eqnarray}
However, the analysis related to the operators $\hat{\Pi}_\pm$ shows us that in contrast with the classical case, 
only  for 
 $\alpha=q$ we  can have a well-defined additional symmetry generators, 
that reveals a  kind of the quantum anomaly in the case of rational non-integer values of $\alpha$.
For $\alpha=q$,
  the corresponding symmetry operators are 
\begin{eqnarray}
&
\label{Ooperators}
\hat{\mathcal{O}}^\pm_{\mu,\nu}=(\hat{\Xi}_\pm)^\mu(\hat{\Pi}_\pm)^\nu\,,\qquad 
\mu+\nu=q\,.
&\\&
\hat{\mathcal{S}}_{\mu',\nu'}^\pm=(\hat{\Xi}_\pm)^{\mu'} (\hat{\Pi}_\pm)^{\nu'}\,,\qquad \mu'+\nu'=2q\,,
&
\end{eqnarray}
being the quantum counterpart of the classical integrals (\ref{ClasicalO}) and (\ref{ClasicalS}) with $k=1$. 
By means of the  
commutation relations
\begin{eqnarray}
\label{RPDJKH1}
&[\hat{\Xi}_\pm,\hat{\Pi}_\mp]=2im\hbar\,,\qquad
[\hat{\Xi}_\pm,\hat{\Pi}_\pm]=0\,,&\\&
[\hat{H}^{(\alpha)},\hat{\Xi}_\pm]=-i\hbar \hat{\Pi}_{\pm}\,,\qquad
[\hat{D},\hat{\Xi}_\pm]=-\frac{i\hbar}{2}\hat{\Xi}_\pm\,,\qquad 
[\hat{J}_0,\hat{\Xi}_\pm]=\pm  \frac{\hbar}{2}\hat{\Xi}_\pm \,, &\\&
[\hat{K},\hat{\Pi}_\pm]=i\hbar\hat{\Xi}_{\pm}\,,\qquad
[\hat{D},\hat{\Pi}_\pm]=\frac{i \hbar}{2}\hat{\Pi}_\pm\,,\qquad 
[\hat{J}_0,\hat{\Pi}_\pm]=\pm\frac{\hbar}{2}\hat{\Pi}_\pm\,,&\label{RPDJKH2}
\end{eqnarray}
and the commutator identity 
$
[\hat{A},\hat{B}^n]=\sum_{j=1}^{n}\hat{B}^{j-1}[\hat{A},\hat{B}]\hat{B}^{n-j}\,,
$
it can be shown that the properties of the  classical nonlinear algebra generated by the integrals 
$\mathcal{O}^\pm_{\mu,\nu}$ and  $\mathcal{S}^\pm_{\mu',\nu'}$
with $ k = 1 $ are preserved at the quantum level.
  
Some important algebraic relations are
\begin{eqnarray}
&
[\hat{p}_\varphi,\hat{\mathcal{O}}_{\mu,\nu}^\pm]=\pm \hbar \hat{\mathcal{S}}_{\mu,\nu}^\pm\,,
 \qquad
[\hat{p}_\varphi,\hat{\mathcal{S}}_{\mu',\nu'}^\pm]=\pm \hbar \hat{\mathcal{S}}_{\mu',\nu'}^\pm\,,\label{OSphi}
&\\& 
[\hat{D},\hat{\mathcal{O}}^\pm_{\mu,\nu}]= i\hbar\frac{\nu-\mu}{2}\hat{\mathcal{O}}^\pm_{\mu,\nu}\,,\qquad
[\hat{D},\hat{\mathcal{S}}^\pm_{\mu,\nu}]= i\hbar\frac{\nu'-\mu'}{2}\hat{\mathcal{S}}^\pm_{\mu',\nu'}\,. \label{OSD}
&
\end{eqnarray}
From (\ref{OSphi}) we learn that the operators $ \hat{\mathcal{O}}^\pm _ {\mu, \nu} $ ($ \hat{\mathcal{S}}^\pm_{ \mu',\nu'} $)
change the angular momentum quantum number by $ \pm 1$ ($ \pm 2 $), and it its clear that
 operators $\hat{\mathcal{O}}^\pm _ {0, q}=(\hat{\Pi}_\pm)^q$ 
 are 
responsible for the infinite degeneracy of the spectrum of the system. 

\section{Harmonic oscillator in a cosmic string background}

\label{SecHar}
To study the dynamics of the harmonic oscillator 
 in
  a cosmic string background, we set the potential term in action (\ref{Lagrangian}) 
to be $V(\vr)=\frac{m\omega^2}{2}g_{ij}x_ix_j=\frac{1}{2}m\alpha^2\omega^2 r^2$. 
 We will see that the particle dynamics  has special characteristics for different values of $\alpha$, 
 and these peculiarities are coherently reflected in  the classical orbits, 
 in the  spectra of the system at the quantum level, 
 and in the construction of the symmetry algebra, 
which again will reveal the  quantum anomaly phenomenon.

We also show that this system is related to the free motion in the conical geometry 
 by means of the conformal bridge transformation, and we 
use this to reconstruct the properties
 of the complete classical symmetry algebra for the case $\alpha=q/k$ and the quantum 
version in
  the case $\alpha=q$.

\subsection{Classical case}
The Hamiltonian of the system corresponds to 
 \begin{eqnarray}
 &H_{\text{os}}^{(\alpha)}=\frac{1}{2m}\left(\frac{p_r^2}{\alpha^2}+\frac{p_{\varphi}^2}{r^2}\right)+
 \frac{m\alpha^2\omega^2}{2}r^2\,,&
 \end{eqnarray}
 and as it happened with the free particle, it
 is directly related to the Hamiltonian 
of the harmonic oscillator in the Euclidean plane,
which  was reviewed in the subsection \ref{AppB}, 
by means of the local canonical transformation (\ref{Canonicaltrnas}). 
From here, the solutions of the corresponding
 equations of motion are immediately obtained,
\begin{eqnarray}
&\label{r(varohi)os}
r^2(\varphi)=\frac{p_\varphi^2}{m H_{\text{os}}^{(\alpha)}}\left(1+\delta\cos(\frac{2}{\alpha}(\varphi-\varphi_*))\right)^{-1}\,,
&\\&
\label{AngularOsalpha}
\varphi(\tau)=\alpha
\arctan(\frac{r_{+}}{r_{-}}\tan(\omega (\tau-\tau_{*})))+\varphi_*\,,
&\\&\label{rDKHalpha}
r^2(\tau)=\frac{2}{m\omega^2}\big(\omega \mathcal{D}\sin(2\omega \tau)+\omega^2
\mathcal{K}\cos(2\omega \tau)+ H_{\text{os}}^{(\alpha)}\sin^2(\omega \tau)\big)\,.&
\end{eqnarray}
Here $\mathcal{D}$ and $\mathcal{K}$ are given by 
\begin{eqnarray}
\label{Kalpha}
&
\mathcal{K}=
\frac{m\alpha^2}{2}r^2 \cos(2\omega \tau)+
\frac{1}{\omega^2 }H_{\text{os}}^{(\alpha)}\sin^2(\omega \tau)
-\frac{1}{2\omega } rp_r \sin(2\omega \tau)\,,&\\&
\label{Dalpha}
\mathcal{D}=
\frac{1}{2}rp_r \cos(2\omega \tau)-
\frac{1}{2\omega}(  H_{\text{os}}^{(\alpha)}-m\omega^2 \alpha^2 r^2)\sin(2\omega \tau)\,.
\end{eqnarray}
The quantities 
\begin{eqnarray}
&
r_{\pm}^{2}=\frac{H_{\text{os}}^{(\alpha)}}{m\omega^2\alpha^2}\left(1\pm\delta \right)\,, \qquad
\delta=\sqrt{1-\left(\frac{\omega\alpha p_{\varphi}}{ H_{\text{os}}^{(\alpha)}}\right)^2}\,,
&
\end{eqnarray}
are identified with the
radial turning points of the trajectory, and 
 $ \varphi_*$ and $ \tau_* $ 
are
 the angular position of one of the radial minima of the trajectory and
 the corresponding moment of time 
when the particle is in that  place.

\begin{figure}[hbt!]
\begin{center}
\hskip0.5cm
\begin{subfigure}[c]{0.27\linewidth}
\includegraphics[scale=0.45]{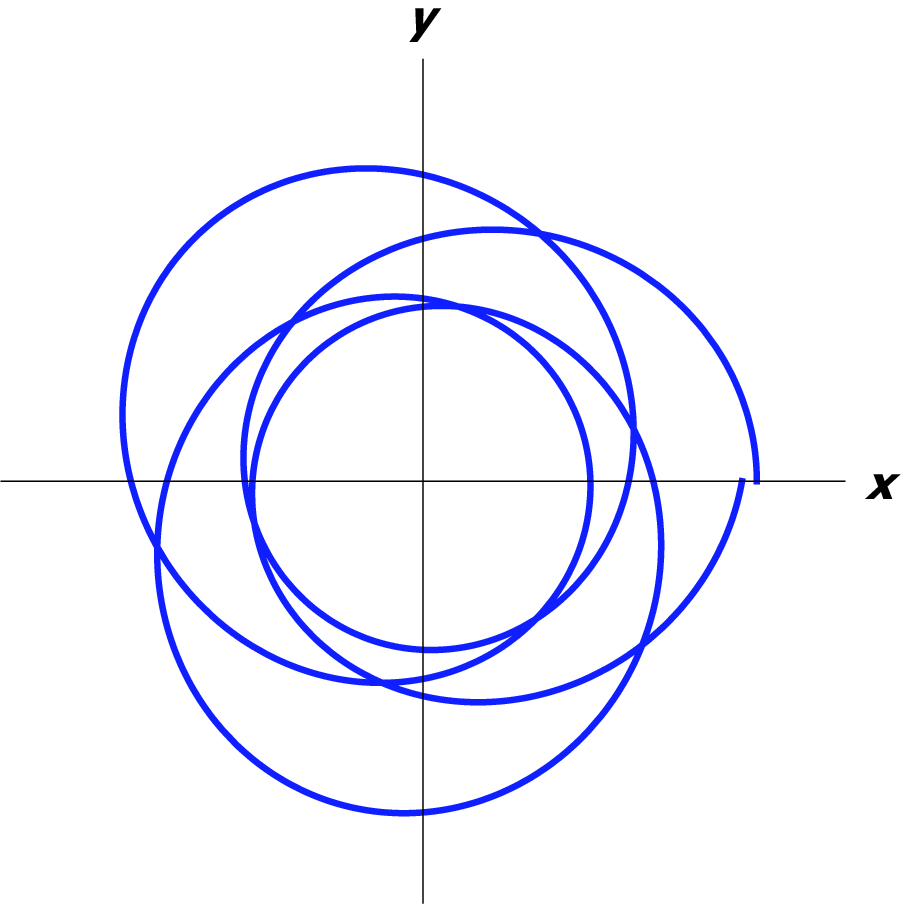}
\caption{\small{$\alpha=e$}}
\end{subfigure}
\begin{subfigure}[c]{0.25\linewidth}
\includegraphics[scale=0.45]{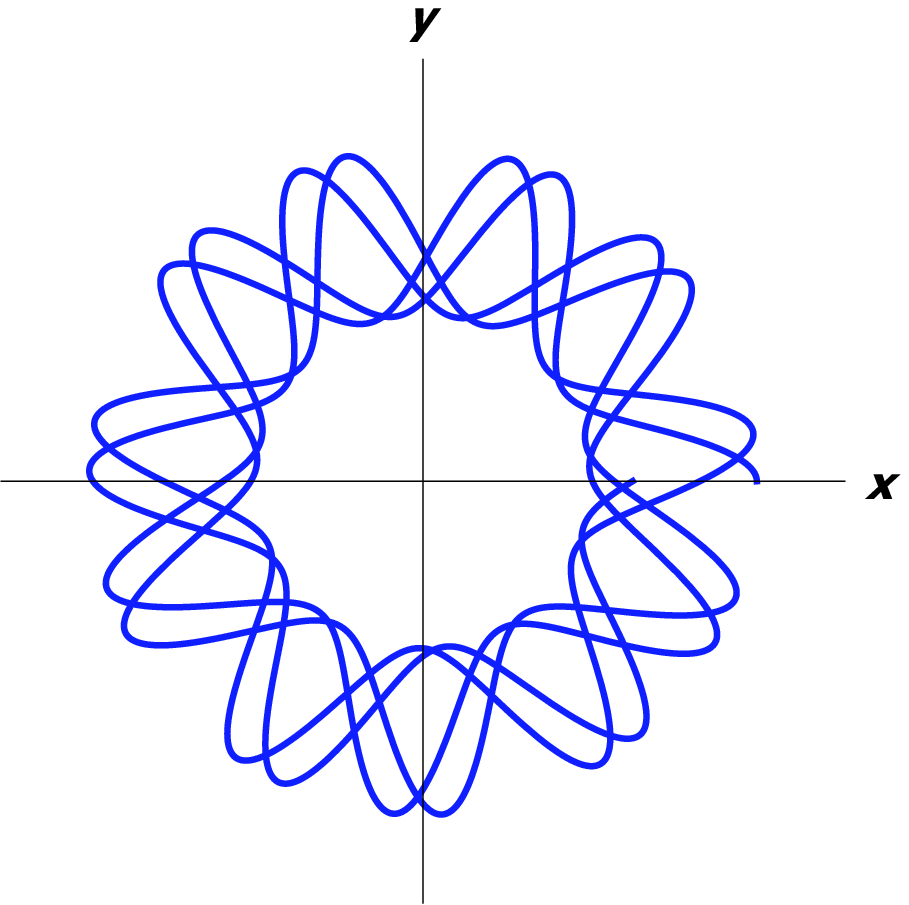}
\caption{\small{$\alpha=1/e$}}
\end{subfigure}

\hskip0.5cm
\begin{subfigure}[c]{0.27\linewidth}
\includegraphics[scale=0.45]{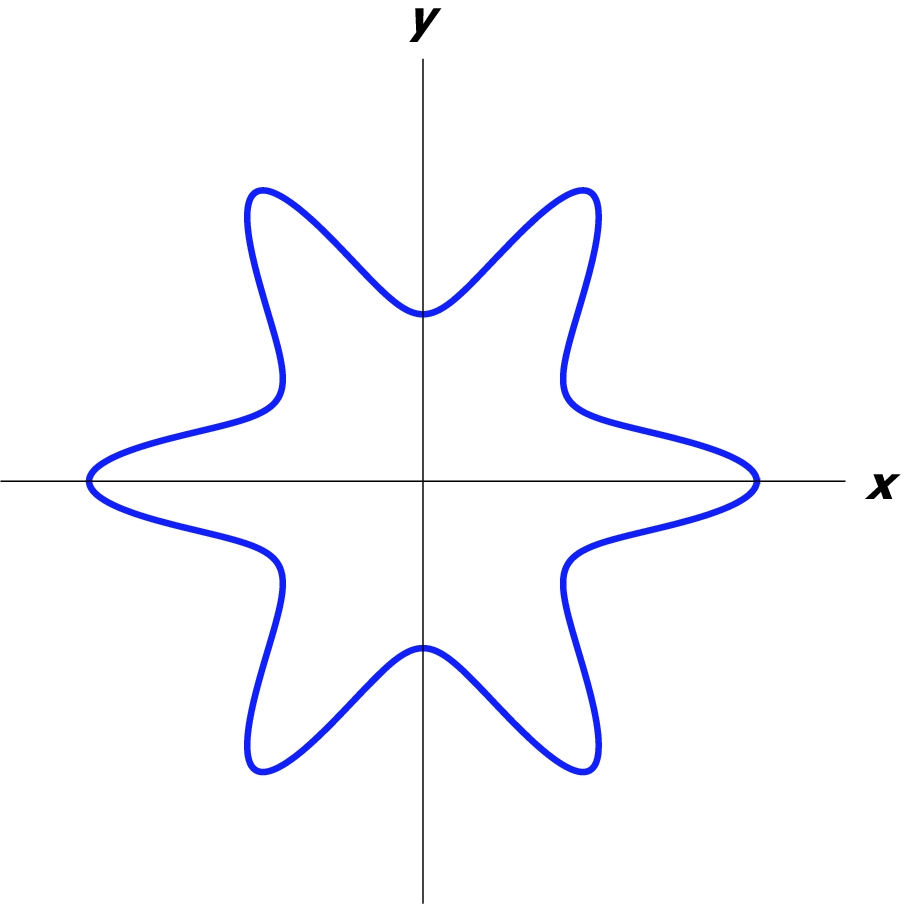}
\caption{\small{$\alpha=1/3$}}
\end{subfigure}
\begin{subfigure}[c]{0.26\linewidth}
\includegraphics[scale=0.45]{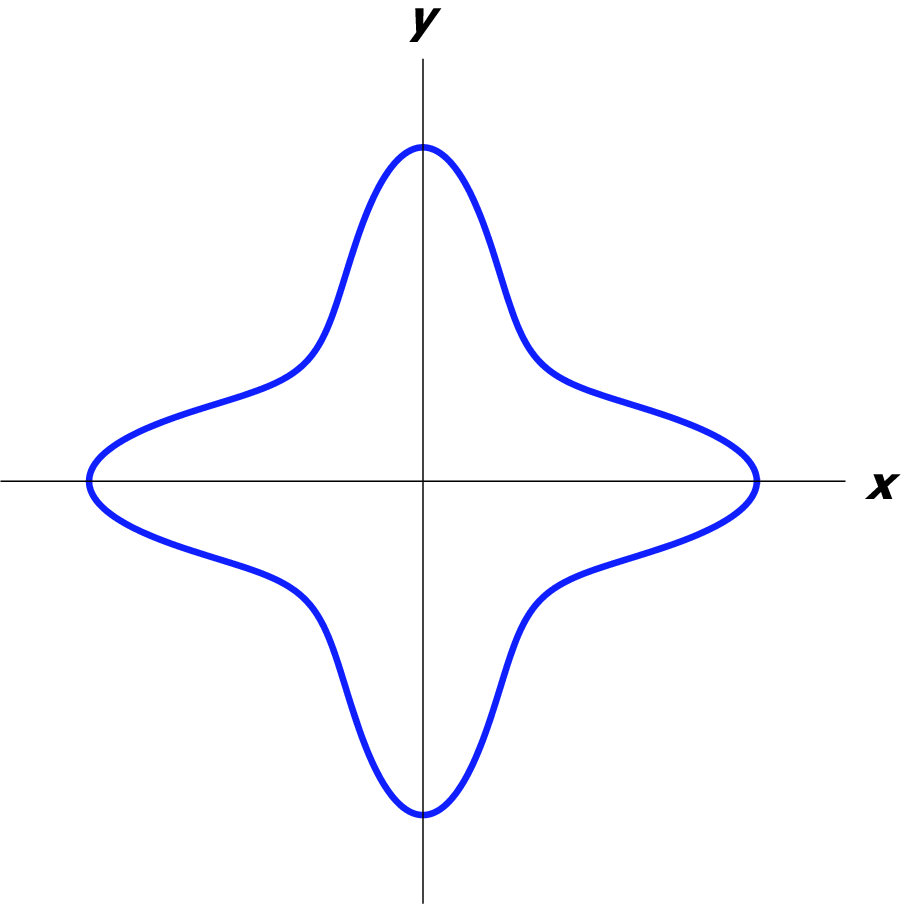}
\caption{\small{$\alpha=1/2$}}
\end{subfigure}
\begin{subfigure}[c]{0.25\linewidth}
\includegraphics[scale=0.45]{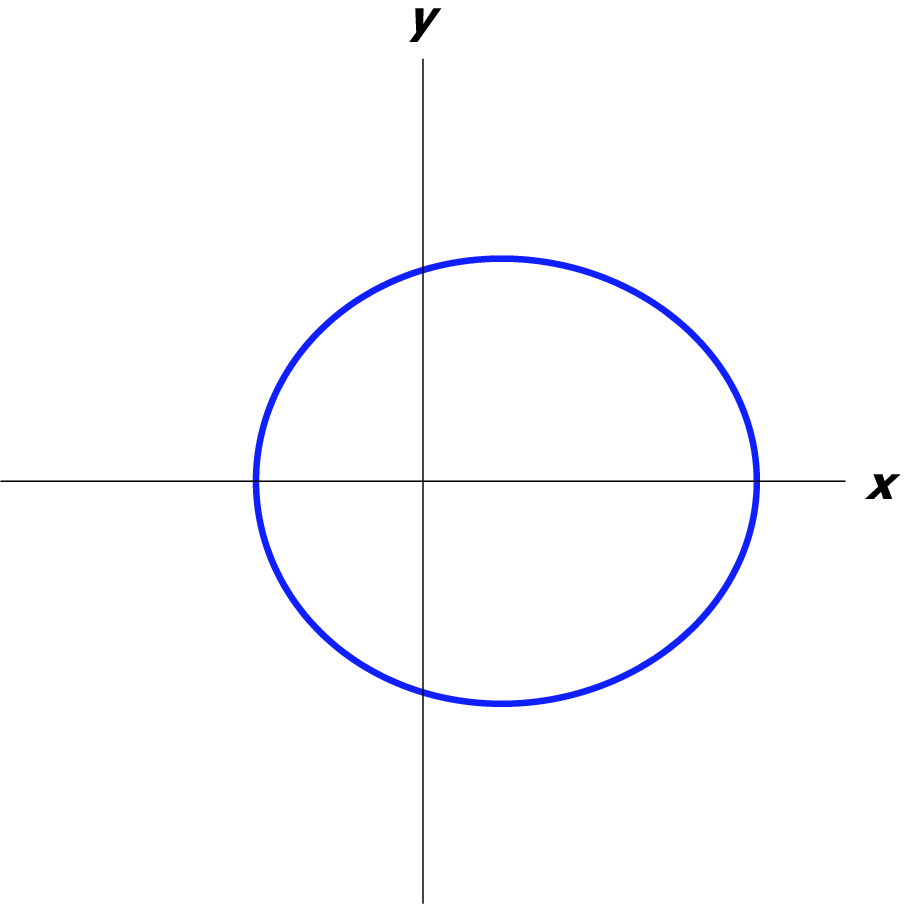}
\caption{\small{$\alpha=2$}}
\end{subfigure}

\hskip0.5cm
\begin{subfigure}[c]{0.27\linewidth}
\includegraphics[scale=0.45]{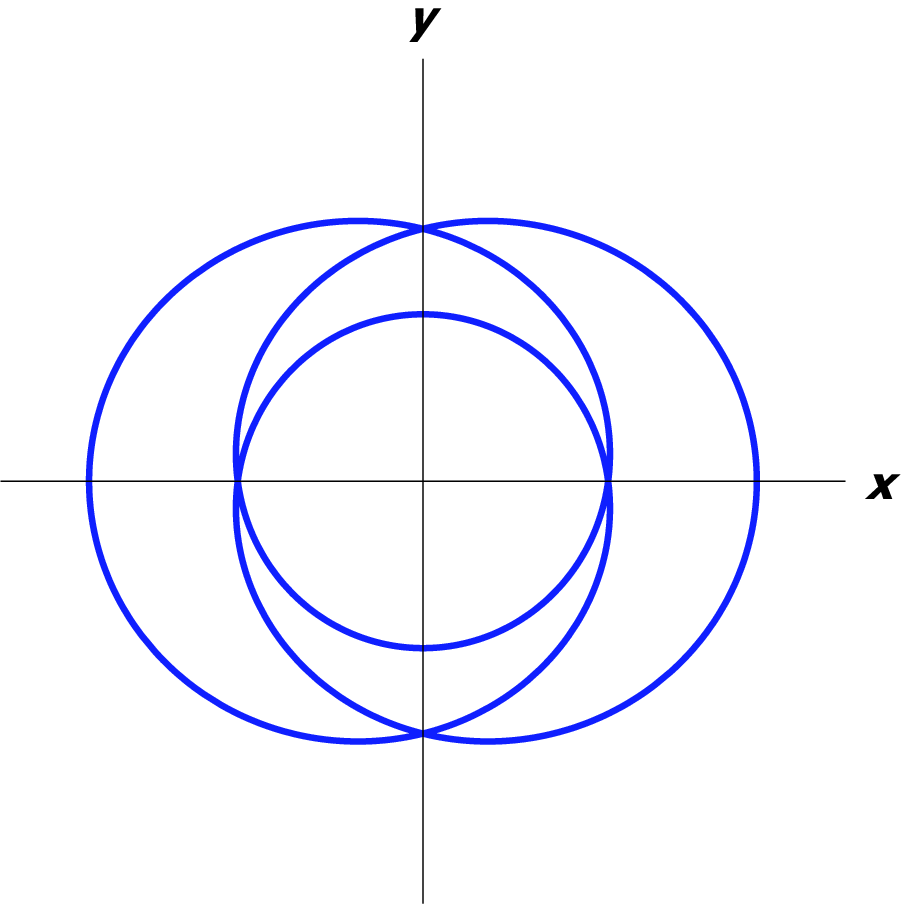}
\caption{\small{$\alpha=3$}}
\end{subfigure}
\begin{subfigure}[c]{0.26\linewidth}
\includegraphics[scale=0.45]{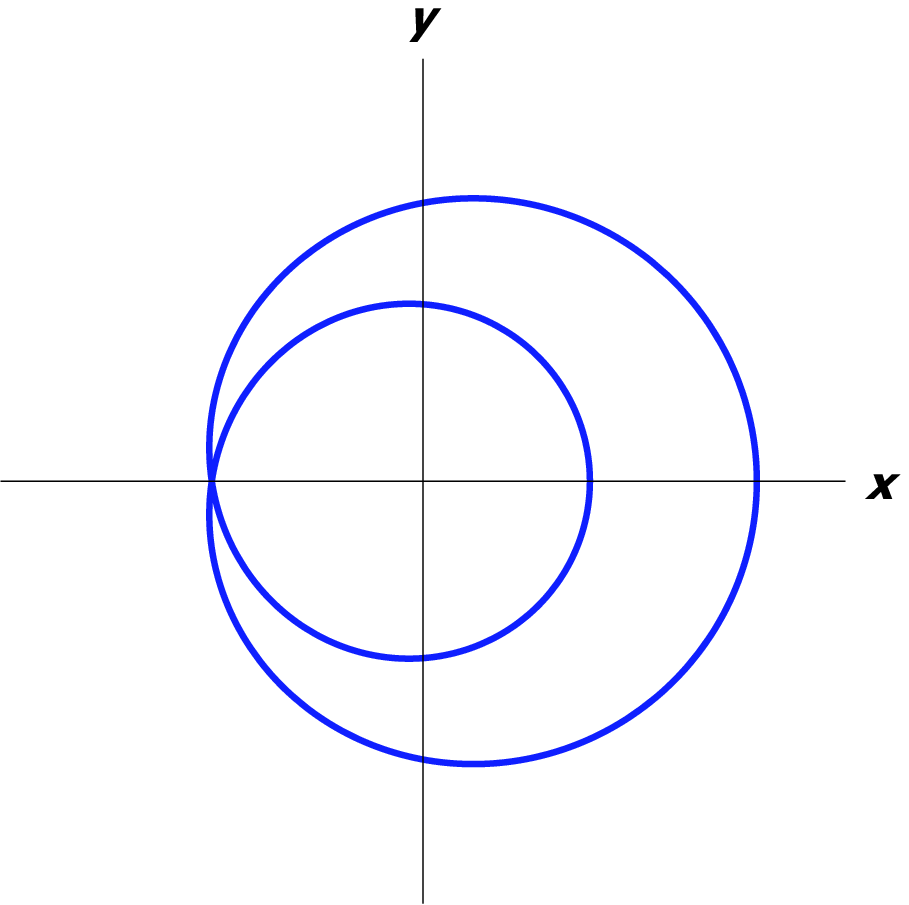}
\caption{\small{$\alpha=4$}}
\end{subfigure}
\begin{subfigure}[c]{0.25\linewidth}
\includegraphics[scale=0.45]{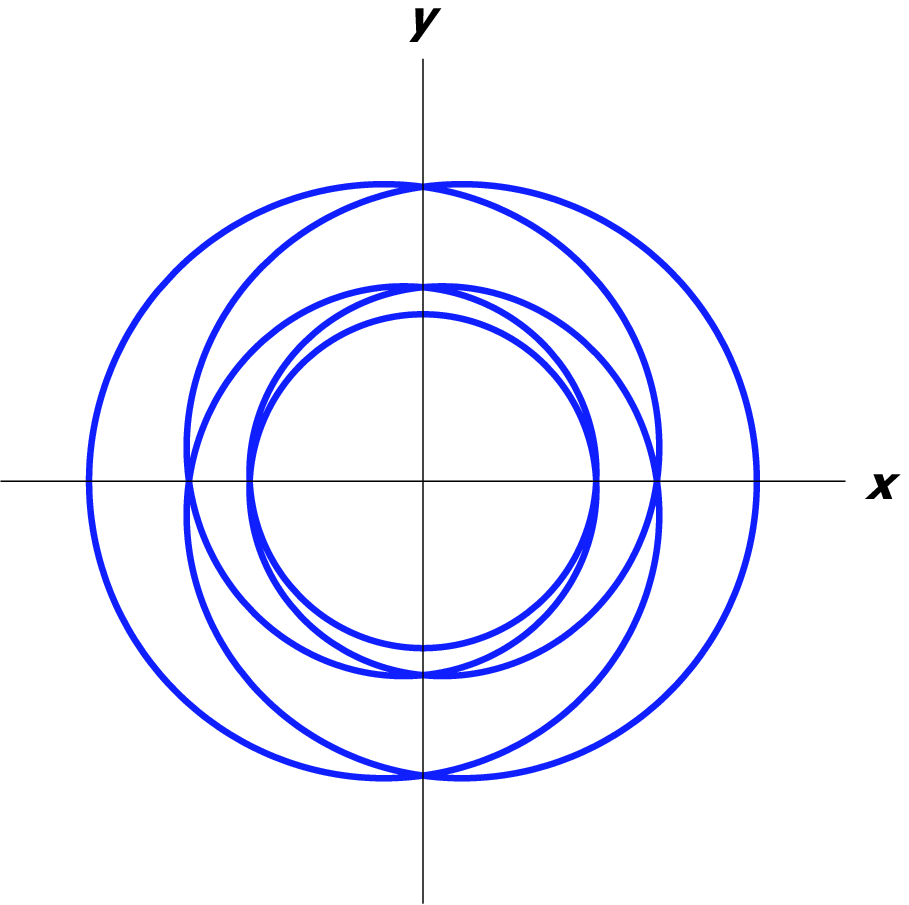}
\caption{\small{$\alpha=5$}}
\end{subfigure}
\end{center}
\caption{\small{Images of the trajectory  for some irrational and rational values of $\alpha$.
From (\ref{r(varohi)os}) one has that $r(\varphi)=r(\varphi+\alpha l \pi)$, 
implying that the number of  radial
 minima $r_-$ and maxima $r_+$ depends on the value of $\alpha$. 
For
$\alpha=1/k$, one has $2k$ minima and $2k$ maxima.
When $\alpha=2n$, there is only  one minimum and one maximum of $r$ on the orbit.
When $\alpha=2n+1$, there are two maxima and two minima, similarly to the case of the 
isotropic harmonic oscillator in the Euclidean plane. 
In general case, for $\alpha=q/k$ the number of maxima/minima is 
$\mathscr{N}_{\text{max}/\text{min}}=k(q\,\text{mod}\,2+1)$. } }
\label{figure3}
\end{figure}

From the corresponding solutions of the equation of motion 
one finds that the trajectory is closed
if and only if $\alpha$ is a rational number.
The images of the orbit 
  for some 
 rational and  irrational values of  $\alpha$ 
are shown in Fig. \ref{figure3}.

By using the canonical transformation (\ref{Canonicaltrnas}), 
one can also obtain the dynamical integrals representing the classical analogues of the ladder operators
in polar coordinates,  which are: 
\begin{eqnarray}
&
\mathfrak{b}_1^-=
\frac{1}{\sqrt{2}}(a_1^--ia_2^-)= \frac{1}{2}e^{i(\omega t-\frac{\varphi}{\alpha})}
\left(
\alpha\sqrt{m\omega}r
+\frac{p_\varphi}{\sqrt{m\omega}r} +\frac{ip_r}{\alpha\sqrt{m\omega}}\right)\,,\qquad 
\mathfrak{b}_1^+=(\mathfrak{b}_1^-)^*\,,&\\&
\mathfrak{b}_2^-= 
\frac{1}{\sqrt{2}}(a_1^-+ia_2^-)= \frac{1}{2}e^{i(\omega t+\frac{\varphi}{\alpha})}
\left(
\alpha\sqrt{m\omega}r
-\frac{p_\varphi}{\sqrt{m\omega}r} +\frac{ip_r}{\alpha\sqrt{m\omega}}\right)\,, \qquad
\mathfrak{b}_2^+=(\mathfrak{b}_2^-)^*\,.
\end{eqnarray}
In resemblance with the free particle system in a conical geometry, we also note that these dynamical integrals 
are formal (locally defined only)
functions in the respective  phase space when $ \alpha $ takes 
values different 
from $1/k$.
However, we can use these formal integrals
 to 
construct the well-defined  generators of the 
 $\mathfrak{sl}(2,\R)\oplus \mathfrak{u}(1)$  symmetry algebra
of the system,
which  
 are given by 
\be
\label{ConeHarGen}
\mathcal{J}_0=\frac{1}{2} \mathfrak{b}_a^+\mathfrak{b}_a^-=\frac{1}{2\omega}H_{\text{os}}^{(\alpha)}\,,\qquad
\mathcal{J}_{\pm}=\mathfrak{b}_1^\pm \mathfrak{b}_2^\pm\,,\qquad
\mathcal{L}_2=\frac{1}{2}(\mathfrak{b}_1^+\mathfrak{b}_1^--\mathfrak{b}_2^+\mathfrak{b}_2^-)=\frac{1}{2}\alpha p_\varphi\,. 
\ee

In the context of the conformal bridge transformation, if we set the generators $H^{(\alpha)}$, $D$ and $K$  as 
those in (\ref{ClasicalH})-(\ref{ClasicalH2}), 
the transformation considered in Sec. \ref{SecConBrid}  produces 
the generators (\ref{ConeHarGen}), in accordance with relations (\ref{Conformal1}),
(\ref{Conformal2})  and the first equation in (\ref{Conformal3}). 
To obtain the complete symmetry algebra for the case $ \alpha=q/k $ we must apply the transformation to the higher-order 
generators (\ref{ClasicalO}) and (\ref{ClasicalS}).
However, it is convenient first to consider the formal relations 
\begin{eqnarray}
&
\mathscr{T}(\tau,\beta,\delta,\gamma,t)(\Pi_-)=-i\sqrt{2m\omega}\mathfrak{b}_1^-\,,\qquad
\mathscr{T}(\tau, \beta,\delta,\gamma, t)(\Pi_+)=-i\sqrt{2m\omega}\mathfrak{b}_2^-\,,
&\\&
\mathscr{T}(\tau, \beta,\delta,\gamma, t)(\Xi_{+})=\sqrt{\frac{2m}{\omega}}\mathfrak{b}_1^+\,,\qquad
\mathscr{T}(\tau, \beta,\delta,\gamma, t)(\Xi_-)=\sqrt{\frac{2m}{\omega}}\mathfrak{b}_2^+\,,
&
\end{eqnarray}
where $\beta$, $\delta$ and $\gamma$  are still given by (\ref{betadeltagamma}), 
and the formal equations (\ref{FormalClaAl1})-(\ref{FormalClaAl2})  are 
employed. 
These equations facilitate the application of the conformal bridge transformation 
to
 the mentioned 
well-defined on  the phase space integrals of motion.
This
yields  us 
\begin{eqnarray}
&
\mathscr{T}(\tau, \beta,\delta,\gamma, t)( \mathcal{O}_{\mu,\nu}^\pm)=
(-i)^\nu (m)^{\frac{q}{2}}(\omega)^{\frac{\mu-\nu}{2}} \mathcal{G}_{\mu,\nu}^\pm\,,
&\\&
\mathscr{T}(\tau, \beta,\delta,\gamma, t)(\mathcal{S}_{\mu',\nu'}^\pm) =
(-i)^\nu (m)^{\frac{q}{2}} (\omega)^{\frac{\mu-\nu}{2}}\mathcal{F}_{\mu',\nu'}^\pm \,,
&
\end{eqnarray}
where 
\begin{eqnarray}
\mathcal{G}_{\mu,\nu}^+=(b_1^+)^\mu(b_2^-)^\nu\,,\quad (\mathcal{G}_{\mu,\nu}^+)^*=\mathcal{G}_{\nu,\mu}^-
\qquad 
\mathcal{F}_{\mu',\nu'}^+= (b_1^+)^{\mu'}(b_2^-)^{\nu'}\,,\quad
(\mathcal{F}_{\mu',\nu'}^+)^*=\mathcal{F}_{\nu',\mu'}^-\,,
\end{eqnarray}
and $\mu,\nu=0,1,\ldots q$, ($\mu',\nu'=0,1,\ldots 2q$) satisfy the restriction $\mu+\nu=q$ ($\mu'+\nu'=2q$). 
The properties of the symmetry algebra  generated by these integrals of motion 
are  retrieved
directly from the symmetry algebra of the free particle moving in the same conical geometry. 
In summary we have:
 \begin{itemize}
 \item The set of generators  $\mathcal{U}_1=\{\mathcal{J}_0,\mathcal{J}_\pm, \mathcal{L}_2, \mathcal{G}_{\mu,\nu}^\pm\}$ 
 produces an ideal nonlinear subalgebra.
  The (in general) dynamical integrals  $\mathcal{G}_{\mu,\nu}^\pm$ are eigenstates of 
  $i\mathcal{L}_2=i\alpha p_\varphi/2$
   with eigenvalue $\lambda=\pm q/2$,
   in the sense 
 of the   Poisson bracket relation
   $\{\mathcal{L}_2, A\}=\lambda A$, and, therefore, 
   they are 
   eigenstates of $ip_\varphi$, with eigenvalue $\pm k$.
   In the same way, they are 
 eigenstates of $H_{\text{os}}^{(\alpha)}$ with eigenvalue $i\omega(\nu-\mu)$. Note that when 
 $ \nu = \mu =q/2$   
and 
 $ q $ is an even number, we have two true 
 (not depending explicitly on time)
 integrals of motion for the system. 
 Finally,  the Poisson bracket action
 of generators $ \mathcal{J}_\pm $ corresponds to 
 \be
 \{\mathcal{J}_{-}, \mathcal{G}_{\mu,\nu}^\pm\}=i\mu \mathcal{G}_{\mu-1,\nu+1}^\pm\,,\qquad
  \{\mathcal{J}_{+}, \mathcal{G}_{\mu,\nu}^\pm\}=i\nu \mathcal{G}_{\mu+1,\nu-1}^\pm\,.
 \ee 
 \item The integrals of the set  $\mathcal{U}_2=\{\mathcal{J}_0,\mathcal{J}_\pm, \mathcal{L}_2, \mathcal{F}_{\mu',\nu'}^\pm\}$ 
 also produce a  nonlinear subalgebra, but
 which
 is not an ideal. In this case the  (in general) dynamical integrals  $\mathcal{F}_{\mu',\nu'}^\pm$ are eigenstates of  $i\mathcal{L}_2$
  with eigenvalue $\lambda=\pm q$, implying that with respect to $ip_\varphi$, the  eigenvalues are $\pm 2k$. 
  These objects 
  are also eigenstates of $ H_{\text{os}}^{(\alpha)} $, with eigenvalue $ i\omega(\nu'-\mu') $, and again, 
  when $ \nu '= \mu' = q $ we have two true
integrals. In contrast with the previous case, these 
  true integrals can be constructed for any natural value of $ q $, 
  but  when $q$ is even, $ q = 2\tilde{q}$, $\tilde{q}=1,\ldots$,
  we have the equalities $ \mathcal{F}_{q, q}^\pm = (\mathcal{G}_{\tilde{q}, \tilde{q}}^\pm) ^ 2 $.
   The  Poisson bracket action of the  integrals
   $ \mathcal{J}_\pm$ is
 \be
 \{\mathcal{J}_{-}, \mathcal{F}_{\mu',\nu'}^\pm\}=i\mu' \mathcal{F}_{\mu'-1,\nu'+1}^\pm\,,\qquad
  \{\mathcal{J}_{+}, \mathcal{F}_{\mu',\nu'}^\pm\}=i\nu' \mathcal{F}_{\mu'+1,\nu'-1}^\pm\,.
 \ee 
 \end{itemize}

When $ \alpha = 1 $, we obtain the harmonic oscillator in the Euclidean plane, 
and the dynamical integrals $ \mathcal{G}_{\mu, \nu}^\pm $ correspond to the classical analogues 
of the ladder operators themselves. 
In this case, the true integrals $\mathcal{F}_{1,1}^\pm$ 
correspond
to the $\mathcal{L}_\pm$
generators of the $\mathfrak{su}(2)$ symmetry of the system, while $\mathcal{F}_{0,2}^-$ and 
$\mathcal{F}_{2,0}^+$ are the mutually 
complex conjugated dynamical integrals $\mathcal{B}^\pm$, see Section \ref{AppB}. 
In the case $ \alpha = 1 / k $, with $ k = 2,3, \ldots $ 
we still have the same symmetry algebra  as for the Euclidean case,
 since the generators  $ \mathfrak{b}_a^\pm $ are the  globally
 well-defined functions in   the phase space.

\subsection{Quantum case}

 Here we study the quantum theory corresponding to the classical system discussed in the previous section. 
As we have seen, at the classical level the system has a large number of dynamical symmetries that 
are eigenstates of the Hamiltonian in the sense of the Poisson 
bracket relation 
$ \{H_{\text{os}}^{(\alpha)}, \mathcal{C} \} = \lambda \mathcal{C} $
 when $\alpha$
is rational.
Each of these integrals were obtained by applying the classical conformal bridge transformation to 
the classical free particle system  symmetry generators in conical geometry, 
and in this section we follow the quantum version of that approach. In this context, 
and remembering that at the quantum level the system of the free particle  in conical space reveals
 a
 quantum anomaly for 
rational, non-integer  values of $ \alpha $, it should not be surprising
 that this anomaly is also present in the harmonically trapped system, and in this section we show how this happens.
  As all the integrals that 
 admit a well-defined quantum extensions must be  eigenstates of the 
 corresponding quantum Hamiltonian $\hat{H}_{\text{os}}^{(\alpha)}$ 
in the sense of  $ [\hat{H}_{\text{os}}^{(\alpha)}, \hat{\mathcal{C}}] = i\hbar\lambda \hat{\mathcal{C}}$, 
one concludes that  the action 
 of these operators at $ \tau = 0 $  on 
 a particular eigenstate of the system (and that actually is what we need to 
 calculate according to Eq. (\ref{DynamocalAction}))  
 will produce another eigenstate. So, for the sake of simplicity, we assume that 
 all the evolution parameters in dynamical integrals are zero from now on.

At the quantum level, the system is governed by the  Hamiltonian operator 
\begin{eqnarray}
\label{HarmonicOcilator}
&\hat{H}_{\text{os}}^{(\alpha)}=
-\frac{\hbar^2}{2m}
\left(\frac{1}{\alpha^2r}\frac{\partial}{\partial r}\left(r\frac{\partial}{\partial r}\right) +
\frac{1}{r^2}\frac{\partial^2}{\partial\varphi}\right)+\frac{\alpha^2 m \omega^2}{2}r^2\,,&
\end{eqnarray} 
whose eigenstates and spectrum are given by
\begin{eqnarray}
&\label{EigenstatesHar}
\psi_{n_r,l}^\pm(r,\varphi) =\left(\frac{m\omega\alpha^2}{\hbar}\right)^{\frac{1}{2}}
\sqrt{\frac{n_r!}{2\pi\alpha\Gamma(n_r+\alpha l+1)}}\,
\zeta^{\alpha l}L_{n_r}^{(\alpha l)}(\zeta^2)
e^{-\frac{\zeta^{2}}{2} \pm i l\varphi}\,,\qquad 
\zeta=\sqrt{\frac{m\alpha^2\omega}{\hbar}}r\,, &\\&
E_{n,l}=\hbar\omega(2n_r+\alpha l+1)\,,\qquad
n_r\,,l=0,1,\ldots\,.
\end{eqnarray} 
These eigenstates are orthonormal, 
 $\braket{\psi_{n_r,l}^\pm}{\psi_{n_r',l'}^\mp}=\delta_{n_r,n_r'}\delta_{l,l'}$,  with 
respect to the scalar product (\ref{scalarp}).

Note that, contrary to the case of 
the free particle, the degeneracy of the energy levels depends on the value of $ \alpha $.
In the particular case of rational values   $\alpha=q/k$  we distinguish two cases: 
\begin{itemize}
\item When  $q$ is even, $q=2\tilde{q}$,  the energy levels satisfy the 
relation
\begin{eqnarray}
\label{EnergiaDegenerada1}
&E_{n_r+\tilde{q}j,l-kj}= E_{n_r,l}\,,\qquad j=-[\frac{n_r}{\tilde{q}}],\ldots, 
[\frac{l}{k}]\,,&
\end{eqnarray}
where $[.]$ indicates the integer part of the quotient.  This 
implies that all 
the eigenstates $\psi_{n_r+\tilde{q}j,l-kj}^\pm$ have the same energy eigenvalue. 
Counting the number of these eigenstates we get the following 
value
for the degeneracy: 
\begin{eqnarray}
&g(N)=2[\frac{N}{\tilde{q}}]+1\,,\qquad N=n_r+\frac{\tilde{q}}{k}l\,. &
\end{eqnarray}
\item When $q$ is odd, we have 
\begin{eqnarray}
\label{EnergiaDegenerada2}
&E_{n_r+qj,l-2kj}= E_{n_r,l}\,,\qquad j=-[\frac{n_r}{q}],\ldots, [\frac{l}{2k}]\,,&
\end{eqnarray}
and the degeneracy is given by 
\begin{eqnarray}
&g(N)=[\frac{N}{q}]+1\,,\qquad N=2n_r+\frac{q}{k}l\,. &
\end{eqnarray}
\end{itemize}
Like  
 the classical case, 
we can use the quantum version of the conformal bridge transformation
to connect this system with the quantum version of the free particle 
in the conical geometry  with the same value of $\alpha$.
As we are interested in the operators at $\tau=0$, we consider
the stationary conformal bridge transformation 
generated by
$\hat{\mathfrak{S}}(0,0)=\hat{\mathfrak{S}}_0$, which produces 
\begin{eqnarray}
&
\hat{\mathfrak{S}}_0(\hat{H}^{(\alpha)})\hat{\mathfrak{S}}_0^{-1}=-\omega\hbar \hat{\mathcal{J}}_-\,,\qquad
\hat{\mathfrak{S}}_0(\hat{iD}_0)\hat{\mathfrak{S}}_0^{-1}=\hbar \hat{\mathcal{J}}_0\,,\qquad
\hat{\mathfrak{S}}_0(\hat{K}_0)\hat{\mathfrak{S}}_0^{-1}=\frac{\hbar}{\omega}\hat{\mathcal{J}}_-\,,&
\end{eqnarray}
where 
\begin{eqnarray}
\label{QuantumSLCone}
&\hat{\mathcal{J}}_0=\frac{1}{2\hbar \omega}\hat{H}_{\text{os}}^{(\alpha)}\,,\qquad
\hat{\mathcal{J}}_\pm=
-\frac{m\omega}{4\hbar }\left(\hat{H}_{\text{os}}^{(\alpha)}-
m\omega^2 \alpha^2 r^2 \pm \hbar\omega\left(r\frac{\partial}{\partial r}+1\right)
\right)\,. &
\end{eqnarray}
Also, when the transformation acts on the formal operators $\hat{\Pi}_\pm$ and $\hat{\Xi}_\pm$,
one gets 
\begin{eqnarray}
&
\hat{\mathfrak{S}}_0(\hat{\Pi}_-)\hat{\mathfrak{S}}_0^{-1}=-i\sqrt{2m\hbar\omega}\hat{\mathfrak{b}}_1^-\,,\qquad
\hat{\mathfrak{S}}_0(\hat{\Pi}_+)\hat{\mathfrak{S}}_0^{-1}=-i\sqrt{2m\hbar\omega}\hat{\mathfrak{b}}_2^-\,,
&\\&
\hat{\mathfrak{S}}_0(\hat{\Xi}_{+})\hat{\mathfrak{S}}_0^{-1}=\sqrt{\frac{2m\hbar}{\omega}}\hat{\mathfrak{b}}_1^+\,,\qquad
\hat{\mathfrak{S}}_0(\hat{\Xi}_-)\hat{\mathfrak{S}}_0^{-1}=\sqrt{\frac{2m\hbar}{\omega}}\hat{\mathfrak{b}}_2^+\,,
&
\end{eqnarray}
where 
\begin{eqnarray}
&
\hat{\mathfrak{b}}_1^-=\frac{1}{2}e^{-i\frac{\varphi}{\alpha}}\sqrt{\frac{m \omega }{\hbar }}
\left(
\alpha r+
\frac{\hbar }{m\omega\alpha}\left(\frac{\partial}{\partial r}
-\frac{i\alpha}{r}\frac{\partial}{\partial \varphi} \right)\right)\,, \qquad \hat{\mathfrak{b}}_1^+=(\hat{\mathfrak{b}}_1^-)^\dagger\,, &\\&
\hat{\mathfrak{b}}_2^-= \frac{1}{2}e^{i\frac{\varphi}{\alpha}}\sqrt{\frac{m \omega }{\hbar }}
\left(
\alpha r+
\frac{\hbar  }{m\omega\alpha}\left(\frac{\partial}{\partial r}
+\frac{i \alpha}{ r}\frac{\partial}{\partial \varphi}\right) \right) \,, \qquad \hat{\mathfrak{b}}_2^+=(\hat{\mathfrak{b}}_2^-)^\dagger\,,
&
\end{eqnarray}
are the  formal dimensionless ladder operators in polar coordinates representation. In terms of them, the generators 
of the conformal algebra (\ref{QuantumSLCone}) and the quantum version of $\mathcal{L}_2$ 
take the form
\begin{eqnarray}
&
\hat{\mathcal{J}}_0=\frac{1}{4}(\{\hat{\mathfrak{b}}_1^+,\hat{\mathfrak{b}}_1^-\}+\{\hat{\mathfrak{b}}_2^-,\hat{\mathfrak{b}}_2^+\})\,,\qquad
 \hat{\mathcal{J}}_\pm=\hat{\mathfrak{b}}_1^\pm\hat{\mathfrak{b}}_2^\pm\,,
 &\\&
 \hat{\mathcal{L}}_2= \frac{1}{2}(\hat{\mathfrak{b}}_1^+\hat{\mathfrak{b}}_1^--\hat{\mathfrak{b}}_2^+\hat{\mathfrak{b}}_2^-)
 =\frac{1}{2\hbar}\hat{p}_\varphi\,.
&
\end{eqnarray}

According to
relation (\ref{EigenEstates}), the 
corresponding 
eigenstates of the form 
 $\braket{\vr}{\lambda}$ we are looking for correspond to 
\begin{equation}
\label{JordanStates}
\Omega_{n_r,l}^\pm(r,\varphi) =r^{2n_r+\alpha l }e^{\pm l \varphi}\,,
\end{equation}
which satisfy the following
set of equations, 
\begin{eqnarray}
&\label{DonOmega}
2i\hat{D}_0\Omega_{n_r,l}^{\pm}(r,\varphi)=\hbar(2n_r+\alpha l+1)\Omega_{n_r,l}^{\pm}(r,\varphi)\,,&\\&
\hat{K}_0\Omega_{n_r,l}^{\pm}(r,\varphi)=\frac{m\alpha^2}{2}\Omega_{n_r+1,l}^{\pm}(r,\varphi)\,, 
&\\&
\label{HKonOmega}
\hat{H}^{(\alpha)}\Omega_{n_r,l}^{\pm}(r,\varphi)= -\frac{2\hbar^2}{m\alpha^2} n_r(n_r+\alpha l)\Omega_{n_r-1,l}^\pm(r,\varphi)\,,&\\&
\label{RPionOmega1}
\hat{\Xi}_\pm\Omega_{n_r,l}^\pm(r,\varphi)=
\alpha m \Omega_{n_r,l+\frac{1}{\alpha}}^\pm(r,\varphi) \,,\qquad
\hat{\Xi}_\pm\Omega_{n_r,l}^\mp(r,\varphi)=
\alpha m\Omega_{n_r+1,l-\frac{1}{\alpha}}^\mp(r,\varphi)\,, &\\&
\hat{\Pi}_\pm\Omega_{n_r,l}^\pm(r,\varphi)=-i\frac{2\hbar n_r}{\alpha}\Omega_{n_r-1,l+\frac{1}{\alpha}}^\pm(r,\varphi)\,,&\\&
\hat{\Pi}_\pm\Omega_{n_r,l}^\mp(r,\varphi)=-i2
(n_r+\alpha l)\frac{\hbar}{\alpha}\Omega_{n_r,l-\frac{1}{\alpha}}^\mp(r,\varphi)\label{RPionOmega2}
\,.&
\end{eqnarray}
Here, the functions
$\Omega_{0,l}^\pm$ are the zero energy eigenstates of  $\hat{H}^{(\alpha)}$, but only the function $\Omega_{0,0}^+=\Omega_{0,0}^-=1$ is 
a physical eigenstate for the free particle system. Functions  $\Omega_{n_r,l}^\pm$ are the
rank $n_r$ Jordan states of zero energy which satisfy 
\begin{eqnarray}
&
(\hat{H}^{(\alpha)})^j\Omega_{n_r,l}^{\pm}(r,\varphi)=\left(-2\frac{\hbar}{m\alpha^2}\right)^{j} (n_r)_{j}
(n_r+\alpha l)_{j}\Omega_{n_r-j,l}^{\pm}(r,\varphi)\,, \qquad j=1,\ldots n_r\,,&\\&
(g)_{j}=\prod_{i=0}^{j-1}(g-i)\,.
&
\end{eqnarray}
Another remarkable 
property of these functions is that the eigenstates of the free particle admit the representation 
\begin{eqnarray}
&
\psi_{\kappa,l}^\pm(r,\varphi)=\sqrt{\frac{\kappa}{2\pi\alpha}}
\sum_{n_r=0}^{\infty}\frac{(-1)^{n_r} (\kappa/2)^{2n_r+\alpha l}}{n_r!\Gamma(n_r+\alpha l+1)}\Omega_{n_r,l}^\pm(r,\varphi)\,.&
\end{eqnarray}
By direct application of the stationary  
conformal bridge operator $\hat{\mathfrak{S}}_0$ 
to  functions (\ref{JordanStates})
one gets 
\begin{eqnarray}
&
\hat{\mathfrak{S}}_{0}\Omega_{n_r,l}^\pm(r,\varphi)= \mathcal{N}_{n_r,l}
\psi_{n_r,l}^\pm(r,\varphi)\,, &\\&
\mathcal{N}_{n_r,l}=(-1)^{n_r}
 \left(\frac{2\hbar}{m\omega \alpha^2}\right)^{n_r+\frac{\alpha l}{2}}\sqrt{2\alpha\pi n_r!\Gamma(n_r+\alpha l+1)}\,.
&
\end{eqnarray}
On the other hand,  the action of the transformation on
 the  free particle eigenstates produces 
 the (non-normalized) coherent states of the 
system corresponding to  
\begin{eqnarray}
\begin{array}{lcl}
\Psi_{\kappa,\ell}^\pm(r,\varphi)&=&\hat{\mathfrak{S}}_{0}\psi_{\kappa,\ell}^\pm(r,\varphi)=2^{\frac{1}{2}}
e^{\mathcal{E}}\psi_{\kappa,\ell}^\pm(\sqrt{2}r,\varphi)e^{-\frac{m\omega r^2}{2\hbar}}\\
&=&
\sqrt{\kappa}\sum_{n_r=0}^{\infty}\frac{\mathcal{E}^{n_r+\frac{\alpha l}{2}}
}{\sqrt{n_r!\Gamma(n_r+\alpha l+1)}}\psi_{n_r,l}^\pm(r,\varphi)\,,
\end{array}
\end{eqnarray}
where $\mathcal{E}=\frac{\hbar\kappa^2}{2m\omega\alpha^2}$. 
These states satisfy the relation
$\hat{\mathcal{J}}_-\Psi_{\kappa,\ell}^\pm(r,\varphi)=-\mathcal{E}\Psi_{\kappa,\ell}^\pm(r,\varphi)$.

From the  equations 
 (\ref{HKonOmega}) we obtain the action of generators $\hat{\mathcal{J}}_\pm$,
\begin{eqnarray}
&
\hat{\mathcal{J}}_+\psi_{n_r,l}^\pm(r,\varphi)=- \sqrt{(n_{r}+1)
(n_{r}+\alpha l+1)} \psi_{n_r+1,l}^\pm(r,\varphi)\,,&\\&
\hat{\mathcal{J}}_-\psi_{n_r,l}^\pm(r,\varphi)=-\sqrt{n_{r}
(n_{r}+\alpha l)} \psi_{n_r-1,l}^\pm (r,\varphi)\,,&
\end{eqnarray}
and from equations (\ref{RPionOmega1})-(\ref{RPionOmega2}) we get the formal relations 
\begin{eqnarray}
&
\hat{\mathfrak{b}}_a^{\pm}\,\psi_{n_r,l}^{(a)}(r,\varphi)=\sqrt{(n_r+\alpha l +\beta_\pm)}\,\psi_{n_r,l\pm \frac{l}{\alpha}}^{(a)}(r,\varphi)\,,&\\&
\hat{\mathfrak{b}}_a^{\pm}\,\psi_{n_r,l}^{(b)}(r,\varphi)=-\sqrt{(n_r+\beta_\pm)}\,\psi_{n_r\pm 1,l\mp \frac{l}{\alpha}}^{(b)}(r,\varphi)\,,&
\end{eqnarray}
where 
$\psi_{n_r,l}^{(1)}=\psi_{n_r,l}^{+}$, 
$\psi_{n_r,l}^{(2)}=\psi_{n_r,l}^{-}$, 
and $\beta_\pm=\frac{1}{2}(1\pm 1)$.
From
these equations it becomes obvious that these operators cannot be physical for arbitrary values of
$\alpha$, since they produce wave-functions outside
the Hilbert space generated by the physical eigenstates $\psi_{n_r,l}^\pm$. 
In fact, this is already observable 
from the equations (\ref{RPionOmega1})-(\ref{RPionOmega2}),
 where it is seen that the produced functions on  the
 right hand side do not satisfy some of the criteria imposed at the end of Sec. \ref{SecConBrid}:
these functions are not single-valued in the angular coordinate, and some of them are singular 
at  $r=0$.

Similarly to
 the free particle system in the conical space, 
we must study carefully the rational case $\alpha=q/k$, and to do that,
it is enough to 
analyze
 the relations
 that imply a decrease in the angular momentum quantum number $l$ 
(the relations in which  
this number is  increasing have no problems). 
First,   consider the relation
\begin{eqnarray}
&
\label{b^q}
(\hat{\mathfrak{b}}_a^{-})^q\,\psi_{n_r,l}^{(a)} (r,\varphi)=
\sqrt{\frac{\Gamma(n_r+(q/k) l+1)}{\Gamma(n_r+(q/k) l-q+1)}}\,\psi_{n_r,l-k}^{(a)}(r,\varphi)\,.
&
\end{eqnarray}
When $l<k$, that is $l-k=-j<0$,
the explicit form of the function 
on  the right hand side 
is 
\begin{eqnarray}
\label{G1}&
\psi_{n_r,-j}^\pm(r,\varphi) =\left(\frac{m\omega q^2}{\hbar k^2}\right)^{\frac{1}{2}}
\sqrt{\frac{kn_r!}{2\pi q \Gamma(n_r-(q/k)j+1)}}\,
\zeta^{-\frac{q}{k}j}L_{n_r}^{(-\frac{q}{k} j)}(\zeta^2)
e^{-\frac{\zeta^{2}}{2} \mp i j\varphi}\,,&
\end{eqnarray}
and because the upper index of the generalized Laguerre polynomial
 is negative and rational, we conclude that 
this function is outside  
the Hilbert space generated by the physical eigenstates, 
implying that the operators can not be observable. 

Now, in the particular case when $k=1$, the upper index in the generalized 
Laguerre polynomial is a negative integer number, 
and  due to the  identity
\be
\label{LagerIdentity}
\frac{(-\eta)^{i}}{i!}L_{n}^{(i-n)}(\eta)=
\frac{(-\eta)^{n}}{n!}L_{i}^{(n-i)}(\eta)\,,\qquad i,n=0,1,\ldots,
\ee
one gets that in the case $n_r\geq ql$, 
\be
\label{K4}
\psi_{n_r,-l}^\pm(r,\varphi)=(-1)^{ql} \psi_{n_r-ql,l}^\mp(r,\varphi)\,,
\ee
while for $n_r<ql$ the right hand side in Eq. (\ref{G1}) vanishes due to 
the poles in the Gamma function.
Therefore, Eq.  (\ref{b^q}) has no problems when $k=1$. 
Note that 
in these cases the normalization factor (\ref{G1}) also equals  zero 
by
the same reason. 

Now,  we consider the relation
\begin{eqnarray}
\label{b+q}&
(\hat{\mathfrak{b}}_a^{+})^{q}\,\psi_{n_r,l}^{(b)}(r,\varphi)=(-1)^q
\sqrt{\frac{\Gamma(n_r+1 +q)}{\Gamma(n_r+1)}}\,\psi_{n_r+ q,l-k}^{(b)}(r,\varphi)\,,
&
\end{eqnarray}
from where we see that the same problems for the cases $l<k$ appear again, and only in 
the case $k=1$ 
 the relation (\ref{K4}) ensures that the operator always 
produces physical eigenstates. 
 
The realized analysis of Eqs. (\ref{b^q}) and  (\ref{b+q}) 
  reveals  
   that, again, we face the problem of the quantum anomaly in the general case of rational
   values of $\alpha$, and 
  we can construct a well-defined symmetry operators only in the case 
  of integer values of $\alpha$, 
 that
 we assume from now on. 

Like 
the classical case, 
we construct the quantum symmetry generators by means of the conformal bridge transformation, obtaining 
\begin{eqnarray}
&
\hat{\mathfrak{S}}_0( \hat{\mathcal{O}}_{\mu,\nu}^\pm)\hat{\mathfrak{S}}_0^{-1}=
(-i)^\nu (m\hbar)^{\frac{q}{2}}(\omega)^{\frac{\mu-\nu}{2}}\hat{\mathcal{G}}_{\mu,\nu}^\pm\,,
&\\&
\hat{\mathfrak{S}}_0(\hat{\mathcal{S}}_{\mu',\nu'}^\pm)\hat{\mathfrak{S}}_0^{-1} =
(-i)^\nu (m\hbar)^{q} (\omega)^{\frac{\mu'-\nu'}{2}}\hat{\mathcal{F}}_{\mu',\nu'}^\pm \,,&
\end{eqnarray}
and the  properties of the classical algebra (with $k=1$) are again preserved at the quantum level.

In conclusion of this discussion, let us
  make some comments with respect to the differences between the cases 
 of even and odd values of  $\alpha=q$.
In the even case $q=2\tilde{q}$, we have the true integrals of motion $\hat{\mathcal{G}}_{\tilde{q},\tilde{q}}^\pm$, 
whose
explicit action on the eigenstates  is
\begin{eqnarray}
&\label{G+-psi+-}
\hat{\mathcal{G}}_{\tilde{q},\tilde{q}}^{\pm}\psi_{n_r,l}^\pm(r,\varphi)=(-1)^{\tilde{q}}\sqrt{\frac{\Gamma(n_r+1)\Gamma(n_r+ 2\tilde{q} l+1+\tilde{q})}
{\Gamma(n_r+1-\tilde{q})\Gamma(n_r+2\tilde{q} l+1)}}\psi_{n_r-\tilde{q},l+1}^\pm(r,\varphi)\,,
&\\&\label{G-+psi+-}
\hat{\mathcal{G}}_{\tilde{q},\tilde{q}}^{\mp}\psi_{n_r,l}^\pm(r,\varphi)=(-1)^{\tilde{q}}\sqrt{\frac{\Gamma(n_r+1+\tilde{q})
\Gamma(n_r+ 2\tilde{q} l+1)}{\Gamma(n_r+1)\Gamma(n_r+2\tilde{q} l+1-\tilde{q})}}\psi_{n_r+\tilde{q},l-1}^\pm(r,\varphi)\,.
&
\end{eqnarray}
From  Eq. (\ref{G+-psi+-}) we learn that 
the operators $\hat{\mathcal{G}}_{\tilde{q},\tilde{q}}^{\pm}$ 
annihilate the eigenstates $\psi_{n_r,l}^\pm$
with $n_r<\tilde{q}-1$. Also,  due to the relation (\ref{K4}), and the fact that $\psi_{n_r,0}^+=\psi_{n_r,0}^-:=\psi_{n_r,0}$,
 both equations (\ref{G+-psi+-}) and (\ref{G-+psi+-}) 
are equivalent  in 
the case $l=0$ and $n_r>\tilde{q}-1$. 
Otherwise, the pole in the Gamma function in (\ref{G-+psi+-}) 
produces zero.  One also
notes that the index in the wave-functions 
that appears  on the right hand 
side of these equations corresponds to 
the same index in the equations (\ref{EnergiaDegenerada1}) for the cases $j=\pm1$. 

On the other hand, for the odd case $q=2\tilde{q}+1$ we need to consider instead the integrals $\mathcal{F}_{q,q}^\pm$ which produce 
\begin{eqnarray}
&\label{F+-psi+-}
\hat{\mathcal{F}}_{q,q}^{\pm}\psi_{n_r,l}^\pm(r,\varphi)=(-1)^{q}
\sqrt{\frac{\Gamma(n_r+1)\Gamma(n_r+ q l+1+q)}{\Gamma(n_r+1-q)\Gamma(n_r+q l+1)}}\psi_{n_r-q,l+2}^\pm(r,\varphi)\,,
&\\&\label{F-+psi+-}
\hat{\mathcal{F}}_{q,q}^{\mp}\psi_{n_r,l}^\pm(r,\varphi)=(-1)^{q}
\sqrt{\frac{\Gamma(n_r+1+q)\Gamma(n_r+ q l+1)}{\Gamma(n_r+1)\Gamma(n_r+q l+1-q)}}\psi_{n_r+q,l-2}^\pm(r,\varphi)\,.
&
\end{eqnarray}
These equations 
reveal the properties similar  to those described  for the even case. 
 The index in the resulting wave-function is
  in correspondence with the 
index in the equation (\ref{EnergiaDegenerada2}). 

Then, we conclude that the existence of the true (not depending explicitly on time) integrals 
$\hat{\mathcal{G}}_{\tilde{q},\tilde{q}}^\pm$ and  $\hat{\mathcal{F}}_{q,q}^\pm$  
reflects 
the degeneracy of the system in the unique anomaly-free cases
of the even and odd values of $\alpha$, respectively.

\section{Discussion and outlook}
\label{SecDisOut}
The premise of our research here was that on the one hand,  geometric properties of space-time must be 
reflected in the intrinsic characteristics of the physical systems that inhabit it, and on the other hand, these 
peculiarities 
must be encoded in a set of well-defined integrals of motion.
Bearing this in mind, we have studied different 
non-relativistic forms of dynamics (in the sense of Dirac  \cite{Dirac}) associated with
the $\mathfrak{so}(2,1)\cong\mathfrak{sl}(2,\R)$ conformal symmetry, namely, the free particle 
and the harmonic oscillator,  on a cosmic string background \cite{Kibble,Vilen,Vilenkin,Dow}. 
This is the analogous problem of 
considering the non-relativistic  
conformal invariant dynamical models
on
 a two-dimensional cone surface, the deficiency/excess angle of which 
is given in terms of the ``geometrical parameter"  
$ \alpha = {1}/(1- \frac{4 \mu G}{c^2})\,, $
where $ \mu $ is the linear mass 
density of the cosmic string  \cite{SokSta,Vilenkin} which can be positive, 
or negative when topological defects in condensed matter physics and 
wormholes are considered  \cite{Visser,Cramer,KatVol,Volovik,Manton}. 
Based on the previous observations related to the shape of the trajectories of the systems 
 in this space  \cite{DesJack,Furtado},
one might assume 
that for some special values of the parameter  $ \alpha $, 
the systems may
have well-defined 
hidden
symmetry generators that adequately describe them. 
Our results confirm this hypothesis.

The main tools used in this investigation  were a local  canonical 
transformation and the conformal bridge transformation \cite{IPW}. 
The first transformation relates the systems under investigation with 
their version in the flat Euclidean plane, while the second one 
allows us to map  integrals from one form of conformal dynamics to another.
The strategy was to start with a free particle in $\R^2$
($\alpha=1$),
from where we obtain, by a local canonical transformation, 
 the solutions of the equations of 
motion and the formal integrals of the free system in the cone.
  These formal conserved  quantities are the images of the Euclidean canonical momenta and the 
  Galilean boosts,
  which  are not globally
   well-defined functions in phase space for arbitrary $ \alpha $ due to their 
   peculiar angular 
   dependence. 
   However, no problems
  appear for $\alpha=1/k$,  $k=2,3,\ldots,$ while 
  in general case of rational values $\alpha=q/k$ with $q,k=1,2,\ldots$, 
these formal conserved quantities can be used to construct well-defined functions on phase space. 
After characterizing the complete symmetry algebra of the free particle in the cone, 
we have proceeded to apply the conformal bridge transformation in order  to obtain 
the integrals  and their
symmetry  algebra 
for the harmonic oscillator in the 
same geometric background. 

In order to know what to expect for systems on the cone, it is instructive to make
 some comments about the models in $\R^2$. 
In the case of the free particle we have ten second-order integrals that 
generate the $ \mathfrak{sp}(4,\R) $ Lie algebra.
Also, 
there are four first-order integrals, namely the canonical momenta and the Galilean boosts, that produce the 
centrally extended  two-dimensional Heisenberg algebra. The 
first order integrals  generate an ideal sub-algebra of the complete
Lie type  symmetry of the system. 
In addition to the principal,
conformal  $\mathfrak{so}(2,1)$ algebra,
produced by the Hamiltonian $H$ and 
 generators of dilatations, $D$, 
and special conformal transformations, $K$,
one also can  identify some other sub-algebraic structures: 
\begin{itemize}
\item  A secondary 
 $ \mathfrak{sl}(2,\R) $ algebra, generated by the angular momentum and two 
 dynamical integrals 
that commute with the dilatation generator.
\item An 
$\mathfrak{su}(2)\oplus\mathfrak{su}(2)$ dynamical symmetry, produced by complex dynamical generators. 
\end{itemize}
When considering  the planar isotropic harmonic oscillator, one also finds 10 second-order symmetry generators. 
They produce an algebraic structure that is isomorphic to the $ \mathfrak{sp} (4,\R) $  algebra 
(up to complex linear combinations of generators). Furthermore, the system possesses an ideal two-dimensional 
Heisenberg sub-algebra, produced by the 
four first-order ladder operators. 
As for the free particle system, here we also 
have 
the principal 
$\mathfrak{sl}(2,\R)$ algebra
of the conformal Newton-Hooke symmetry
  generated by the Hamiltonian and the second order radial ladder operators.
  Other 
 sub-algebraic structures 
 which we would like to highlight are:
\begin{itemize}
\item The very well known $ \mathfrak{su}(2) $ symmetry, generated by the angular momentum and two other 
true 
integrals of motion that commute with the corresponding Hamiltonian. 
\item The  $ \mathfrak {sl} (2,\R)\oplus\mathfrak{sl}(2,\R)$ symmetry,
whose generators can  be identified with  the integrals of
 the associated Landau problem in  symmetric gauge \cite{IPW}. 
\end{itemize}
After applying the conformal bridge transformation to the Euclidean free particle 
system, we find  that the principal 
$\mathfrak{so}(2,1)$ conformal symmetry is mapped into 
 the 
$\mathfrak{sl}(2,\R)$ algebra of the conformal Newton-Hooke 
symmetry of the harmonic oscillator. 
In a similar way, 
the mentioned secondary 
 $ \mathfrak {sl} (2, \R) $  symmetry  of the free dynamics is mapped
  to the above-mentioned 
  $ \mathfrak{su} (2) $ symmetry of the
  harmonically trapped  particle; the complex algebra 
  $\mathfrak{su}(2)\oplus \mathfrak{su}(2)$  of the free case
    is transformed into the 
     $\mathfrak{sl}(2,\R)\oplus\mathfrak{sl}(2,\R)$ sub-symmetry of the harmonically 
    confined system,
     and there 
    is a non-unitary correspondence between the Heisenberg algebras of both models. 
    In the general case  
    the conformal bridge transformation maps sub-algebras of one system into 
    sub-algebras of the another. This
    happens  due to its nature of the complex canonical transformation 
    at the classical level, and of the non-unitary similarity  transformation at the quantum 
    level, which is  the   non-unitary 
    automorphism of  the  
    conformal $\mathfrak{so}(2,1)\cong \mathfrak{sl}(2,\R)$ algebra
    \cite{IPW}.

It is important to reinforce the fact that in both cases the Lie algebraic structure
described 
here is greater than the Schr\"odinger symmetry presented in Niederer's early works 
\cite{Nied1,Nied2}. 
There, only the generators of the conformal symmetry, the Heisenberg symmetry and rotations were considered.
Let us note here 
that at the quantum level, the ten second-order generators in both systems
 can be built by taking anti-commutators of
the linear integrals.  
Then,
if we consider 
a nonlocal operator 
of the rotation in $ \pi $, $ \mathscr{R}=\exp(i\pi\hat{p}_\varphi)$, 
due to relations 
$ \mathscr{R} x_i = -x_i \mathscr{R} $ and 
$ \mathscr{R}^2=1$,
it can be identified as a grading operator. 
As a result, the Lie symmetry algebras we considered 
can be reinterpreted as Lie superalgebras $\mathfrak{osp}(1\vert 4,\R)$ of the corresponding 
two-dimensional systems with no fermionic degrees of freedom
\cite{CroRit}. 
This corresponds to the 
so-called systems with 
the bosonized (hidden)  supersymmetry, see 
\cite{HiddenSUSY1,HiddenSUSY+,HiddenSUSY++,HiddenSUSY2,HidConf,HiddenSUSY3}
and  references therein. 
In one-dimensional quantum case,
 the origin of such  type of hidden supersymmetries 
 can be understood in terms of reduction of  
supersymmetric systems with fermionic degrees of freedom \cite{HiddenSUSY2,HiddenSUSY3}.
It would be interesting to 
 investigate in a similar way
 the origin of the indicated
two-dimensional hidden (bosonized) supersymmetry.

The conformal   and rotational symmetries
of the free particle on the  cone 
 are described by the integrals 
which are well-defined phase space functions 
for arbitrary value of the geometric parameter $ \alpha>0$,
and generate
 the $ \mathfrak{so} (2,1) \oplus  \mathfrak{u}(1)$ algebra.
They are quadratic in the above-mentioned 
formal in general case basic integrals.  
For the case of rational $ \alpha = q/k $ ($q,k=1,2,\ldots$ with no common divisors),
the formal basic generators can be used to build the new well-defined integrals of motion:
  \begin{itemize}
  \item When $ q = 1 $ and $ k = 2,3, \ldots $ the symmetry algebra 
  is the same as for the free particle in the flat Euclidean plane ($\alpha=1$).
  \item When $ q = 2,3 \ldots $ and $ k = 1,2, \ldots $, it is possible to construct two different sets of higher-order 
  integrals of motion: a set $ \mathcal{U}_1 $, consisting of $ 2q + 2 $ integrals of order $ q $, and the other set
   $ \mathcal{U}_2 $, which contains $ 4q + 2 $ integrals of order $ 2q $. We have verified that both sets 
   generate independent finite  nonlinear algebras, and together they produce a larger finite nonlinear algebraic structure. 
   We have also shown that the nonlinear algebra  generated by $ \mathcal {U}_1 $ is an ideal sub-algebra of the complete
    nonlinear symmetry.
\end{itemize}   

After analyzing the classical system, we have considered the quantum case.
 The system can be quantized for arbitrary values of $ \alpha $, and one could  expect that  
 for the rational case, the corresponding quantum versions of the (in general) higher-order hidden symmetry 
 integrals   
 will be the spectrum generating operators.
 However,
  we have revealed  a quantum anomaly, since only in the case 
  of integer values $\alpha=q=2,3,\ldots$ such
  spectrum generating  integrals indeed can  be constructed,
  while in the case of rational non-integer values of $\alpha$ 
  the quantum analogs of the classically well-defined hidden symmetry generators 
  take out the states from the physical Hilbert space.

 Here, there is  a couple of open interesting questions.
 First,
 knowing that via the corresponding non-relativistic limit \cite{Hagen,Barut,Son},
 one can relate conformal symmetry of the free  particle 
 on the cone with the corresponding Killing and conformal  Killing  
 vector fields of the cosmic string space-time background,
 a natural question is what geometrical objects correspond 
 to the considered hidden symmetry generators. 
We speculate  that they  can be related 
to  some Killing and  conformal Killing tensors \cite{Cariglia,Crampin}
of  the cosmic string space-time.
Second,  it would be interesting to look  what happens with  the quantum anomaly
under perspective of unconventional boundary conditions, which were 
considered for quantum systems in the cone  in   \cite{Kay}.

Finally, we reconstruct the complete information on the symmetry algebra for the harmonic 
oscillator at the classical and quantum levels through the conformal bridge transformation.
From here we have  learned that the generators of the $ \mathfrak{sl} (2, \R) \oplus\mathfrak{u}(1) $ 
 algebra are well-defined for any value of $ \alpha $, but only 
 in the rational case $  \alpha = q / k $ there
 is an extended set of the well-defined integrals of motion.
 This reflects the peculiarity of the geometry of the  trajectories of the harmonically trapped particle
 on the cone:  the trajectories are closed only  when $\alpha=q/k$.
The number of radial minima/maxima  on the trajectory is given by 
$\mathscr{N}_{\text{min}/\text{max}}=k(q\,\text{mod}\,2+1)$. 
In particular, 
for $k=1$ one has  $\mathscr{N}_{\text{min}/\text{max}}=1$ when $\alpha$ is even and 
 $\mathscr{N}_{\text{min}/\text{max}}=2$ 
when $\alpha$ is odd.  At the quantum level it is seen that the 
value of $ \alpha $ explicitly determines the 
  degeneracy of the energy spectrum of the system, while 
the symmetry operators
   and their action on the corresponding eigenstates of the system are obtained directly 
   from those of the free particle 
   by employing  the quantum version of the conformal bridge transformation. 
   From there, the quantum anomaly is revealed automatically: only when $ \alpha = q = 2,3, \ldots $,
    it is possible 
   to have the well-defined higher-order differential 
   operators corresponding to hidden symmetries,
  which reflect the  spectral degeneracy.
  
In conclusion, let us indicate some other problems for which 
the results and ideas employed in this article can be used.

 First, we note that  the local 
 canonical transformation that relates the cone with the flat Euclidean plane  
 can be applied for the analysis of other 
central potentials on the conical background. 
One can  expect that the  well-defined hidden symmetry generators 
in such systems can appear only at  special values of $\alpha$, and 
that quantum anomaly can also emerge  there.
In particular, all the analysis presented here can immediately be transferred and 
generalized for the case of conformal mechanics on the cone
as it was done in \cite{IPW} in the flat Euclidean plane.
In the same vein, the results of \cite{IPW}  can be employed and generalized
immediately for the Landau problem on the cone. 

Second, it would be  interesting  to employ  the conformal bridge transformation 
for the systems in different geometries, such as the Lobachevsky plane and 
the non-commutative plane. 
These both geometries are used  in the description of anyons 
\cite{Torsion,NonCom2,NonComPLB},
and on the other hand, one has to bare in mind that anyons can  directly be related 
with the cone geometry \cite{LeiMyr,LeiMyr+,MKWil}.  

Finally, as it was shown in \cite{IPW}, 
the non-unitary generator $\hat{\mathfrak{S}}$ of the conformal bridge transformation is 
 the fourth order root of the space reflection operator  $\mathcal{P}$, 
 which in the present two-dimensional case has to be  substituted for
 the above-mentioned non-local operator $ \mathscr{R}=\exp(i\pi\hat{p}_\varphi)$.  
 At the same time,   it is easy to see from the explicit form (\ref{QCB1}) 
 of $\hat{\mathfrak{S}}$
in terms of the $ \mathfrak{sl} (2, \R)$ generators of conformal symmetry
that it is $\mathcal{PT}$ symmetric, 
where $\mathcal{T}$ corresponds to 
the time reversal (anti-linear) transformation 
\cite{Bender,Mostafazadeh,Bender2007,DCT,Fring}.
Taking also into account that its classical analog 
generates complex canonical transformations 
in the phase space, and that conformal symmetry 
plays important role in some  $\mathcal{PT}$-symmetric
systems with peculiar properties \cite{DCT,MatMik1,MatMik2}, 
it would be very interesting  to look at the conformal bridge transformations in the 
light  of  the  $\mathcal{PT}$ symmetry.
     
\vskip0.3cm

\noindent {\large{\bf Acknowledgements} } 
\vskip0.1cm

The work was partially supported by the FONDECYT Project 1190842 
and the DICYT Project 042131P\_POSTDOC.


\begin{thebibliography}{99}



\bibitem{Cariglia}
M.~Cariglia,
\emph{``Hidden symmetries of dynamics in classical and quantum physics,"}
\href{https://journals.aps.org/rmp/abstract/10.1103/RevModPhys.86.1283}{Rev. Mod. Phys. \textbf{86} (2014)  1283 } 
\href{https://arxiv.org/abs/1411.1262}{\textcolor{magenta}{[arXiv: 1411.1262 [math-th]]}}.

\bibitem{nonlinear2}
J. de Boer, F. Harmsze and T. Tjin, 
\emph{``Nonlinear finite W symmetries and applications
in elementary systems,''}
\href{https://www.sciencedirect.com/science/article/abs/pii/0370157395000755?via%3Dihub}{Phys. Rept. {\bf 272} (1996) 139}
\href{https://arxiv.org/abs/hep-th/9503161}{\textcolor{magenta}{[arXiv: hep.th/9503161]}}.



\bibitem{nonlinear3}
 J. Beckers, Y. Brihaye and N. Debergh, \emph{``On realizations of `nonlinear' Lie algebras
by differential operators,"} 
\href{https://iopscience.iop.org/article/10.1088/0305-4470/32/15/008/meta}{J. Phys. A \textbf{32} (1999) 2791}
\href{https://arxiv.org/abs/hep-th/9803253}{\textcolor{magenta}{[arXiv: hep.th/9803253]}}.


\bibitem{Pauli}
W. Pauli, 
\emph{``\"Uber das wasserstoffspektrum vom standpunkt der neuen quantenmechanik,"}
\href{https://link.springer.com/article/10.1007%2FBF01450175}{Z. Physik {\bf 36} (1926) 336}.

 \bibitem{Frad}
D. M. Fradkin, 
\emph{``Three-dimensional isotropic harmonic oscillator and SU$_3$,''}
\href{https://aapt.scitation.org/doi/10.1119/1.1971373}{Am. J. Phys. \textbf{33} (1965) 207}.



\bibitem{PlyWipf}
M.~S.~Plyushchay and A.~Wipf,
\emph{``Particle in a self-dual dyon background: hidden free nature, and exotic superconformal symmetry,''}
\href{https://journals.aps.org/prd/abstract/10.1103/PhysRevD.89.045017}{Phys. Rev. D \textbf{89} (2014)  045017}
\href{https://arxiv.org/abs/1311.2195}{\textcolor{magenta}{[arXiv: 1311.2195 [hep-th]]}}.


\bibitem{IPW2}
L.~Inzunza, M.~S.~Plyushchay and A.~Wipf,
\emph{``Hidden symmetry and (super)conformal mechanics in a monopole background,''}
\href{https://link.springer.com/article/10.1007/JHEP04(2020)028}{JHEP \textbf{04} (2020) 028}
\href{https://arxiv.org/abs/2002.04341}{\textcolor{magenta}{[arXiv: 2002.04341 [hep-th]]}}.



\bibitem{Woj}
S. Wojciechowski,
\emph{``Superintegrability of the Calogero-Moser system,''}
\href{https://www.sciencedirect.com/science/article/abs/pii/037596018390018X}{Phys. Lett. A {\bf 95} (1983) 279}.

\bibitem{Kuz}
V. B. Kuznetsov, 
\emph{``Hidden symmetry of the quantum Calogero-Moser system,''}
\href{https://www.sciencedirect.com/science/article/abs/pii/0375960196004215?via%3Dihub}{Phys. 
Lett. A {\bf 218}  (1996) 212} 
\href{https://arxiv.org/abs/solv-int/9509001}{\textcolor{magenta}{[arXiv: solv-int/9509001]}}.



\bibitem{Hawking}
S.~W.~Hawking,
\emph{``Black hole explosions,''}
\href{https://www.nature.com/articles/248030a0}{Nature \textbf{248} (1974) 30}.


\bibitem{Unruh}
W.~G.~Unruh,
\emph{``Notes on black hole evaporation,''}
\href{https://journals.aps.org/prd/abstract/10.1103/PhysRevD.14.870}{Phys. Rev. D \textbf{14} (1976) 870}.

\bibitem{ConformalBH0}
R.~Britto-Pacumio, J.~Michelson, A.~Strominger and A.~Volovich,
\emph{``Lectures on superconformal quantum mechaics and multi-black hole moduli spaces,"}
\href{https://link.springer.com/chapter/10.1007%2F978-94-011-4303-5_6}{NATO Sci. Ser. C {\bf 556} (2000) 255}
\href{https://arxiv.org/pdf/hep-th/9911066.pdf}{\textcolor{magenta}{[arXiv: hep-th/9911066]}}.


\bibitem{ConformalBH1} 
  P.~Claus, M.~Derix, R.~Kallosh, J.~Kumar, P.~K.~Townsend and A.~Van~Proeyen,
  \emph{ ``Black holes and superconformal mechanics,''}
  \href{https://journals.aps.org/prl/abstract/10.1103/PhysRevLett.81.4553}{Phys.\ Rev.\ Lett.\  {\bf 81} (1998) 4553}
   \href{https://arxiv.org/abs/hep-th/9804177}{\textcolor{magenta}{[arXiv: hep-th/9804177]}}.
 
  
\bibitem{ConformalBH2}
J.~A.~de~Azc\'arraga, J.~M.~Izquierdo, J.~C.~P\'erez~Bueno and P.~K.~Townsend,
\emph{``Superconformal mechanics, black holes, and nonlinear realizations,"}
\href{https://journals.aps.org/prd/abstract/10.1103/PhysRevD.59.084015}{Phys. Rev. D \textbf{59} (1999)  084015}
\href{https://arxiv.org/abs/hep-th/9810230}{\textcolor{magenta}{[arXiv: hep-th/9810230]}}.

\bibitem{ConformalBH3}
  G.~W.~Gibbons and P.~K.~Townsend,
  \emph{``Black holes and Calogero models,''}
  \href{https://www.sciencedirect.com/science/article/pii/S037026939900266X?via%3Dihub}{Phys.\ Lett.\ B {\bf 454} (1999) 187}
  \href{https://arxiv.org/abs/hep-th/9812034}{\textcolor{magenta}{[arXiv: hep-th/9812034]}}.



\bibitem{AFF}
V.~de~Alfaro, S.~Fubini and  G.~Furlan,
\emph{``Conformal invariance in quantum mechanics,"}
\href{https://link.springer.com/article/10.1007%2FBF02785666}{Nuovo Cim. A {\bf 34} (1976) 569}.


\bibitem{GPL}
A. Galajinsky, K.  Polovnikov  and O. Lechtenfeld,
\emph{``${\cal N}{=}\,4$  superconformal Calogero models,''}
\href{https://iopscience.iop.org/article/10.1088/1126-6708/2007/11/008/meta}{JHEP {\bf 11} (2007) 008}
\href{https://arxiv.org/abs/0708.1075}{\textcolor{magenta}{[arXiv: 0708.1075 [hep-th]]}}.

\bibitem{Kozyrev}
N.~Kozyrev, S.~Krivonos, O.~Lechtenfeld and A.~Sutulin,
\emph{``SU(2$|$1) supersymmetric mechanics on curved spaces,''}
\href{https://link.springer.com/article/10.1007%2FJHEP05%282018%29175}{JHEP \textbf{05} (2018) 175}
\href{https://arxiv.org/abs/1712.09898v1}{\textcolor{magenta}{[arXiv: 1712.09898 [hep-th]]}}.




\bibitem{Carter} 
  B.~Carter,
  \emph{``Axisymmetric black hole has only two degrees of freedom,''}
  \href{https://journals.aps.org/prl/abstract/10.1103/PhysRevLett.26.331}{Phys. Rev. Lett.  {\bf 26} (1971) 331}.


\bibitem{Gibbons}
G.~W.~Gibbons, R.~H.~Rietdijk and J.~W.~van Holten,
\emph{``SUSY in the sky,''}
\href{https://www.sciencedirect.com/science/article/abs/pii/0550321393904722?via%3Dihub}{Nucl. Phys. B \textbf{404} (1993) 42 }
\href{https://arxiv.org/abs/hep-th/9303112v1}{\textcolor{magenta}{[arXiv: hep-th/9303112]}}.




\bibitem{Cariglia2}
M.~Cariglia,
\emph{``Quantum mechanics of Yano tensors: Dirac equation in curved spacetime,''}
\href{https://iopscience.iop.org/article/10.1088/0264-9381/21/4/022}{Class. Quant. Grav. \textbf{21} (2004) 1051}
\href{https://arxiv.org/abs/hep-th/0305153v3}{\textcolor{magenta}{[arXiv: hep-th/0305153]}}. 




\bibitem{Frolov}
V.~P.~Frolov and D.~Kubiznak,
\emph{``Hidden symmetries of higher dimensional rotating black holes,''}
\href{https://journals.aps.org/prl/abstract/10.1103/PhysRevLett.98.011101}{Phys. Rev. Lett. \textbf{98} (2007) 011101}
\href{https://arxiv.org/abs/gr-qc/0605058v2}{\textcolor{magenta}{[arXiv: gr-qc/0605058]}}.

\bibitem{Frolov2}
V.~P.~Frolov and D.~Kubiznak,
\emph{``Higher-dimensional black holes: Hidden symmetries and separation of variables,''}
\href{https://iopscience.iop.org/article/10.1088/0264-9381/25/15/154005}{Class. Quant. Grav. \textbf{25} (2008) 154005}
\href{https://arxiv.org/abs/0802.0322v2}{\textcolor{magenta}{[arXiv: 0802.0322 [hep-th]]}}.

\bibitem{Frolov3}
V.~Frolov, P.~Krtous and D.~Kubiznak,
\emph{``Black holes, hidden symmetries, and complete integrability,''}
\href{https://link.springer.com/article/10.1007%2Fs41114-017-0009-9}{Living Rev. Rel. \textbf{20} (2017) no.1, 6}
\href{https://arxiv.org/abs/1705.05482}{\textcolor{magenta}{[arXiv: 1705.05482 [gr-qc]]}}.


\bibitem{tHooft}
G.'t Hooft, \emph{``Non-perturbative 2 particle scattering amplitudes in 2+1 dimensional quantum gravity,"}
\href{https://link.springer.com/article/10.1007/BF01218392}{Commun. Math. Phys. {\bf 117} (1988) 685}.


\bibitem{DesJack}
 S. Deser and R. Jackiw, 
  \emph{``Classical and quantum scattering on a cone,"}
\href{https://link.springer.com/article/10.1007/BF01466729}{Comm.
Math. Phys {\bf 118} (1988) 495}.


\bibitem{Kay}
 B.S. Kay, and  U.M. Studer,  \emph{``Boundary conditions for quantum mechanics on cones and fields around cosmic strings," }
\href{https://link.springer.com/article/10.1007%2FBF02102731}{Commun. Math. Phys. {\bf 139} (1991) 103}.

\bibitem{Furtado}
C.~Furtado and F.~Moraes,
\emph{``Harmonic oscillator interacting with conical singularities,''}
\href{https://iopscience.iop.org/article/10.1088/0305-4470/33/31/306}{J. Phys. A \textbf{33} (2000) 5513}.

\bibitem{Coelho}
J.~L.~A.~Coelho and R.~L.~P.~G.~Amaral,
\emph{``Coulomb and quantum oscillator problems in conical spaces with arbitrary dimensions,''}
\href{https://iopscience.iop.org/article/10.1088/0305-4470/35/25/307}{J. Phys. A \textbf{35} (2002) 5255}
\href{https://arxiv.org/abs/gr-qc/0111114}{\textcolor{magenta}{[arXiv: gr-qc/0111114]}}.


\bibitem{Barros}
C.~C.~Barros Jr,
\emph{``Quantum mechanics in curved space-time,"}
\href{https://link.springer.com/article/10.1140/epjc/s2005-02252-7}{Eur. Phys. J. C \textbf{42} (2005) 119}
\href{https://arxiv.org/abs/physics/0409064}{\textcolor{magenta}{[arXiv: 0409064 [physics.gen-ph]]}}.



\bibitem{Example1}
G.~De A.Marques, V.~B.~Bezerra and S.~G.~Fernandes,
\emph{``Exact solution of the Dirac equation for a Coulomb and scalar potentials in the gravitational field of a cosmic string,"}
 \href{https://www.sciencedirect.com/science/article/abs/pii/S0375960105005761}{Phys. Lett. A \textbf{341} (2005) 39}.
 

\bibitem{Kowalski}
K.~Kowalski and J.~Rembieli\'nski,
\emph{``On the dynamics of a particle on a cone,''}
\href{https://linkinghub.elsevier.com/retrieve/pii/S0003491612001649}{Annals of Physics \textbf{329} (2013) 146}
\href{https://arxiv.org/abs/1304.4412v2}{\textcolor{magenta}{[arXiv: 1304.4412 [quant-ph]]}}.

 
 
 
\bibitem{Example2}
F.~M.~Andrade and E.~O.~Silva,
\emph{``Effects of spin on the dynamics of the 2D Dirac oscillator in the magnetic cosmic string background,''}
\href{https://link.springer.com/article/10.1140/epjc/s10052-014-3187-6}{Eur. Phys. J. C \textbf{74} (2014)  3187}
\href{https://arxiv.org/abs/1403.4113}{\textcolor{magenta}{[arXiv: 1403.4113 [hep-th]]}}.


\bibitem{Example3}
M.~Hosseinpour, F.~M.~Andrade, E.~O.~Silva and H.~Hassanabadi, 
\emph{``Scattering and bound states for the Hulth\'en potential in a cosmic string background,"} 
\href{https://link.springer.com/article/10.1140/epjc/s10052-017-4834-5}{ Eur. Phys. J. C \textbf{77} (2017) 373}
\href{https://arxiv.org/abs/1608.03558}{\textcolor{magenta}{[arXiv: 1608.03558 [hep-th]]}}.

\bibitem{Example4}
F.~Ahmed,
\emph{``Relativistic quantum dynamics of spin-0 massive charged particle in the presence of external fields in 4D curved space-time with a cosmic string,''}
\href{https://link.springer.com/article/10.1140/epjp/s13360-020-00199-w}{Eur. Phys. J. Plus \textbf{135} (2020)  108}.
\href{https://arxiv.org/abs/1910.12700v1}{\textcolor{magenta}{[arXiv:1910.12700 [hep-th]]}}.


\bibitem{Kibble}
T.~Kibble,
\emph{``Topology of cosmic domains and strings,''}
\href{https://iopscience.iop.org/article/10.1088/0305-4470/9/8/029/meta}{J.\ Phys.\ A \textbf{9} (1976) 1387}.

\bibitem{Vilen}
A. Vilenkin, \emph{``Gravitational field of vacuum domain walls and strings,"}
\href{https://journals.aps.org/prd/abstract/10.1103/PhysRevD.23.852}{Phys. Rev. D {\bf 23} (1981) 852}.


\bibitem{Vilenkin}
A.~Vilenkin,
\emph{``Cosmic strings and domain walls,''}
\href{https://www.sciencedirect.com/science/article/abs/pii/037015738590033X}{Phys.\ Rept.\  \textbf{121} (1985)  263}.

\bibitem{Dow}
J. S. Dowker, 
\emph{``Quantum field theory on a cone,"}
\href{https://iopscience.iop.org/article/10.1088/0305-4470/10/1/023/meta}{J. Phys. A: Math. Gen. {\bf 10} (1977) 115}.

 
\bibitem{Visser}
M. Visser, 
 \emph{``Traversable wormholes: Some simple examples,"}
 \href{https://doi.org/10.1103/PhysRevD.39.3182}{Phys. Rev. D {\bf 39} (1989)  3182}.
 
 
\bibitem{Cramer}
J. G. Cramer, R. L. Forward, M. S. Morris, M. Visser, G. Benford, and G. A. Landis,
 \emph{``Natural wormholes as gravitational lenses,"}
\href{https://doi.org/10.1103/PhysRevD.51.3117}{Phys. Rev. D {\bf 51} (1995)  3117 }. 
 
 \bibitem{KatVol}
M. O. Katanaev and I. V. Volovich,
 \emph{``Theory of defects in solids and three-dimensional gravity,"}
 \href{https://www.sciencedirect.com/science/article/abs/pii/0003491652900407}{Annals Phys.
{\bf 216} (1992) 1}.

 
\bibitem{Volovik}
G. E. Volovik, \textit{The universe in a helium droplet} 
\href{https://oxford.universitypressscholarship.com/view/10.1093/acprof:oso/9780199564842.001.0001/acprof-9780199564842}{(Oxford Science Publications, 2003)}. 
 
\bibitem{Manton}
N. S. Manton, 
 \emph{``Five vortex equations,"}
 \href{https://iopscience.iop.org/article/10.1088/1751-8121/aa5f19/meta}{
J. Phys. A {\bf 50}  (2017)  125403} 
\href{https://arxiv.org/abs/1612.06710}{\textcolor{magenta}{[arXiv: 1612.06710 [hep-th]]}}.


  \bibitem{SokSta}
 D. D. Sokolov  and A. A. Starobinsky,
  \emph{``On the structure of curvature tensor on conical singularities,"} 
\href{http://www.mathnet.ru/php/archive.phtml?wshow=paper&jrnid=dan&paperid=41048&option_lang=eng}{Dokl. Akad. Nauk 
{\bf 234} 
 (1977) 1043 [Sov. Phys. - Dokl. {\bf  22}  (1977) 312]}.

\bibitem{Aryal}
M.~Aryal, L.~H.~Ford and A.~Vilenkin,
\emph{``Cosmic strings and black holes,''}
\href{https://journals.aps.org/prd/abstract/10.1103/PhysRevD.34.2263}{Phys. Rev. D \textbf{34} (1986)  2263}.


\bibitem{LandCon}
C. Furtado,  B. G. C. da Cunha, F.  Moraes,   E. R. B. de Mello,  V. B.  Bezzerra, 
\emph{``Landau levels in the presence of disclinations,''}
\href{https://www.sciencedirect.com/science/article/abs/pii/0375960194904324}{ Phys.  Lett.  A  {\bf 195}  (1994) 90}.  

\bibitem{Solod}
S. N. Solodukhin,
\emph{``Conical singularity and quantum corrections to the entropy of a black hole,''}
\href{https://journals.aps.org/prd/abstract/10.1103/PhysRevD.51.609}{Phys. Rev. D {\bf 51} (1995)  609}
\href{https://arxiv.org/abs/hep-th/9407001}{\textcolor{magenta}{[arXiv: hep-th/9407001]}}.



\bibitem{Germano}
M. G. Germano, V. B. Bezerra, and E.~R.~B.~de Mello,
\emph{``Gravitational effects due to a cosmic string in Schwarzschild spacetime,"} 
\href{https://iopscience.iop.org/article/10.1088/0264-9381/13/10/006/meta}{Class. Quant. Grav. \textbf{13}  (1996) 2663}.


\bibitem{deMello}
E.~R.~B.~de Mello and A.~A.~Saharian,
\emph{``Vacuum polarization induced by a cosmic string in anti-de Sitter spacetime,''}
\href{https://iopscience.iop.org/article/10.1088/1751-8113/45/11/115402}{J. Phys. A \textbf{45} (2012)  115002}
\href{https://arxiv.org/abs/1110.2129}{\textcolor{magenta}{[arXiv: 1110.2129 [hep-th]]}}.


\bibitem{Nied2}
U. Niederer, 
 \emph{``The maximal kinematical invariance group of the harmonic oscillator,"}
\href{https://www.e-periodica.ch/digbib/view?pid=hpa-001:1973:46::960#201}{Helv. Phys. Acta {\bf 46} (1973) 191}.

\bibitem{NewHook}
P. D. Alvarez, J. Gomis, K.  Kamimura,  and M. S. Plyushchay,
 \emph{``(2+1)D Exotic Newton-Hooke symmetry, duality and projective phase,"}
 \href{https://www.sciencedirect.com/science/article/abs/pii/S0003491607000425?via%3Dihub}{Annals Phys. {\bf 322} (2007) 1556}
  \href{https://arxiv.org/abs/hep-th/0702014}{\textcolor{magenta}{[arXiv: hep-th/0702014]}}.


  \bibitem{NH1}
 A.~Galajinsky,
 \emph{ ``Conformal mechanics in Newton-Hooke spacetime,''
 }
 \href{https://www.sciencedirect.com/science/article/pii/S0550321310001173?via%3Dihub}{Nucl.\ Phys.\ B {\bf 832} (2010) 586}
    \href{https://arxiv.org/abs/1002.2290}{\textcolor{magenta}{[arXiv: 1002.2290 [hep-th]]}}.

  
 \bibitem{NH2} 
 K.~Andrzejewski,
 \emph{ ``Conformal Newton-Hooke algebras, Niederer's transformation and Pais-Uhlenbeck oscillator,''}
  \href{https://www.sciencedirect.com/science/article/pii/S037026931400731X?via%3Dihub}{Phys.\ Lett.\ B {\bf 738} (2014) 405}
  \href{https://arxiv.org/abs/1409.3926}{\textcolor{magenta}{[arXiv: 1409.3926 [hep-th]]}}.

    

\bibitem{IPW}
L. Inzunza, M.  S. Plyushchay, and A. Wipf,
\emph{``Conformal bridge between asymptotic freedom and confinement,"}
\href{https://journals.aps.org/prd/abstract/10.1103/PhysRevD.101.105019}{Phys. Rev. D {\bf 101}  (2020)  105019}
  \href{https://arxiv.org/abs/1912.11752}{\textcolor{magenta}{[arXiv: 1912.11752 [hep-th]]}}.


\bibitem{Nied1}
U. Niederer,
\emph{``Maximal kinematical invariance group of the free Schr\"odinger equation,"}
\href{https://www.e-periodica.ch/digbib/view?pid=hpa-001:1972:45::1246#808}{Helv. Phys. Acta {\bf 45} (1972) 802}.


\bibitem{Hagen}
C. R. Hagen, (1972). 
 \emph{``Scale and conformal transformations in Galilean-covariant field Theory,"} 
\href{https://journals.aps.org/prd/abstract/10.1103/PhysRevD.5.377}{Phys. Rev. D  {\bf 5} (1972) 377}. 

\bibitem{Barut}
A. O. Barut, 
 \emph{``Conformal group $\rightarrow$ Schr\"odinger group
 $\rightarrow$
 dynamical group - the maximal kinematical group of the massive
Schr\"odinger particle,"} 
\href{https://www.e-periodica.ch/digbib/view?pid=hpa-001:1973:46::981#506}{Helv. Phys. Acta {\bf 46} (1973) 496}.

\bibitem{LeiPly}
C. Leiva and M. S. Plyushchay,
 \emph{``Conformal symmetry of relativistic and nonrelativistic systems and AdS/CFT correspondence,"} 
 \href{https://www.sciencedirect.com/science/article/abs/pii/S0003491603001180?via%3Dihub}{Annals Phys. {\bf 307} (2003) 372}
   \href{https://arxiv.org/abs/hep-th/0301244}{\textcolor{magenta}{[arXiv: hep-th/0301244]}}.

\bibitem{HenUnter}
M. Henkel and J. Unterberger,
\emph{``Schr\"odinger invariance and space-time symmetries,"}
\href{https://www.sciencedirect.com/science/article/abs/pii/S0550321303002529?via%3Dihub}{Nucl. Phys. B {\bf 660} (2003) 407}
   \href{https://arxiv.org/abs/hep-th/0302187}{\textcolor{magenta}{[arXiv: hep-th/0302187]}}.



\bibitem{Son}
D. T. Son,
 \emph{``Toward an AdS/cold atoms correspondence: 
A geometric realization of the Schr\"odinger symmetry,"} 
\href{https://journals.aps.org/prd/abstract/10.1103/PhysRevD.78.046003}{ Phys. Rev. D  {\bf 78}  (2008)  046003}
 \href{https://arxiv.org/abs/0804.3972}{\textcolor{magenta}{[arXiv: 0804.3972 [hep-th]]}}.

\bibitem{BagGop}
A. Bagchi and R. Gopakumar, 
 \emph{``Galilean conformal algebras and AdS/CFT,"} 
\href{https://iopscience.iop.org/article/10.1088/1126-6708/2009/07/037/meta}{JHEP {\bf 0907} (2009) 037}
\href{https://arxiv.org/abs/0902.1385}{\textcolor{magenta}{[arXiv: 0902.1385 [hep-th]]}}.



 
\bibitem{sl2R}
M. S. Plyushchay,
  \emph{``Quantization of the classical $SL(2,R)$ system and representations of $\overline{SL(2,R)}$ group,"}
  \href{https://aip.scitation.org/doi/10.1063/1.530016}{J. Math. Phys. {\bf 34} (1993) 3954 }. 

\bibitem{Barg}
V. Bargmann,
  \emph{``Irreducible unitary representations of the Lorentz group,"}
  \href{https://www.jstor.org/stable/1969129?seq=1}{ Annals of Mathematics, {\bf 48} (1947) 568}.



\bibitem{Dirac}
P. A. M. Dirac,
 \emph{``Forms of relativistic dynamics,"}
  \href{https://journals.aps.org/rmp/abstract/10.1103/RevModPhys.21.392}{Rev. Mod. Phys. {\bf 21} (1949) 392}.


\bibitem{CJP}
F. Correa, V. Jakubsky and M. S. Plyushchay, 
 \emph{``PT-symmetric invisible defects and
confluent Darboux-Crum transformations,"}
\href{https://journals.aps.org/pra/abstract/10.1103/PhysRevA.92.023839}{Phys. Rev. A {\bf 92} (2015) 023839}
\href{https://arxiv.org/abs/1506.00991}{
\textcolor{magenta}{[arXiv: 1506.00991 [hep-th]]}}.


\bibitem{InzPly3}
L.~Inzunza and M.~S.~Plyushchay,
\emph{``Klein four-group and Darboux duality in conformal mechanics,''}
\href{https://journals.aps.org/prd/abstract/10.1103/PhysRevD.99.125016}{Phys. Rev. D \textbf{99} (2019)  125016}
\href{https://arxiv.org/abs/1902.00538}{
\textcolor{magenta}{[arXiv: 1902.00538 [hep-th]]}}.


\bibitem{CroRit}
M. de Crombrugghe and  V. Rittenberg, 
\emph{``Supersymmetric quantum mechanics,''}
\href{https://www.sciencedirect.com/science/article/abs/pii/0003491683903160?via%3Dihub}{Annals Phys. 
{\bf 151} (1983) 99}.


\bibitem{HiddenSUSY1}
M.~S.~Plyushchay,
\emph{``Deformed Heisenberg algebra, fractional spin fields and supersymmetry without fermions,''}
\href{https://www.sciencedirect.com/science/article/abs/pii/S0003491696900123?via%3Dihub}{Annals Phys. \textbf{245} (1996) 339}
\href{https://arxiv.org/abs/hep-th/9601116}{\textcolor{magenta}{[arXiv: hep-th/9601116]}}.


\bibitem{HiddenSUSY+}
M.~S.~Plyushchay,
\emph{``Hidden nonlinear supersymmetries in pure parabosonic systems,''}
\href{https://www.worldscientific.com/doi/abs/10.1142/S0217751X00001981}{Int. J. Mod. Phys. A {\bf 15}
 (2000) 3679} 
 \href{https://arxiv.org/abs/hep-th/9903130}{\textcolor{magenta}{[arXiv: hep-th/9903130]}}.


\bibitem{HiddenSUSY++}
F. Correa and  M. S. Plyushchay, 
\emph{``Hidden supersymmetry in quantum bosonic systems,''}
\href{https://www.sciencedirect.com/science/article/abs/pii/S0003491606002831?via%3Dihub}{ Annals Phys. {\bf 322} (2007) 2493}
\href{https://arxiv.org/abs/hep-th/0605104}{\textcolor{magenta}{[arXiv: hep-th/0605104]}}.


\bibitem{HiddenSUSY2}
V.~Jakubsky, L.~M.~Nieto and M.~S.~Plyushchay,
\emph{``The origin of the hidden supersymmetry,''}
\href{https://www.sciencedirect.com/science/article/pii/S0370269310008270?via%3Dihub}{Phys. Lett. B \textbf{692} (2010) 51}
\href{https://arxiv.org/abs/1004.5489}{\textcolor{magenta}{[arXiv: 1004.5489 [hep-th]]}}.

\bibitem{HidConf}
R. Bonezzi, O. Corradini, E. Latini, and A. Waldron,
\emph{``Quantum mechanics and hidden superconformal symmetry,''}
\href{https://journals.aps.org/prd/abstract/10.1103/PhysRevD.96.126005}{Phys. Rev. D {\bf 96}  (2017) 126005 }
\href{https://arxiv.org/abs/1709.10135}{\textcolor{magenta}{[arXiv: 1709.10135 [hep-th]]}}.

\bibitem{HiddenSUSY3}
L.~Inzunza and M.~S.~Plyushchay,
\emph{``Hidden superconformal symmetry: Where does it come from?,''}
\href{https://journals.aps.org/prd/abstract/10.1103/PhysRevD.97.045002}{Phys. Rev. D \textbf{97} (2018) 045002}
\href{https://arxiv.org/abs/1711.00616}{\textcolor{magenta}{[arXiv: 1711.00616 [hep-th]]}}.


\bibitem{Crampin}
M. Crampin, \emph{``Hidden symmetries and Killing tensors," }
\href{https://www.sciencedirect.com/science/article/abs/pii/0034487784900697}{Reports on Math. Phys \textbf{20}  (1984) 31}.


\bibitem{Torsion}
M. S. Plyushchay,
\emph{``The model of relativistic particle with torsion,''}
\href{https://www.sciencedirect.com/science/article/abs/pii/055032139190555C?via%3Dihub}{Nucl. Phys. B {\bf 362} (1991) 54 }.


\bibitem{NonCom2}
P.~A.~Horvathy and M.~S.~Plyushchay,
\emph{``Non-relativistic anyons, exotic Galilean symmetry and noncommutative plane,''}
\href{https://iopscience.iop.org/article/10.1088/1126-6708/2002/06/033}{JHEP \textbf{06} (2002) 033}
\href{https://arxiv.org/abs/hep-th/0201228}{\textcolor{magenta}{[arXiv: hep-th/0201228]}}.


\bibitem{NonComPLB}
P.~A.~Horvathy and M.~S.~Plyushchay,
\emph{``Anyon wave equations and the noncommutative plane,''}
\href{https://www.sciencedirect.com/science/article/pii/S0370269304008378?via%3Dihub}{Phys. Lett. B {\bf  595} (2004) 547}
\href{https://arxiv.org/abs/hep-th/0404137}{\textcolor{magenta}{[arXiv: hep-th/0404137]}}.


\bibitem{LeiMyr} 
  J.~M.~Leinaas and J.~Myrheim,
 \emph{``On the theory of identical particles,''}
  \href{https://link.springer.com/article/10.1007%2FBF02727953}{Nuovo Cim.\ B {\bf 37} (1977) 1}.
   
   \bibitem{LeiMyr+}
   J.~M.~Leinaas and J.~Myrheim,
  \emph{ ``Intermediate statistics for vortices in superfluid films,''}
   \href{https://journals.aps.org/prb/abstract/10.1103/PhysRevB.37.9286}{Phys.\ Rev.\ B {\bf 37}  (1988)  9286}.


\bibitem{MKWil} 
  R.~MacKenzie and F.~Wilczek,
  \emph{ ``Peculiar spin and statistics in two space dimensions,''}
   \href{https://www.worldscientific.com/doi/abs/10.1142/S0217751X88001181}{Int.\ J.\ Mod.\ Phys.\ A {\bf 3}  (1988)  2827}.


\bibitem{Bender}
C.~M.~Bender and S.~Boettcher,
\emph{``Real spectra in non-Hermitian Hamiltonians having PT symmetry,''}
\href{https://journals.aps.org/prl/abstract/10.1103/PhysRevLett.80.5243}{Phys. Rev. Lett. {\bf 80}  (1998)  5243}
\href{https://arxiv.org/abs/physics/9712001}{\textcolor{magenta}{[arXiv: physics/9712001].}}


\bibitem{Mostafazadeh}
A.~Mostafazadeh,
\textit{``Pseudo-Hermiticity versus PT symmetry. The necessary condition for the reality of the spectrum,''}
\href{https://aip.scitation.org/doi/10.1063/1.1418246}{J. Math. Phys. \textbf{43} (2002) 205}
\href{https://arxiv.org/abs/math-ph/0107001}{\textcolor{magenta}{
[arXiv: math-ph/0107001]}}.


\bibitem{Bender2007}
C.~M.~Bender,
\textit{``Making sense of non-Hermitian Hamiltonians,''}
\href{https://iopscience.iop.org/article/10.1088/0034-4885/70/6/R03}{Rept. Prog. Phys. \textbf{70} (2007) 947}
\href{https://arxiv.org/abs/hep-th/0703096}{\textcolor{magenta}{[arXiv: hep-th/0703096 [hep-th]]}}.


\bibitem{Fring}
A.~Fring,
\textit{``PT-symmetric deformations of integrable models,''}
\href{https://royalsocietypublishing.org/doi/10.1098/rsta.2012.0046}{Phil. Trans. Roy. Soc. Lond. A \textbf{371} (2013)  20120046}
\href{https://arxiv.org/abs/1204.2291}{\textcolor{magenta}{[arXiv: 1204.2291 [hep-th]]}}.




\bibitem{DCT}
P. Dorey, C. Dunning and  R. Tateo, 
\textit{``From PT-symmetric quantum mechanics to conformal field theory,''}
\href{https://link.springer.com/article/10.1007%2Fs12043-009-0114-8}{Pramana {\bf 73} (2009) 217}
\href{https://arxiv.org/abs/0906.1130}{\textcolor{magenta}{[arXiv: 0906.1130 [hep-th]]}}.   

\bibitem{MatMik1}
J.~Mateos Guilarte and M.~S.~Plyushchay,
\textit{``Perfectly invisible $\mathcal{PT}$-symmetric zero-gap systems, conformal 
field theoretical kinks, and exotic nonlinear supersymmetry,''}
\href{https://link.springer.com/article/10.1007/JHEP12(2017)061}{JHEP {\bf 12} (2017)  061 }
\href{https://arxiv.org/abs/1710.00356}{\textcolor{magenta}{[arXiv: 1710.00356 [hep-th]]}}.

\bibitem{MatMik2}
J.~Mateos Guilarte and M.~S.~Plyushchay,
\textit{`` Nonlinear symmetries of perfectly invisible PT-regularized conformal and superconformal 
mechanics systems,''}
\href{https://link.springer.com/article/10.1007%2FJHEP01%282019%29194}{JHEP {\bf 01} (2019) 194} 
\href{https://arxiv.org/abs/1806.08740}{\textcolor{magenta}{[arXiv: 1806.08740 [hep-th]]}}.



\end{thebibliography}
\end{document}